\documentclass{emulateapj}
\usepackage{amsmath,amssymb}
\usepackage{xspace}
\usepackage{graphicx}
\usepackage{epstopdf}  
\usepackage{graphicx}
\usepackage{epsfig}
\usepackage{verbatim}
\usepackage{morefloats} % somehow need this to have lots of figures
\usepackage[colorlinks,urlcolor=blue,citecolor=black,linkcolor=blue]{hyperref}
\usepackage{CJK} 

\def\beq{\begin{equation}}
\def\eeq{\end{equation}}
\def\gammaint{\Gamma_{\rm int}}
\def\gammaeff{\Gamma_{\rm eff}}
\def\ergcms{erg~cm$^{-2}$~s$^{-1}$}
\def\ergs{erg~s$^{-1}$}
\def\nh{N_{\rm H}}
\def\fx{f_{\rm 0.5-8\ keV}}
\def\fxint{f_{\rm 0.5-8\ keV,corr}}
\def\lx{L_{\rm 0.5-8\ keV}}
\def\lxint{L_{\rm 0.5-8\ keV,corr}}
\def\msun{M_{\odot}}
\def\lsun{L_{\odot}}
\def\fagn{f_{\rm AGN}}
\def\cdfs{\mbox{CDF-S}} % \mbox prevents line break
\def\cdfn{\mbox{CDF-N}} 
\def\ecdfs{\mbox{E-CDF-S}}
\def\chandra{{\it Chandra}}
\def\xmm{{XMM-{\it Newton}}}

\def\degree{^{\circ}}
\def\leq{\leqslant}
\def\geq{\geqslant}
\def\micron{$\mu$m}
\def\percm2{{\rm cm^{-2}}}
\def\peryr{yr$^{-1}$}
\def\xray{\mbox{X-ray}}
\def\fbrange{\hbox{0.5--8.0~keV}}
\def\rri{r_{\rm RI}}

\slugcomment{}
\shorttitle{}
\shortauthors{}

\begin{document}

\begin{CJK*}{UTF8}{gbsn}

%\title{X-ray Properties of the Submillimeter Galaxies Seen by ALMA\\
%  in the Extended {\it Chandra} Deep Field-South}

% per Ian
\title{An ALMA Survey of Submillimeter Galaxies in the Extended
  Chandra Deep Field-South: The AGN Fraction and X-ray Properties of
  Submillimeter Galaxies}

% leading authors
\author{S.~X.~Wang (王雪凇)\altaffilmark{1}}
\author{W.~N.~Brandt\altaffilmark{1,2}}
\author{B.~Luo\altaffilmark{1,2}}
\author{I.~Smail\altaffilmark{3}}
% comments and data contribution
\author{D.~M.~Alexander\altaffilmark{3}}
\author{A.~L.~R.~Danielson\altaffilmark{3}}
\author{J.~A.~Hodge\altaffilmark{4}}
\author{A.~Karim\altaffilmark{3,5}}
\author{B.~D.~Lehmer\altaffilmark{6,7}}
\author{J.~M.~Simpson\altaffilmark{3}}
\author{A.~M.~Swinbank\altaffilmark{3}}
\author{F.~Walter\altaffilmark{4}}
\author{J.~L.~Wardlow\altaffilmark{8}}
\author{Y.~Q.~Xue\altaffilmark{9}}
% comments
\author{S.~C.~Chapman\altaffilmark{10,11}}
\author{K.~E.~K.~Coppin\altaffilmark{12}}
\author{H.~Dannerbauer\altaffilmark{13}}
\author{C.~De Breuck\altaffilmark{14}}
\author{K.~M.~Menten\altaffilmark{15}}
\author{P.~van der Werf\altaffilmark{16}}

\altaffiltext{1}{Department of Astronomy \& Astrophysics, 525 Davey
  Lab, The Pennsylvania State University, University Park, PA 16802, USA;
  Send correspondence to xxw131@psu.edu and niel@astro.psu.edu}
\altaffiltext{2}{Institute for Gravitation and the Cosmos, The
  Pennsylvania State University, University Park, PA 16802, USA}
\altaffiltext{3}{Institute for Computational Cosmology, Durham
  University, South Road, Durham, DH1 3LE, UK} 
\altaffiltext{4}{Max-Planck Institute for Astronomy, K\"onigstuhl 17,
  D-69117 Heidelberg, Germany}
\altaffiltext{5}{Argelander-Institute of Astronomy, Bonn University,
  Auf dem H\"ugel 71, D-53121 Bonn, Germany} 
\altaffiltext{6}{The Johns Hopkins University, Homewood Campus,
  Baltimore, MD 21218, USA}
\altaffiltext{7}{NASA Goddard Space Flight Center, Code 662,
  Greenbelt, MD 20771, USA}
\altaffiltext{8}{Department of Physics \& Astronomy, University of
  California, Irvine, CA 92697, USA}
\altaffiltext{9}{Key Laboratory for Research in Galaxies and
  Cosmology, Center for Astrophysics, Department of Astronomy,
  University of Science and Technology of China, Chinese Academy of
  Sciences, Hefei, Anhui 230026, China} 
\altaffiltext{10}{Institute of Astronomy, University of Cambridge,
  Madingley Road, Cambridge CB3 0HA, UK} 
\altaffiltext{11}{Department of Physics and Atmospheric Science,
  Dalhousie University, Coburg Road Halifax, B3H 4R2, Canada} 
\altaffiltext{12}{Centre for Astrophysics, Science \& Technology
  Research Institute, University of Hertfordshire, Hatfield AL10 9AB, UK} 
\altaffiltext{13}{Universit\"at Wien, Institute f\"ur Astrophysik,
  T\"urkenschanzstra\ss e 17, 1180 Wien, Austria}
\altaffiltext{14}{European Southern Observatory, Karl-Schwarzschild
  Stra\ss e 2, D-85748 Garching, Germany} 
\altaffiltext{15}{Max-Planck-Institut f\"ur Radioastronomie, Auf dem
  H\"ugel 69, D-53121 Bonn, Germany}
\altaffiltext{16}{Leiden Observatory, Leiden University, PO Box 9513,
  NL - 2300 RA Leiden, The Netherlands}

%%%%%%%%%%%%%%%%%%%%%%%%%%%%%%%%%%%%%%%%%%%%%%%%%%%%%%%%%%%%%%%%%%%%%%%%%%%%%%%%%%%%%%%%%%%%%%%%%%%%
\begin{abstract}

The large gas and dust reservoirs of submm galaxies (SMGs) could potentially 
provide ample fuel to trigger an Active Galactic Nucleus (AGN), but previous 
studies of the AGN fraction in SMGs have been controversial largely due to 
the inhomogeneity and limited angular resolution of the available submillimeter 
surveys. Here we set improved constraints on the AGN fraction and \xray\ 
properties of the SMGs with ALMA and \chandra\ observations in the Extended 
\chandra\ Deep Field-South (\ecdfs). This study is the first among similar works 
to have unambiguously identified the \xray\ counterparts of SMGs; this is
accomplished using the fully submm-identified, statistically reliable SMG 
catalog with 99 SMGs from the ALMA LABOCA \ecdfs\ Submillimeter Survey (ALESS). 
We found 10 \xray\ sources associated with SMGs (median redshift $z=2.3$), 
of which 8 were identified as AGNs using several techniques that enable
cross-checking. The other 2 \xray\ detected SMGs have levels of \xray\ 
emission that can be plausibly explained by their star-formation activity. 
6 of the 8 SMG-AGNs are moderately/highly absorbed, with $\nh > 10^{23}\ \percm2$. 
An analysis of the AGN fraction, taking into account the spatial variation 
of \xray\ sensitivity, yields an AGN fraction of $17^{+16}_{-6}\%$ for AGNs 
with rest-frame \hbox{0.5--8~keV} absorption-corrected luminosity 
$\geq 7.8 \times 10^{42}$ \ergs; we provide estimated AGN fractions as a 
function of \xray\ flux and luminosity. ALMA's high angular resolution also 
enables direct \xray\ stacking at the precise positions of SMGs for the first 
time, and we found 4 potential SMG-AGNs in our stacking sample.

\end{abstract}  

%%%%%%%%%%%%%%%%%%%%%%%%%%%%%%%%%%%%%%%%%%%%%%%%%%%%%%%%%%%%%%%%%%%%%%%%%%%%%%%%%%%%%%%%%%%%%%%%%%%%
\section{Introduction}\label{sec:intro}

% introduction to SMGs for the X-ray astronomers
Over the past 15 yr, submillimeter (submm) and millimeter surveys have
discovered a population of far-infrared (FIR) luminous,
dust-enshrouded galaxies at $z>1$ (e.g., \citealt{Smail1997,
  Ivison1998, Ivison2000, Coppin2006, Weiss2009,
  Austermann2010}). Multiwavelength follow-up observations of these
submm galaxies (e.g., \citealt{Valiante2007, Pope2008,
  Menendez-Delmestre2007, Menendez-Delmestre2009}) have revealed that
they are among the most luminous objects in the Universe
\citep[e.g.,][]{Ivison2002,Chapman2002,Kovacs2006}, and that they
contribute significantly to the total cosmic star formation around
$z\sim 2$ (e.g., \citealt{Hughes1998, Barger1998, Perez-Gonzalez2005,
  Aretxaga2007, Hopkins2010}). These submm galaxies (SMGs) typically
have infrared (IR) luminosities of $\sim 10^{12}\ \lsun$ or even
greater, and their star formation rates (SFR) are estimated to be
\mbox{$\sim 100$--$1000\ \msun$\peryr}\ (e.g., \citealt{Kovacs2006,
  Coppin2008, Magnelli2012}). They are massive galaxies with stellar
mass $M_*\sim 10^{11}\ \msun$ or greater (e.g.,
\citealt{Borys2005,Xue2010,Hainline2011}) and with large reservoirs of
cold gas ($\gtrsim 10^{10}\ \msun$; e.g., \citealt{Bothwell2013}).

% Compare with ULIRGs
Most commonly found around $z\sim 2$--$3$, the volume density of SMGs
is $\sim1000$ times larger (e.g., \citealt{Chapman2003, Chapman2005,
  Wardlow2011}) than that of the local ultraluminous infrared galaxies
(ULIRGs), which are relatively rare in the local universe (e.g.,
\citealt{Sanders1996, Lonsdale2006}). Also qualified as ULIRGs
($L_{\rm IR}>10^{12}\ L_\odot$; \citealt{Sanders1996}), SMGs are often
considered as the ``distant cousins" of local ULIRGs, typically
exhibiting similarly high SFR and IR luminosity. However, they also
differ in some important ways. The more strongly star-forming SMGs are
not simply the ``scaled up" versions of local ULIRGs --- for example,
it appears that the star formation in SMGs occurs on a larger scale
within the galaxy instead of being concentrated at the core like for
the local ULIRGs \citep[e.g.,][]{Chapman2004,Coppin2012}.

% introduction to AGN-SMG links to the non-AGN astronomers
Believed to be the progenitors of large local elliptical galaxies
(e.g., \citealt{Lilly1999, Smail2004, Chapman2005}) and often involved
in mergers (e.g., \citealt{Tacconi2008, Engel2010, Magnelli2012}),
SMGs present a unique opportunity for studying the co-evolution of
galaxies and their central supermassive black holes (SMBHs; $M\geq
10^6\msun$).
The cosmic star formation rate and active galactic nucleus (AGN)
activity both peak around $z\sim 2$ \citep{Connolly1997,
  Merloni2004, Hopkins2007, Cucciati2012}, and they appear to be
related as suggested by the observed correlations between the
properties of central SMBHs and their host galaxies
(e.g., the $M$-$\sigma$ and the $M$-$L$ relation;
\citealt{Ferrarese2000, Gebhardt2000, Haring2004, Gultekin2009}).
Moreover, simulations of galaxy evolution and SMBH growth show that
merger events can trigger both star-formation activity and the onset
of powerful AGN, with the peak of the AGN activity (possibly a quasar
phase) coming shortly after the peak epoch of star formation (e.g.,
\citealt{Hopkins2008, Narayanan2010}). Observationally, recent studies
suggest that luminous AGNs are more prevalent in massive galaxies
(e.g., \citealt{Xue2010, Mullaney2012}) and star-forming galaxies
(e.g., \citealt{Rafferty2011, Santini2012, Rosario2013, Chen2013}),
and a very high fraction of local ULIRGs exhibit AGN activity as
indicated by line-ratio diagnostics (see the review by
\citealt{Alonso-Herrero2013} and references therein).

% previous studies, overall
AGN activity in SMGs has been identified in previous studies through
mid-IR spectroscopy (e.g., \citealt{Valiante2007, Pope2008, Menendez-Delmestre2007,
  Menendez-Delmestre2009, Coppin2010}) or
\xray\ \citep[e.g.,][]{Alexander2005a, Alexander2005b, Pope2006, Laird2010,
  Lutz2010, Georgantopoulos2011, Gilli2011, Hill2011, Bielby2012,
  Johnson2013} observations. For moderate-to-high \xray\ luminosity
AGNs, the \xray\ emission is arguably the best AGN indicator as the
hard \mbox{X-rays} (rest-frame energies of 2--30 keV) can penetrate through
obscuration ($\nh \lesssim$$10^{24}$ cm$^{-2}$) and also suffer less
from host-galaxy contamination. However, for less \xray\ luminous
sources, the contribution from high mass \xray\ binaries (HMXBs) in the
host galaxies cannot be neglected, especially for extreme starburst
galaxies like SMGs (e.g., \citealt{Alexander2005b}). The studies of
\cite{Alexander2005a, Alexander2005b}, \cite{Pope2006}, \cite{Laird2010},
\cite{Georgantopoulos2011}, and \cite{Johnson2013} have all found that
SMGs have a high \xray\ detection rate, and a significant fraction of
the \xray\ detected SMGs are AGN-dominated in the \xray\ band (though
the exact fraction is under debate) while some are
consistent with the \xray\ emission being powered purely by the
starburst.

% previous studies, the important four
All focusing on \xray\ AGNs, \cite{Alexander2005a, Alexander2005b},
\cite{Laird2010}, \cite{Georgantopoulos2011}, and \cite{Johnson2013}
reported AGN fractions among SMGs that are consistent with each other
within their 1$\sigma$ error bars. The pioneering work by
\cite{Alexander2005a, Alexander2005b} studied the submm sources
discovered by SCUBA \citep{Holland1999} in the \chandra\ Deep Field
North (\mbox{CDF-N}), which were matched to radio counterparts and
spectroscopically identified \citep{Chapman2005}. They estimated the
\xray\ AGN fraction among SMGs to be
$>38^{+12}_{-10}\%$. \cite{Laird2010}, also using submm sources in the
CDF-N but with \textit{Spitzer} IR counterparts identified by
\cite{Pope2006}, reported an \xray\ AGN fraction of 29\%$\pm 7\%$ (or
20\% if being conservative about AGN
classification). \cite{Georgantopoulos2011} studied the submm sources
in the Extended \chandra\ Deep Field South (\ecdfs) detected by the
LABOCA \ecdfs\ Submm Survey (LESS; \citealt{Weiss2009}), which were
matched to 2 Ms \cdfs\ \citep{Luo2008} and 250 ks
\ecdfs\ \citep{Lehmer2005} sources and also {\it Spitzer} MIPS sources
\citep{Magnelli2009}, and they found an \xray\ AGN fraction of $18\pm
7\%$ among the SMGs. \cite{Johnson2013} performed a direct matching
between submm sources (detected at 1.1 mm by AzTEC;
\citealt{Wilson2008}) and \xray\ sources instead of first matching
SMGs to IR or radio counterparts, and they found that, for SMGs in the
\cdfs\ and CDF-N, the AGN fraction is about 28\%.

% Problems with previous studies
Though previous studies were thorough with their statistical analyses
on the reliability of counterpart matching and used supplementary IR
or radio catalogs, they were largely limited by the uncertainties in
finding the true \xray\ counterparts of the SMGs. The submm source
catalogs used in \cite{Alexander2005a, Alexander2005b},
\cite{Laird2010}, \cite{Georgantopoulos2011}, and \cite{Johnson2013}
are all from single-dish submm surveys, which have a typical angular
resolution of $\sim 10\arcsec$--$20\arcsec$ (e.g.,
\citealt{Chapman2005,Weiss2009}). This poses great challenges for
matching submm sources to the IR/radio/\xray\ sources, especially when
multiple multiwavelength counterparts are found within the large
search apertures. Furthermore, a large fraction of the single-dish
detected submm sources are actually found to resolve into multiple
sources, either physically unrelated or due to the clustering of SMGs,
when observed with higher angular resolution instruments such as the
Submm Array (SMA) and the Atacama Large Millimeter/submm Array (ALMA;
e.g., \citealt{Wang2011, Barger2012, Hodge2013}).

% advantage of our work; intro to LESS and ALESS
In this paper, we present the \xray\ properties and the AGN fraction
of the SMGs in the \ecdfs\ detected by the ALMA LABOCA \ecdfs\ Submm
Survey (ALESS; \citealt{Hodge2013,Karim2013}). The ALESS is an ALMA
Cycle 0 survey at 870~\micron\ to follow up 122 of the original 126
submm sources detected by LESS, which is the largest and the most
homogeneous 870 $\mu$m survey to date \citep{Weiss2009}. With the
exquisite angular resolution and great sensitivity of ALMA ($\sim
1.5\arcsec$ and 3$\times$ deeper than LESS; \citealt{Hodge2013}),
ALESS provides the \emph{first} fully submm-identified sample of SMGs based
on a large, contiguous, and well-defined survey (LESS), and this
enables robust counterpart matching at other wavelengths. Pairing with
the powerful ALESS catalog, we use the deep \chandra\ data in the
\ecdfs\ region (\citealt{Lehmer2005}; L05), including the most
sensitive \xray\ survey to date, the 4~Ms \cdfs\ survey
(\citealt{Xue2011}; X11). Combining the power of \chandra\ and ALMA,
we have unambiguously identified the \xray\ counterparts by matching
the \xray\ sources directly onto the submm positions, which is the
first among similar studies.

% clarification of what is AGN and luminosity-dependent fAGN analyses
% per Ian and Niel's comments
The AGN fractions in SMGs presented in this work are in the form of
cumulative fractions as a function of \xray\ flux/luminosity (i.e.,
the fraction of SMGs hosting AGN with \xray\ flux/luminosity larger
than or equal to a given value). Here we define an AGN as an
accreting SMBH with any level of \xray\ luminosity. Identification of
an AGN inside an SMG does not mean the AGN is the main power source of
the SMG or contributes significantly to the galaxy's energy
budget. Though some SMGs are quasar powered, much evidence has shown
that in the majority of SMGs, star formation is the dominant energy
source \citep[e.g.,][]{Chapman2004, Alexander2005b, Pope2006}. SMGs with AGN
signatures (e.g., in the \xray\ or IR bands) are ULIRG-AGN composites
in terms of their spectral energy distributions (SEDs). Since our
cumulative AGN fraction is calculated as a function of
\xray\ flux/luminosity, we focus on the AGNs that dominate in the
\xray\ band because we can measure their \xray\ luminosity reliably without
disentangling the contribution from host-galaxy star formation.

% structure of the paper and cosmology
The paper is structured as follows: we first describe our
\xray\ counterpart matching for the SMGs in Section~\ref{sec:cat}, and
then present our analyses of their \xray\ properties and also some
relevant multiwavelength properties in Section~\ref{sec:prop}. We have
used several approaches to distinguish the \xray\ AGNs from the SMGs
that are star formation dominated in the
\xray\ (Section~\ref{sec:class}). Then we calculate the AGN fraction
among the SMGs for various \xray\ flux/luminosity limits
(Section~\ref{sec:fagn}). Stacking analyses with the \xray\ undetected
SMGs are described in Section~\ref{sec:stack}. In
Section~\ref{sec:discuss} we compare with previous studies, discuss
our results and outlines the possible future work.

% nomenclatures
Throughout the paper, we assume a $\Lambda$CDM cosmology with
$H_0=70.4$ km s$^{-1}$ Mpc$^{-1}$, $\Omega_m=0.27$, and
$\Omega_\Lambda=0.73$ \citep{Komatsu2011}. Whenever galaxy stellar
mass and SFR are involved, we assume a Salpeter initial mass function
(IMF), and we have converted the quantities quoted from other works to
be consistent with the Salpeter IMF whenever necessary. We use the
conversion factor of $M_\star$ (Salpeter IMF) $=1.8 \times M_\star$
(Kroupa or Chabrier IMF). We adopt a Galactic column density of
$\nh=8.8\times 10^{19}$ cm$^{-2}$ for the line of sight to the
\ecdfs\ region (e.g., \citealt{Stark1992}), and all reported
\xray\ quantities are corrected for Galactic extinction.

%%%%%%%%%%%%%%%%%%%%%%%%%%%%%%%%%%%%%%%%%%%%%%%%%%%%%%%%%%%%%%%%%%%%%%%%%%%%%%%%%%%%%%%%%%%%%%%%%%%%
\section{Matching X-ray Sources and Submm Sources}\label{sec:cat}

% Overview
We first aim to find secure \xray\ counterparts for the ALESS
SMGs. Section~\ref{sec:submmcat} describes briefly the ALESS submm
catalog from \cite{Hodge2013}. Section~\ref{sec:xraycat} describes the
\xray\ catalogs used for finding the \xray\ counterparts for the ALESS
SMGs, which include additional sources beyond the L05 and X11
catalogs. Section~\ref{sec:match} contains our methodology for
counterpart matching (likelihood-ratio matching) and summarizes the
results.

%---------------------------------------------------------------------------------------------------
\subsection{The Submm Catalog}\label{sec:submmcat}

% Hodge 2013
We use the ALESS SMG
catalog presented in \cite{Hodge2013} (see also \citealt{Karim2013})
based on ALMA follow-up observations on the submm sources detected by
LESS \citep{Weiss2009}. The main-source catalog in \cite{Hodge2013}
contains 99 SMGs that are within the primary beam of ALMA, with low
axial ratio ($<2$), low RMS ($<0.6$~mJy) and high S/N ($>3.5$). This
catalog is the first fully submm-identified, statistically reliable catalog of
SMGs \citep{Hodge2013, Karim2013}.

\begin{figure*}[!th]
  \center
\includegraphics[angle=0.,scale=0.9]{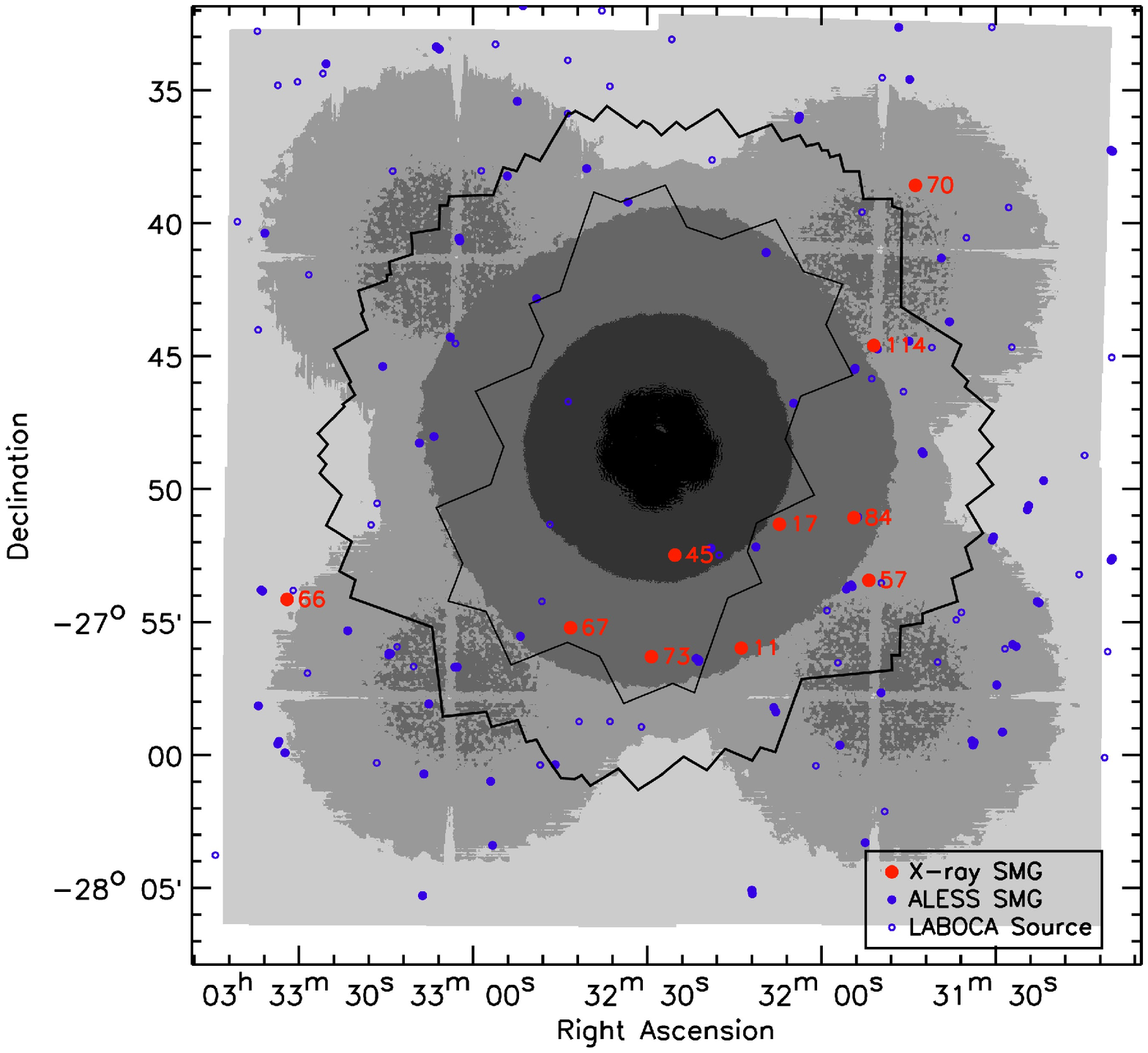}  
\caption{Full-band \xray\ sensitivity map for the \ecdfs\ region. The gray-scale
  levels, from black to light gray, represent areas with flux limits
  of $<4.0\times 10^{-17}$, $4.0 \times 10^{-17}$ to $10^{-16}$,
  $10^{-16}$ to $3.3\times 10^{-16}$, $3.3\times 10^{-16}$ to
  $10^{-15}$, and $>10^{-15}$ \ergcms, respectively. For the
  overlapping region of the \cdfs\ and \ecdfs\ where each sensitivity map
  reports a different flux limit, the smaller one (representing the
  best sensitivity) was used when creating the merged sensitivity map.
  The large red dots mark the \xray\ detected SMGs, while the blue dots
  are other SMGs in the ALESS main catalog. \xray\ detected SMGs are
  labeled with their short LESS IDs (e.g., ALESS 11.1 is labeled as
  ``11"). The same labeling convention also applies to all plots following.
  The small open circles are the LABOCA submm sources \citep{Weiss2009}
  that were followed up by ALMA but whose fields do not contain any
  ALESS main-catalog source (57 such sources; see \citealt{Hodge2013} for
  details).
  The inner thin solid line shows the GOODS-S region
  \citep{Giavalisco2004}, which is also approximately the combined
  coverage for {\it Hubble} WFC3 Early Release Science and CANDELS
  \citep{Grogin2011} in this region. The outer thick solid line marks
  the region for the 4~Ms \cdfs\ (X11). The LABOCA region is roughly a
  square whose edges are $\sim$ 2--3$\arcmin$ outside the
  \ecdfs\ boundaries.
  The average exposure time of the \xray\ detected SMGs is 2.2~Ms,
  while for the \xray\ undetected SMGs it is 0.8~Ms. As also discussed
  in Section~\ref{sec:fagn}, some of the non-detections are simply due
  to shallower \xray\ coverage.
  \label{fig:map}}
\end{figure*}

% the map figure
Figure~\ref{fig:map} shows the positions of the 99 ALESS main-catalog
SMGs and the combined \xray\ exposure maps for both the \chandra\ 4~Ms
\cdfs\ and 250 ks \ecdfs\ in gray scale. 91 of these 99 SMGs lie
within the \chandra\ 250 ks \ecdfs\ region and 44 in the 4~Ms
\cdfs\ region. We identified 10 SMGs with \xray\ counterparts (large
red dots), and below we detail the \xray\ catalog used and our
matching method.

%---------------------------------------------------------------------------------------------------
\subsection{The X-ray Catalog}\label{sec:xraycat}
% \xray\ catalogs, benefits of having additional wavedetect sources
The \xray\ catalog used for matching to the ALESS SMGs consists of two
catalogs: one derived from the 4~Ms \cdfs\ data, and the other from
the 250 ks \ecdfs\ data. The \cdfs\ catalog includes (1) 776 \cdfs\ 4
Ms main and supplementary catalog sources (X11); and (2)
116 additional sources from a {\tt WAVDETECT} catalog with a
false-positive probability threshold of $<10^{-5}$ (higher than used
for selecting the main and supplementary catalogs). The
\ecdfs\ catalog includes: (1) 795 \ecdfs\ 250 ks main and
supplementary catalog sources (L05); and (2) 290
additional sources from a {\tt WAVDETECT} catalog with a
false-positive probability threshold of $10^{-5}$. The {\tt WAVDETECT}
catalogs were used by X11 and L05 as master catalogs, from which they
further selected sources and derived the published 4~Ms \cdfs\ and 250 ks \ecdfs\
catalogs, respectively. Despite their relatively lower
significance, the additional sources from the {\tt WAVDETECT} catalogs
are likely to be real \xray\ sources if identified with submm
counterparts, given the low density of SMGs on the sky and the
excellent available positions. This has
enabled us to recover genuine \xray\ counterparts to the SMGs down to
a lower \xray\ flux limit.

% Duplicate \xray\ sources
Duplicate sources that are in both the \cdfs\ and
\ecdfs\ \xray\ catalogs were removed. For sources in the main and
supplementary catalogs of both fields, X11 has noted all duplicate
sources in their published 4~Ms catalog; for the additional {\tt
  WAVDETECT} sources, duplicate sources were identified by performing
closest-counterpart matching between the two catalogs with a search
radius of $1.5 \arcsec$. In total, our \xray\ catalog contains 892
sources in the 4~Ms \cdfs\ region (with 116 from the {\tt WAVDETECT}
lower-significance catalog), and 762 sources in the 250 ks
\ecdfs\ region but not in the \cdfs\ (with 255 from the {\tt
  WAVDETECT} lower-significance catalog).

%---------------------------------------------------------------------------------------------------
\subsection{Source Matching}\label{sec:match}
% Maching method - likelihood matching
We adopted a likelihood-ratio matching method to find secure \xray\
counterparts for the ALESS SMGs (e.g., \citealt{Ciliegi2003,
  Luo2010}). This method takes into account the positional
uncertainties for both catalogs, as well as the expected flux
distribution of the counterparts. Briefly, we computed the likelihood
ratios, defined as the ratio between the probabilities of the SMG
being the true counterpart and being just a background source, for all
SMGs within $5\arcsec$ of an \xray\ source. Then we iterate to find a
likelihood-ratio cut that maximizes the sum of the matching
completeness and reliability (see \citealt{Luo2010} for details). We found
secure \xray\ counterparts for 10 ALESS SMGs with a false-match
probability of 3\% (i.e., an expected number of false matches of
0.3). The same 10 \xray\ SMGs were recovered when a simple
closest-counterpart matching method with a matching radius of $1.5
\arcsec$ was adopted.

\begin{figure*}[!th]
  \center
\includegraphics[angle=0.,scale=0.6]{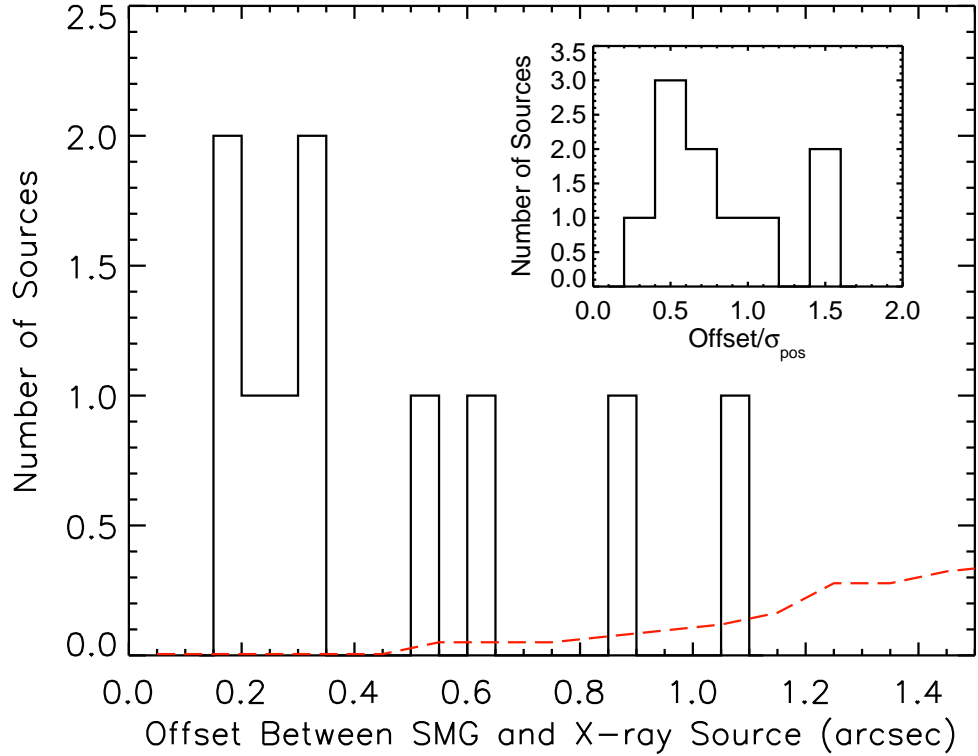}  
\caption{Histogram of the positional offsets between the SMGs and
  their \xray\ counterparts. The red dashed line is the average number
  of expected false matches, estimated for the search radius shown on
  the $x$-axis. For a matching radius of $1.5\arcsec$, the expected
  number of false match is 0.3 (consistent with the estimate given the
  likelihood-ratio method). The inset plot shows the histogram of
  offset$/\sigma_{\rm pos}$, where $\sigma_{\rm pos}$ is the
  quadrature sum of the positional error of each SMG and that of its
  matched X-ray source. None of our sources has an offset of over
  2$\sigma_{\rm pos}$. See Section~\ref{sec:match} for
  details. \label{fig:offset}}
\end{figure*}

\begin{figure*}[!th]
  \center
\includegraphics[angle=0.,scale=0.28]{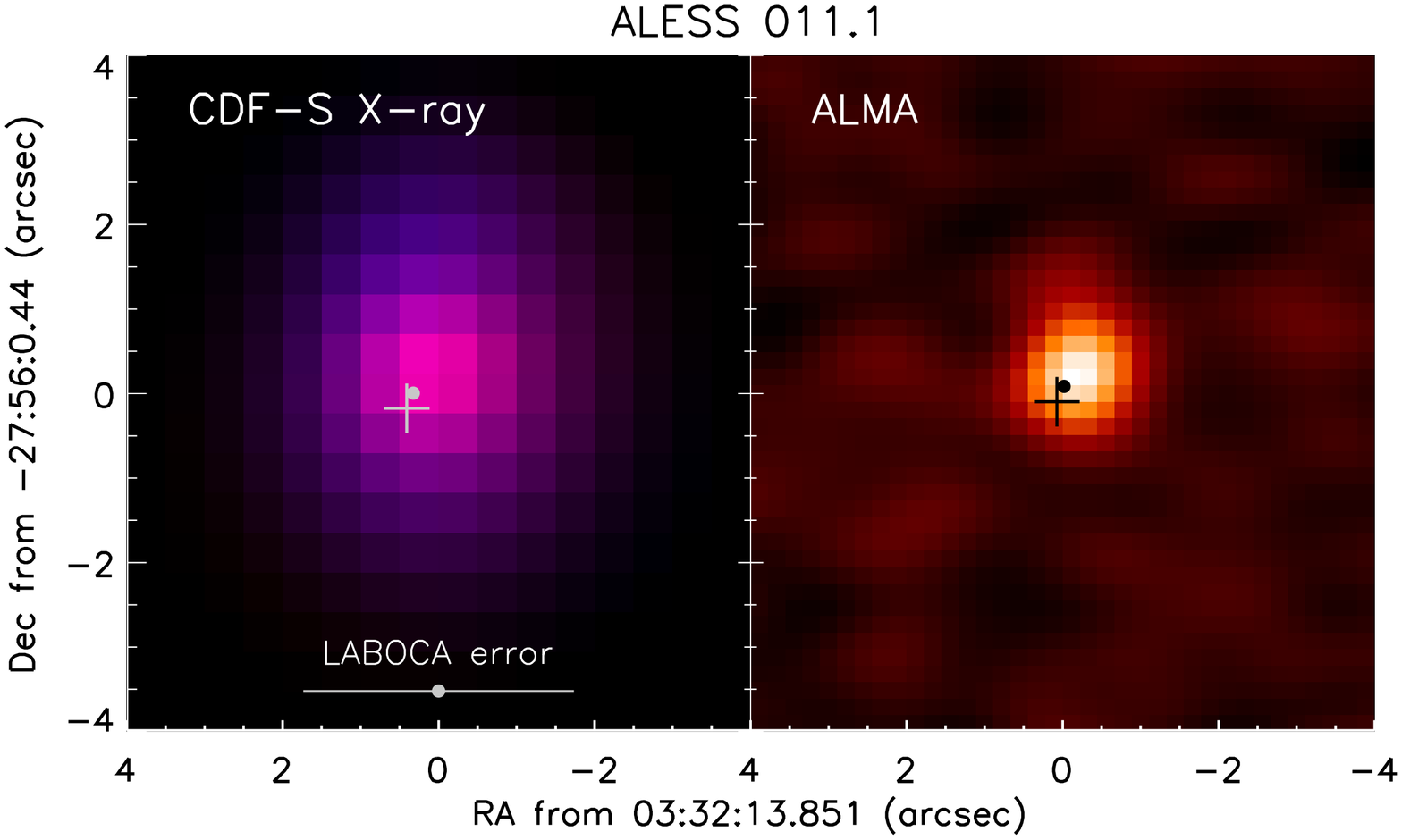}
\includegraphics[angle=0.,scale=0.28]{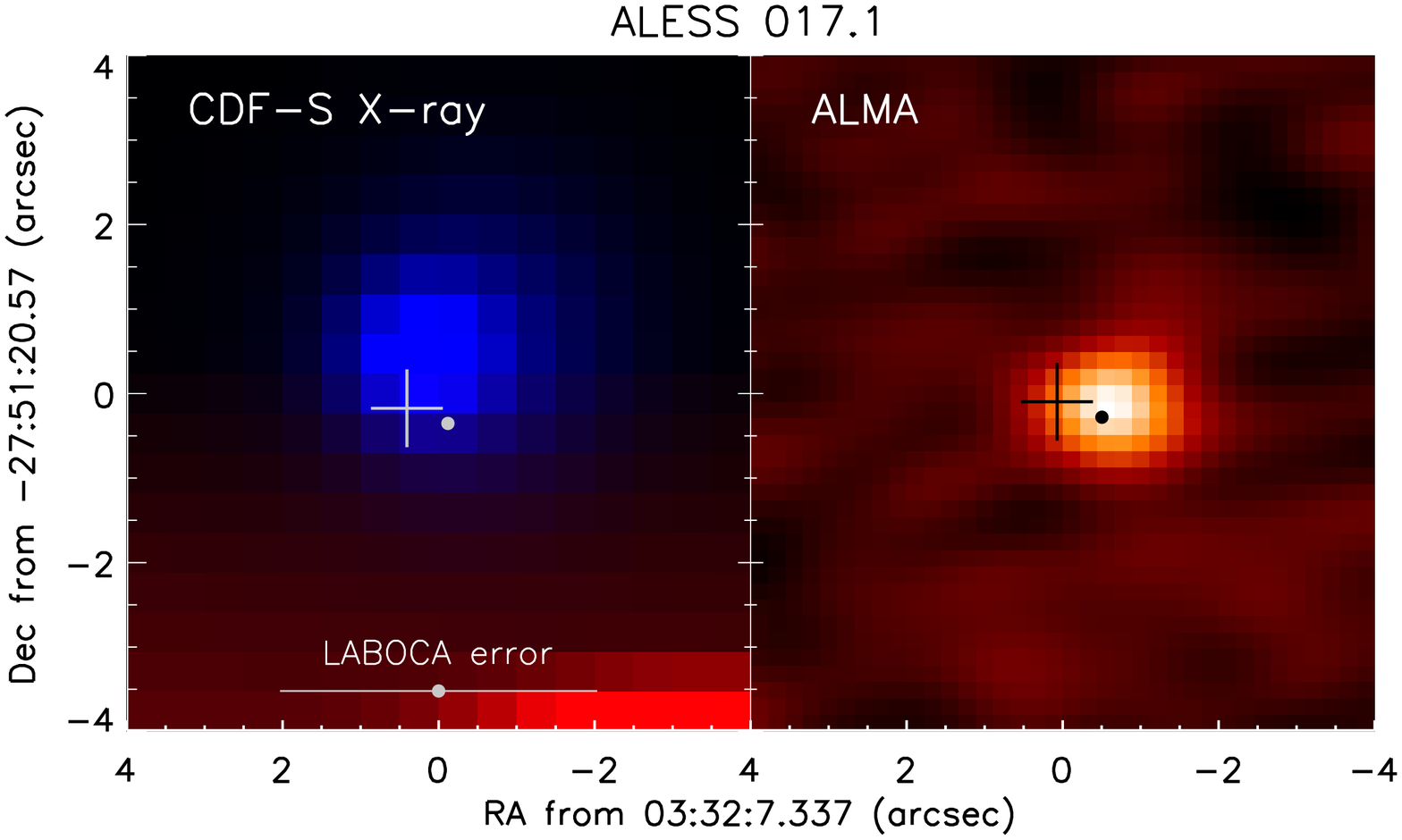}
\includegraphics[angle=0.,scale=0.28]{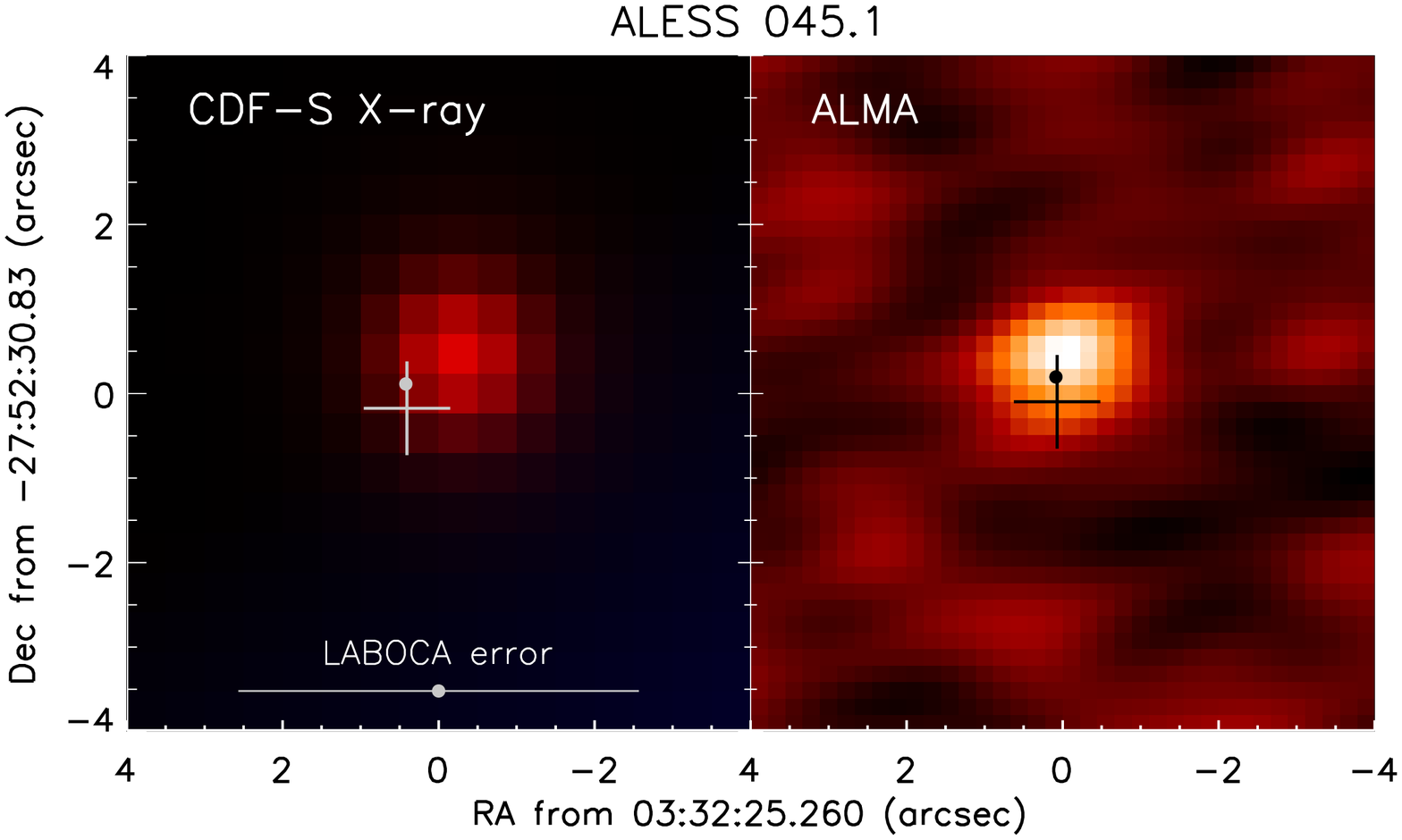}
\includegraphics[angle=0.,scale=0.28]{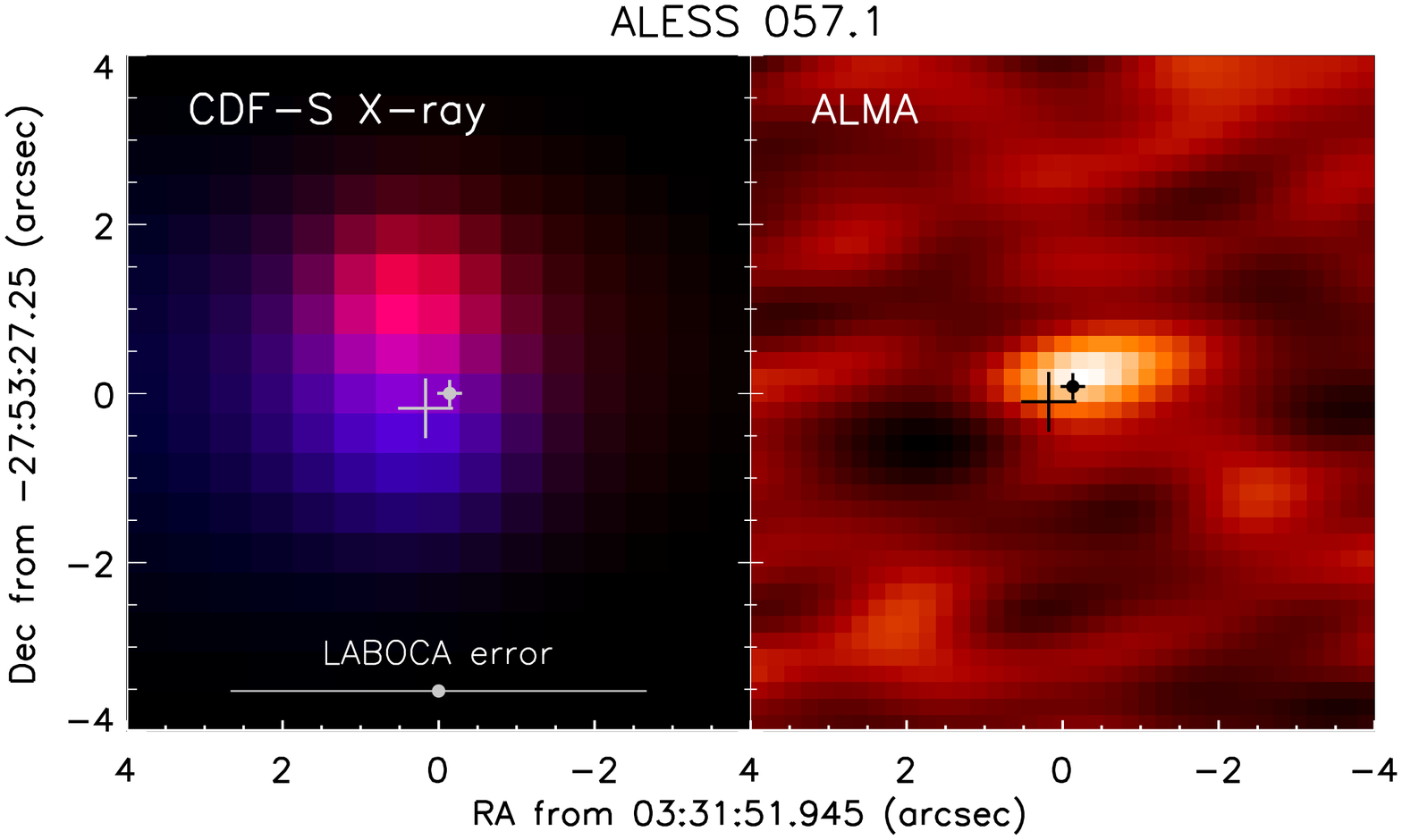}
\includegraphics[angle=0.,scale=0.28]{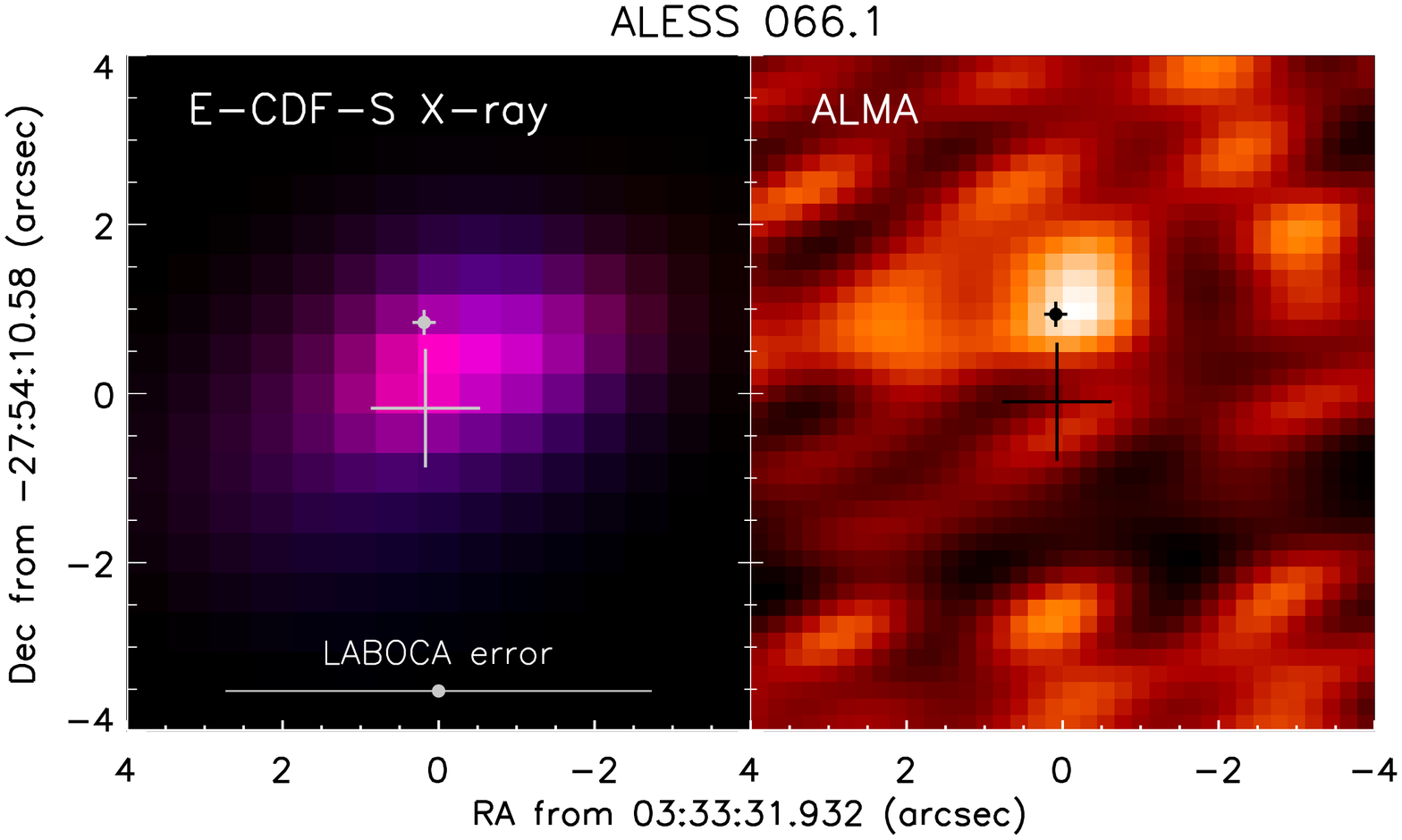}
\includegraphics[angle=0.,scale=0.28]{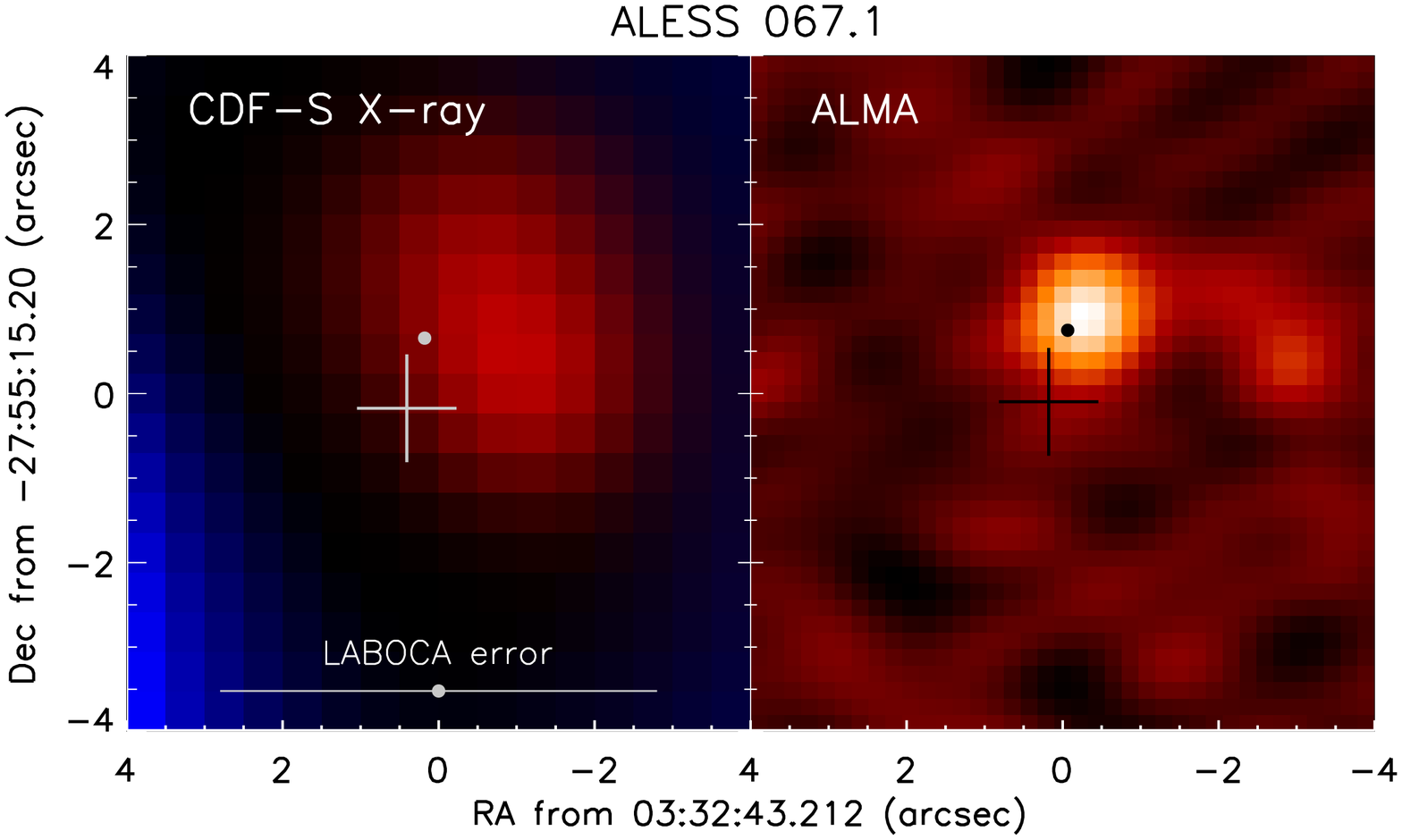}
\includegraphics[angle=0.,scale=0.28]{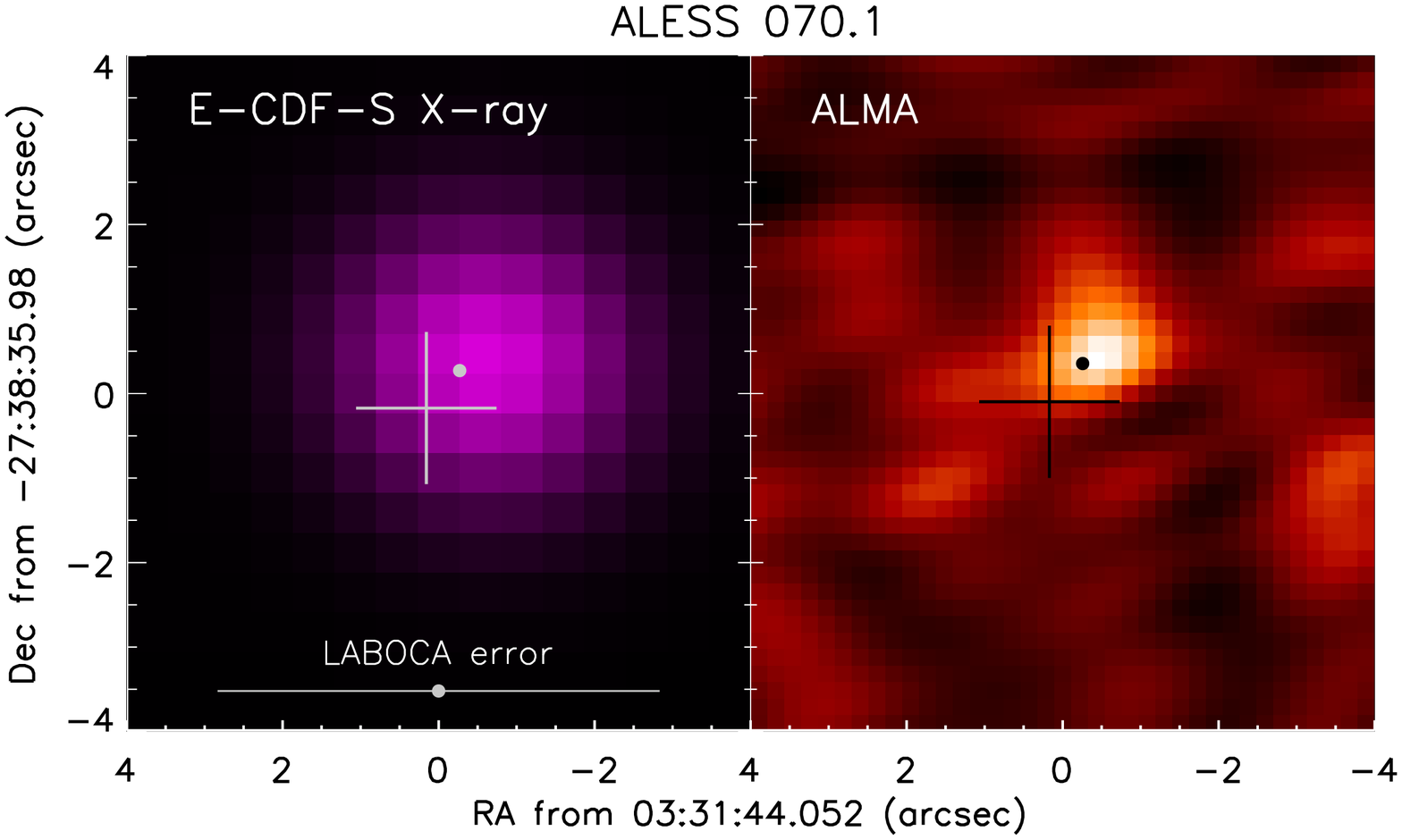}
\includegraphics[angle=0.,scale=0.28]{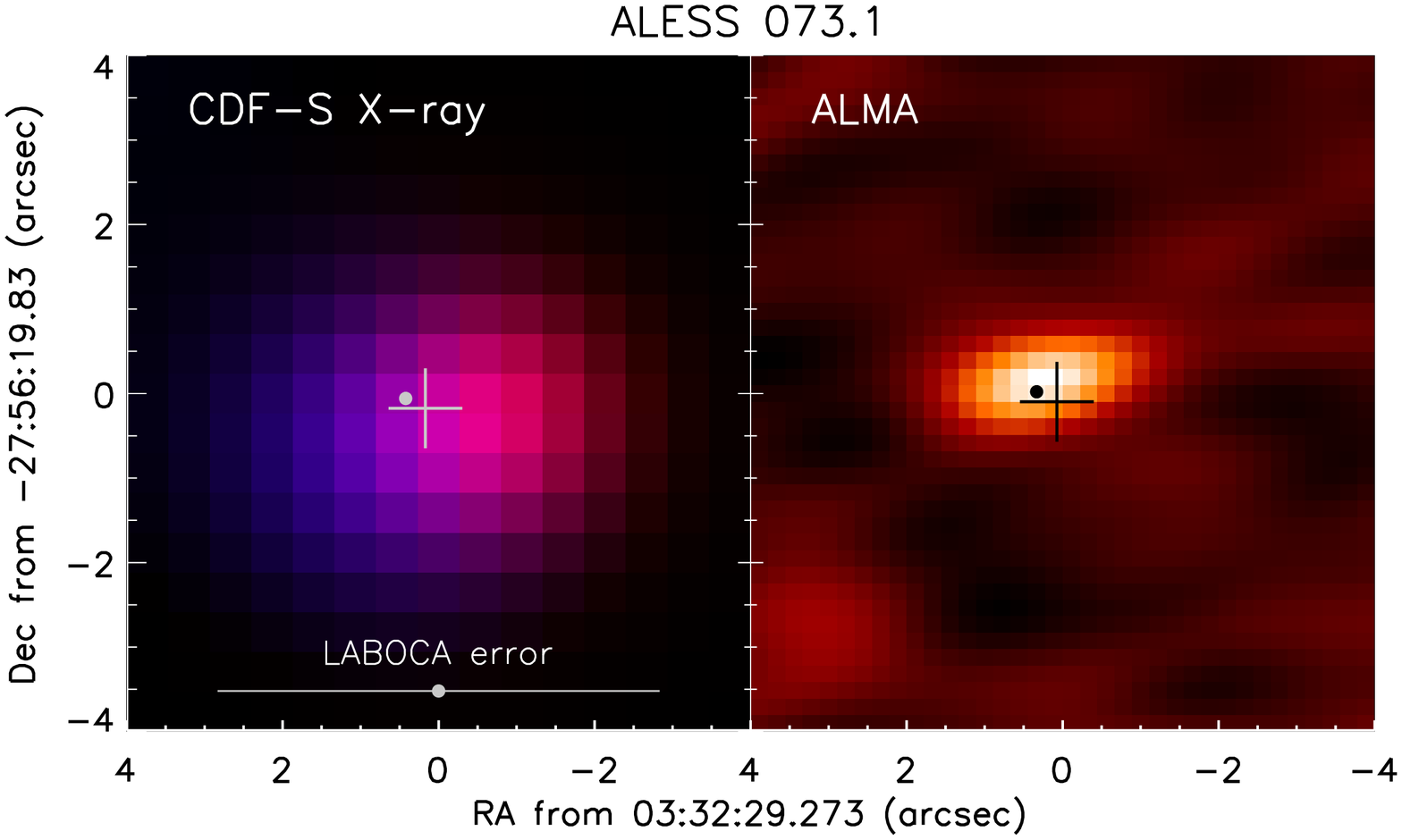}
\includegraphics[angle=0.,scale=0.28]{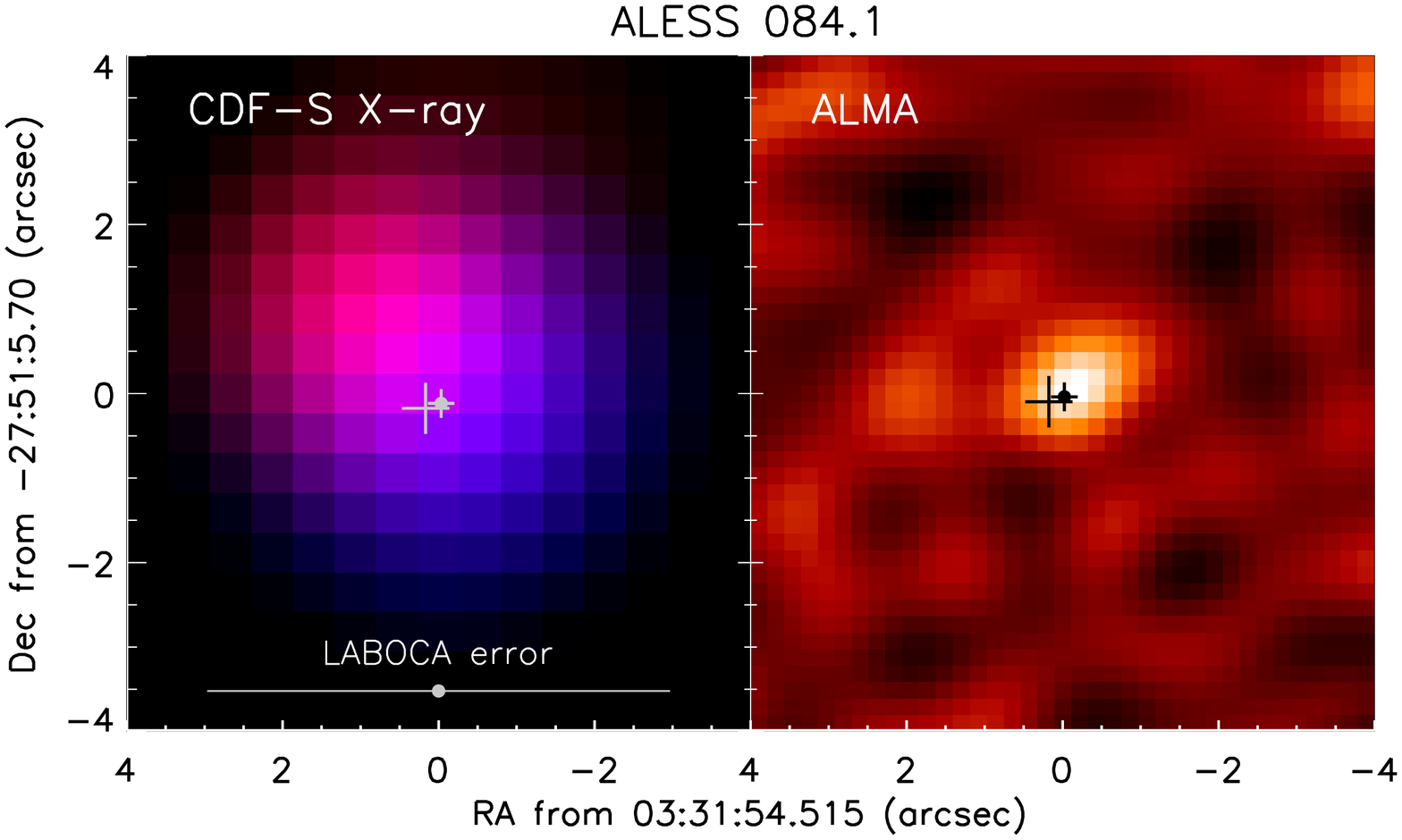}
\includegraphics[angle=0.,scale=0.28]{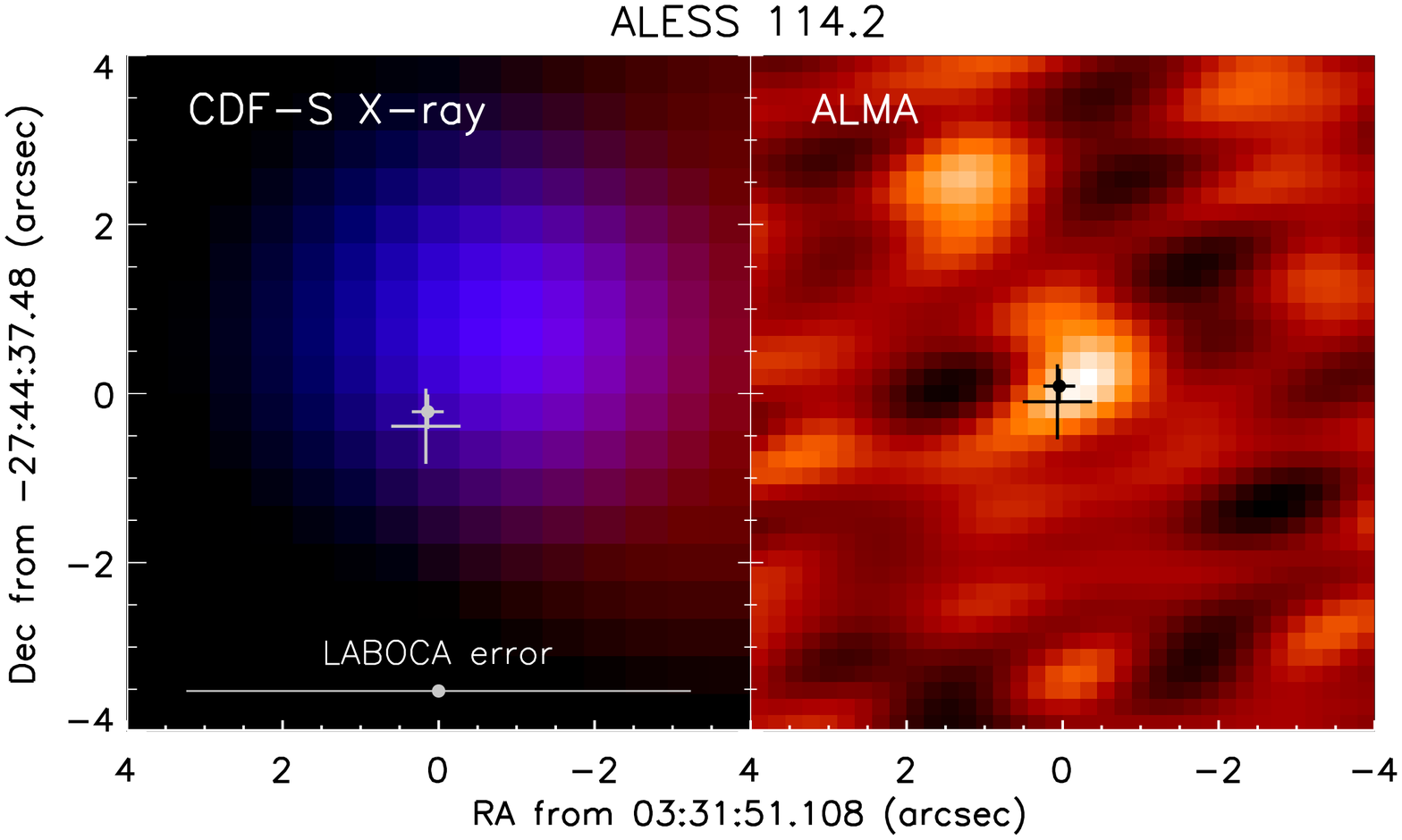}
\caption{X-ray and submm images of the \xray\ detected SMGs (central
  R.A. and Dec. given in the $x$- and $y$-axis labels). The title gives
  the ALESS short ID for each source. The dot (with a plus sign) marks
  the submm position and the larger plus sign marks the
  \xray\ position, with the sizes being their respective $\pm 1\sigma$
  positional errors. Most of the submm positional error bars are too
  small to see ($<0.1\arcsec$).
  The LABOCA $1\sigma$ positional error bar is illustrated near the
  bottom of each \xray\ image panel.
  The \xray\ image (smoothed) is color-coded so that the 0.5--2 keV
  soft band image is red, while the 2--8 keV hard band image is blue.
  For sources at large off-axis angles, the \xray\ positions are
  determined with a matched-filter technique to account for the
  complex PSFs (for sources at $>8\arcmin$ in the \cdfs\ and
  $>6\arcmin$ in the \ecdfs; see L05 and X11 for details). Therefore,
  for sources at large off-axis angles, the position may appear
  shifted from the centroid of the smoothed image.
  The submm images are from \cite{Hodge2013} and \cite{Karim2013}; \xray\ images
  are from X11; and LABOCA positional errors are from \cite{Biggs2011}.
  \label{fig:stamp}}
\end{figure*}

% Figure offset and the curve for the average number of false matches
Figure~\ref{fig:offset} shows the histogram for the positional offsets
between the 10 SMGs and their \xray\ counterparts. The red dashed line
is the estimated number of false matches as a function of the adopted
matching radius for the closest-counterpart matching method. The
number of false matches for a certain matching radius $r_{\rm s}$ was
estimated by manually shifting the \xray\ catalogs in RA and Dec by
$\pm 10$--$60\arcsec$ in $10\arcsec$ increments and re-matching with
the SMGs within $r_{\rm s}$. Then the number of false matches for
$r_{\rm s}$ is just the average number of matches for these shifted
catalogs. As shown in Figure~\ref{fig:offset}, the number of false
matches is much smaller than the actual number of \xray\ matched SMGs
at all distances $\leq 1.5\arcsec$ and is only 0.3 at
$1.5\arcsec$. The inset plot of Figure~\ref{fig:offset} shows the
histogram of offset$/\sigma_{\rm pos}$, where $\sigma_{\rm pos}$ is
the quadrature sum of the positional error of each SMG and that of its
matched X-ray source (i.e., $\sqrt{\sigma_{\rm submm}^2+\sigma_{\rm
    X-ray}^2}$). There is no SMG and X-ray source pair whose
positional offset exceeds 2$\sigma_{\rm pos}$. \mbox{X-ray} and submm
thumbnail images with illustrated positional error bars are in
Figure~\ref{fig:stamp}.

% Emphasize how much better we are - Niel
As discussed in Section~\ref{sec:intro}, when identifying
\xray\ counterparts for SMGs, previous studies had to invoke large
search radii and/or cross-identification with radio/IR counterparts,
which suffer from larger uncertainties and incompleteness (e.g., see
Section~5.5 of \citealt{Hodge2013}). The \xray\ counterparts of SMGs
in our study are of high robustness, and our estimated false-match
probability is more reliable and realistic. Our matching procedure
does not require the assumption that sources detected in other bands
such as radio or IR are very likely to be physically associated with
SMGs, which is often assumed by previous studies as their search radii
for counterpart matching are large. Moreover, our matching results are
robust against the clustering/blending of SMGs thanks to the
fully-identified ALESS SMG catalog.

\begin{figure*}[!th]
  \center
\includegraphics[angle=0.,scale=0.7]{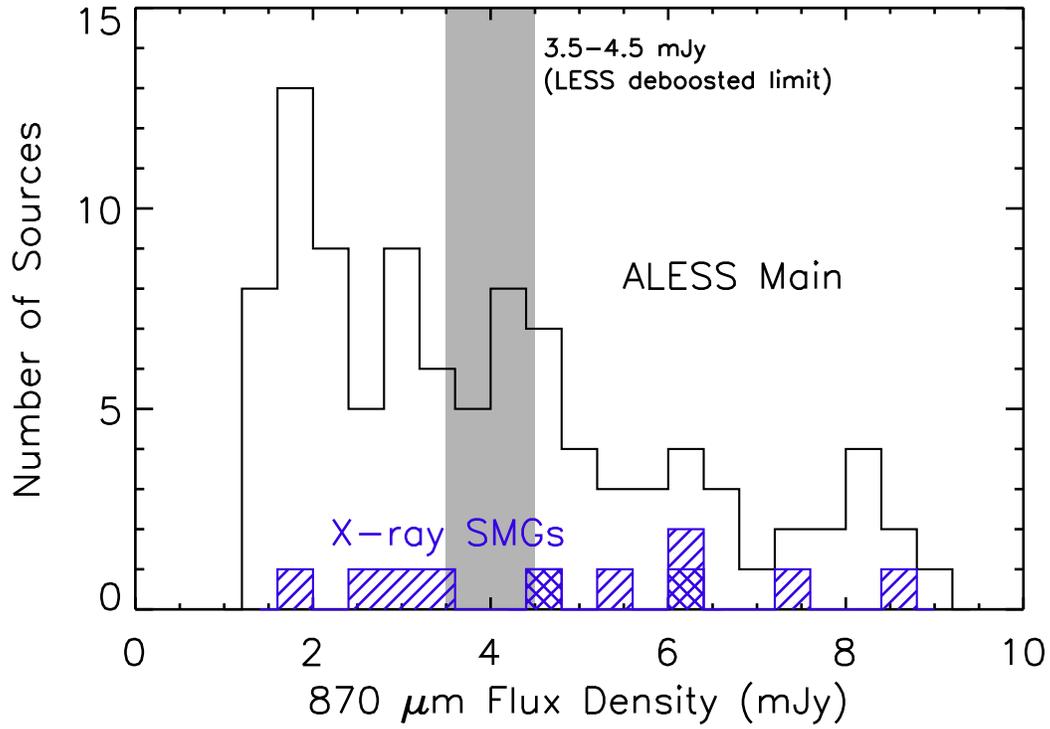}
\caption{Histograms of the 870 $\mu$m flux density for the ALESS
  main-catalog sources (solid line) and for the \xray\ detected SMGs
  (hatched). The two sources plotted in cross-hatched histogram are
  ALESS 45.1 and 67.1, which are not classified as AGNs in
  Section~\ref{sec:class}. The gray region marks the deboosted flux
  limit of the LESS survey, whose submm sensitivity is about a factor
  of three poorer than that of ALESS \citep{Weiss2009}. The flux
  distribution of all ALESS SMGs does not appear statistically
  different from that of the \xray\ detected SMGs, consistent with the
  result of a K-S test ($p=0.39$; see
  Section~\ref{sec:match}). \label{fig:hist}}
\end{figure*}

% summary of matching results
The basic properties of the 10 \xray\ detected SMGs are listed in
Table~\ref{tab:basic}. 8 of them are in the 4~Ms \cdfs\ region, and 9
have spectroscopic redshifts. As shown in Figure~\ref{fig:hist}, their
submm flux distribution (shaded blue) does not appear to differ from
the distribution for all SMGs (black solid line). We performed a
Kolmogorov-Smirnov (K-S) test with these two distributions and the
result suggests that they share the same parent distribution, with
$p=0.39$.

%%%%%%%%%%%%%%%%%%%%%%%%%%%%%%%%%%%%%%%%%%%%%%%%%%%%%%%%%%%%%%%%%%%%%%%%%%%%%%%%%%%%%%%%%%%%%%%%%%%%
\section{Properties of X-ray Detected SMGs}\label{sec:prop}

In this section, we detail our analyses and results on the
\xray\ properties of the \xray\ detected SMGs, and other
multiwavelength properties that we use in the AGN classification
process (Section~\ref{sec:class}) and other following sections. We
first detail the origin of the redshifts for the SMGs in
Section~\ref{sec:redshift}. We then describe our analyses on the
\xray\ properties and present the results in Section~\ref{sec:xspec}:
first for the more directly observed quantities, $\gammaeff$
(effective photon index) and $\lx$ (rest-frame apparent luminosity),
and then for the derived rest-frame intrinsic properties, $\gammaint$
(intrinsic photon index), $\nh$ (absorption column density) and
$\lxint$ (absorption corrected luminosity). In
Section~\ref{sec:multiv}, we describe the origins of some selected
multiwavelength properties that are relevant for this work.

%------------------------------------------------------------------------------------------
\subsection{Redshifts}\label{sec:redshift}

% xray smgs
Except for ALESS 45.1, all \xray\ detected SMGs have spectroscopic
redshifts either from the redshift follow-up survey zLESS
\citep{Danielson2013} or from the literature. The origins of the
spectroscopic reshifts are listed in a footnote of
Table~\ref{tab:spec}. ALESS 45.1 has a photometric redshift (photo-z)
from \cite{Simpson2013}, which is based on optical-NIR (with
photometric data from MUSYC $U,\ B,\ V,\ R,\ I,\ z,\ J,\ H,\ K$, and
VIMOS $U$, HAWK-I $J$, TENIS $J,\ K_s$, and IRAC 3.6--8.0~\micron) SED
fitting using the code {\sc Hyperz} \citep{Bolzonella2000}. The
photo-z estimate for ALESS 45.1, $z=2.34^{+0.26}_{-0.67}$, is
consistent with that from \cite{Xue2012} derived from optical-NIR SED
fitting using {\sc ZEBRA} \citep{Feldmann2006}. The median redshift
for the \xray\ detected SMGs is $z=2.3$.

% other smgs
Whenever reshifts are needed for the \xray\ undetected SMGs in the
\ecdfs, we adopt the photo-z values from \cite{Simpson2013}. For the
91 SMGs in the \ecdfs, 77 have detections in $\geq 3$ wavebands and
thus have SED fits and photo-z estimates, with a median redshift of
$z=2.3$. For the remaining 14 SMGs with detections only in 0--3
wavebands, their redshifts are drawn from the likely redshift
distributions estimated from simulations by \cite{Simpson2013}. The
median redshift for sources with detections in 0/1 waveband (2/3
wavebands) is $z\sim 3.5$ ($z\sim 4.5$). \cite{Simpson2013} estimated
a median redshift of $z=2.5\pm0.2$ for their complete sample of 96
SMGs (3 of the 99 ALESS SMGs only have IRAC coverage and are not
included in their sample).

%------------------------------------------------------------------------------------------
\subsection{X-ray Properties}\label{sec:xspec}

% section summary
We present the \xray\ properties of the 10 \xray\ detected SMGs in this
section. Our goal is to derive basic quantities that describe their
spectral characteristics, such as the intrinsic power-law photon index $\gammaint$
and the intrinsic absorption column density (neutral Hydrogen
equivalent) $\nh$ for each source, with the hope that they will help us
understand the origin of the \xray\ emission (see the classification of the
sources in Section~\ref{sec:class}). Their \xray\ spectral properties
are summarized in Table~\ref{tab:spec}.

% extraction of spectra
The \xray\ spectral analyses were done using spectra within the energy
range 0.5 to 8 keV, following L05 and X11. The spectra for sources
within the 4~Ms \cdfs\ region were extracted by X11 using {\tt ACIS
  Extract} (AE; \citealt{Broos2010}). The details of the AE run can be
found in X11. The spectra for the two sources that are only in the
\ecdfs\ region, i.e. counterparts for ALESS 66.1 and ALESS 67.1, were
extracted and combined for different epochs using the {\tt CIAO}
(version $4.4.1$; \citealt{Fruscione2006}) tools {\tt specextract} and
{\tt combine\_spectra}. Their raw data were downloaded from the
\chandra\ Data Archive and were reprocessed using the {\tt CIAO} tool
               {\tt chandra\_repro}. The source extraction radii for
               them are twice the 90\% encircled-energy aperture radii
               at their off-axis angles, and the background counts are
               estimated using 48 round regions around the source with
               similar or larger sizes to ensure good statistical
               measurements of the background counts.

%------------------------------------------------------------------------
\subsubsection{Hardness Ratio, Effective Photon Index $\gammaeff$, and
  Rest-Frame 0.5--8.0~keV Apparent Luminosity $\lx$}\label{sec:xspec1}
% simple and direct spectral properties, hardness ratios
We first derive two simple and direct spectral characteristics: the hardness
ratio, defined as the ratio of the photon count rates in the hard band (2--8~keV)
and the soft band (0.5--2~keV), and the effective photon
index, $\gammaeff$, for a power-law model with Galactic
absorption. The hardness ratios were derived using the Bayesian
Estimation of Hardness Ratios (BEHR) package by
\cite{Park2006}. This package computes the Bayesian posterior
distribution for the hardness ratios without requiring detections in
both energy bands, and it is especially useful for cases with low photon
counts (5 of the \xray\ counterparts have less than 100 net photon
counts in the \fbrange\ full band). The median of the posterior
distribution is taken as the best-estimate value for the hardness ratio,
and the error bars reported in Table~\ref{tab:spec} are the 68.3\%
(``$1\sigma$") posterior confidence interval (CI). When the hardness
ratio (or its inverse) has a posterior median of essentially zero
($<0.01$), we adopt the upper (or lower) limit value defined by the
90\% posterior CI.

% gamma_eff and Lx
The effective photon index $\gammaeff$ is then derived from the
hardness ratio following the methods described in L05 and X11. The
error bars on $\gammaeff$ are estimated by converting all hardness
ratios in the Bayesian posterior distribution into corresponding
$\gammaeff$ values then taking the 68.3\% CI, as listed in
Table~\ref{tab:spec}. As $\gammaeff$ values were derived from hardness
ratios and are less directly related to the observed quantities, they
are harder to constrain and therefore, following L05 and X11, for
sources having low counts (see L05 and X11 for definitions) in the
soft (or hard) band, we adopted the 90\% CI upper (or lower) limits
for $\gammaeff$. For sources with low counts in both bands, we fixed
$\gammaeff$ to $1.4$ (following X11). Using $\gammaeff$,
redshift, and the observed full-band flux $\fx$ as listed in
Table~\ref{tab:basic}, we derived the rest-frame \fbrange\ apparent
luminosity (with no intrinsic absorption correction; denoted as $\lx$
throughout this paper), for each source following the equation
$\lx=4\pi d_{\rm L}^2 \fx (1+z)^{\gammaeff-2}$ (e.g., X11).

%------------------------------------------------------------------------
\subsubsection{Intrinsic Photon Index $\gammaint$, Intrinsic Absorption
  Column Density $\nh$, and Rest-Frame 0.5--8.0~keV Absorption-Corrected 
  Luminosity $\lxint$}\label{sec:xspec2}

% intrinsic properties
We then estimated the intrinsic photon index, $\gammaint$, the
intrinsic absorption column density, $\nh$, and the rest-frame
\fbrange\ absorption-corrected luminosity (denoted as $\lxint$
throughout this paper), for each source. We used {\tt XSPEC}
\citep{Arnaud1996} for spectral fitting and modeling. The basic model
we adopted was {\tt wabs*zwabs*zpow} in {\tt XSPEC}, where {\tt wabs}
represents the Galactic absorption, {\tt zwabs} represents the
rest-frame intrinsic absorption ($\nh$ being one of its parameters),
and {\tt zpow} is a power-law model (with index $\gammaint$) in the
source rest-frame.

\begin{figure*}[!th]
  \center
\includegraphics[angle=0.,scale=0.35]{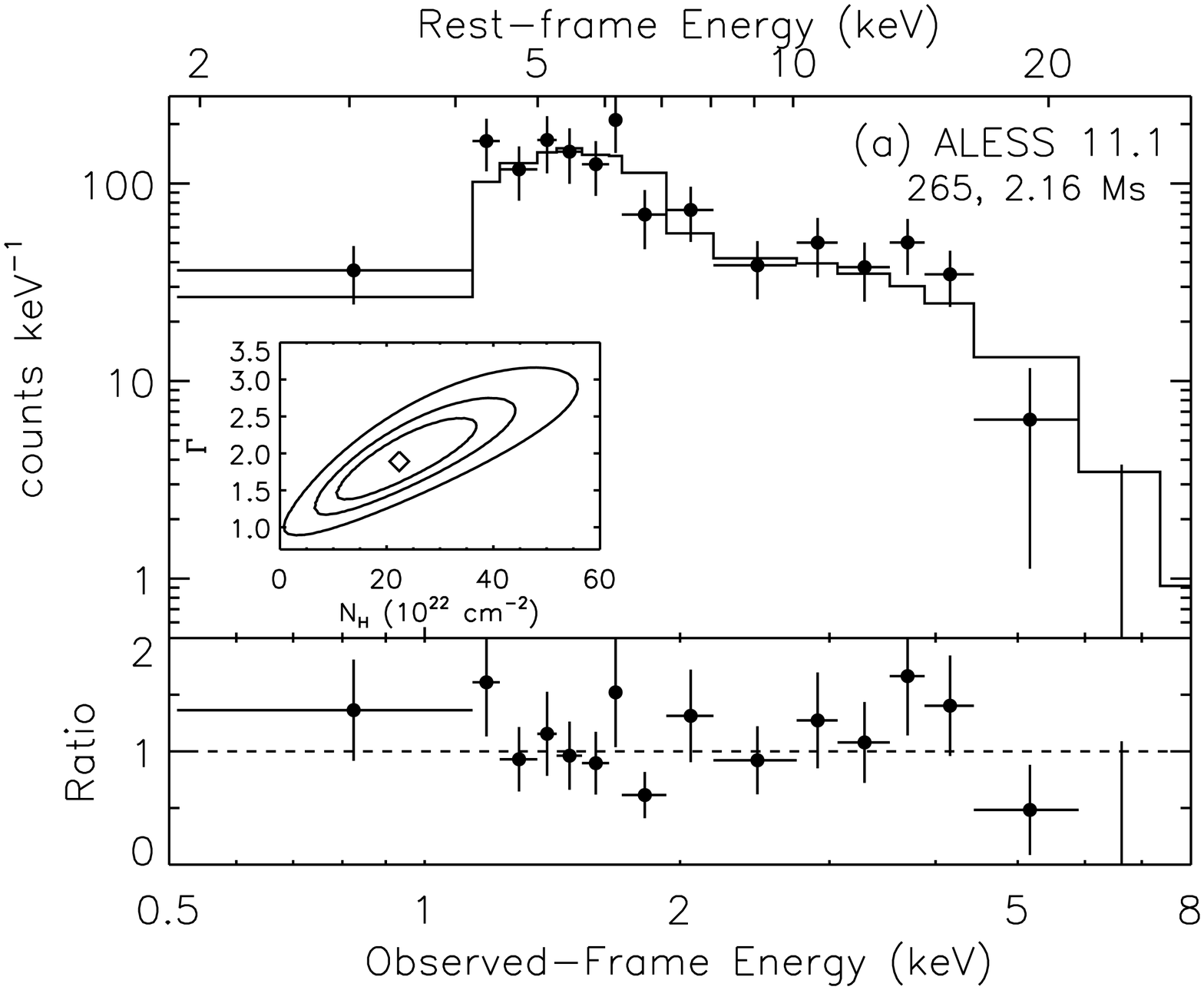}
\includegraphics[angle=0.,scale=0.35]{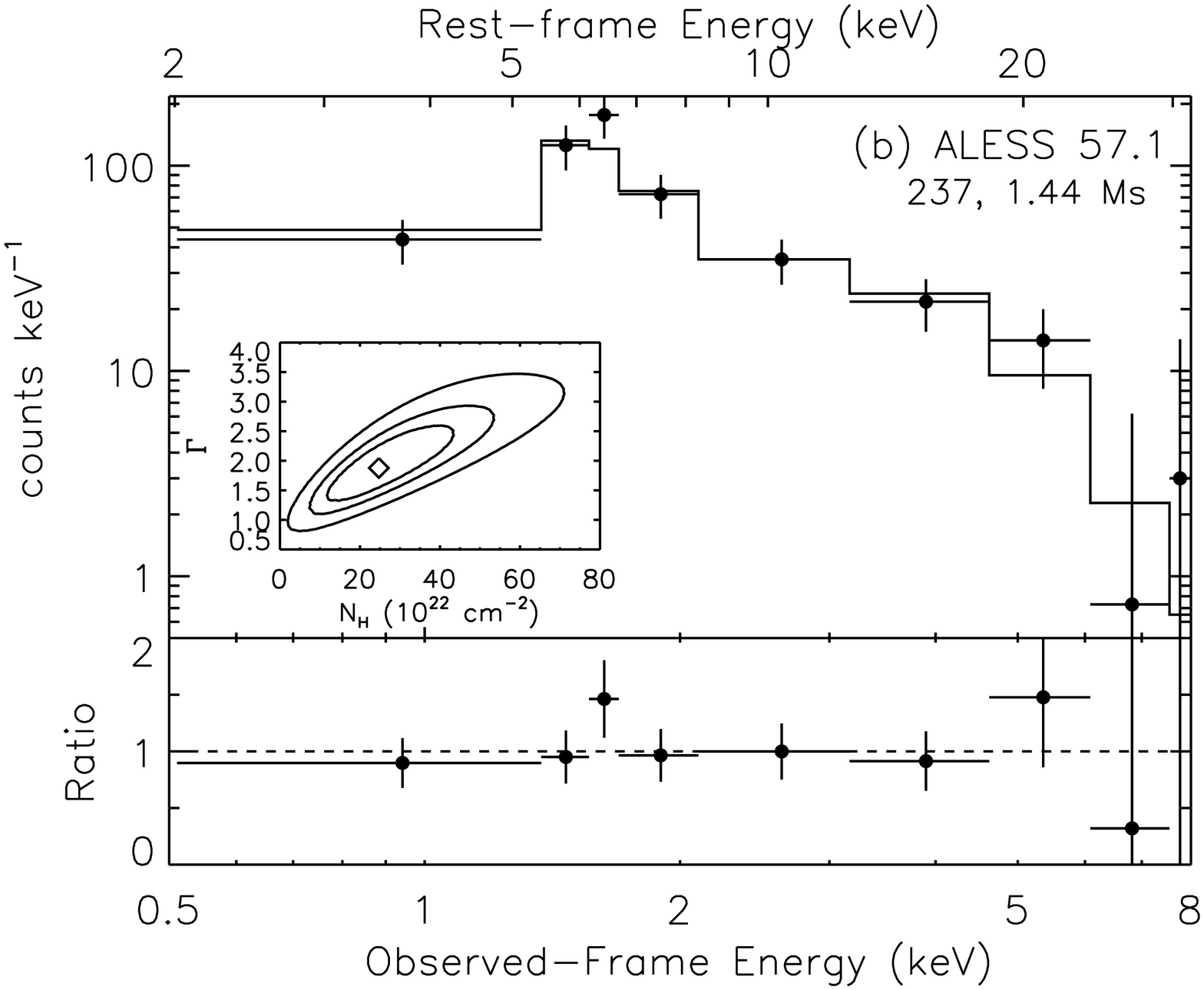}
\includegraphics[angle=0.,scale=0.35]{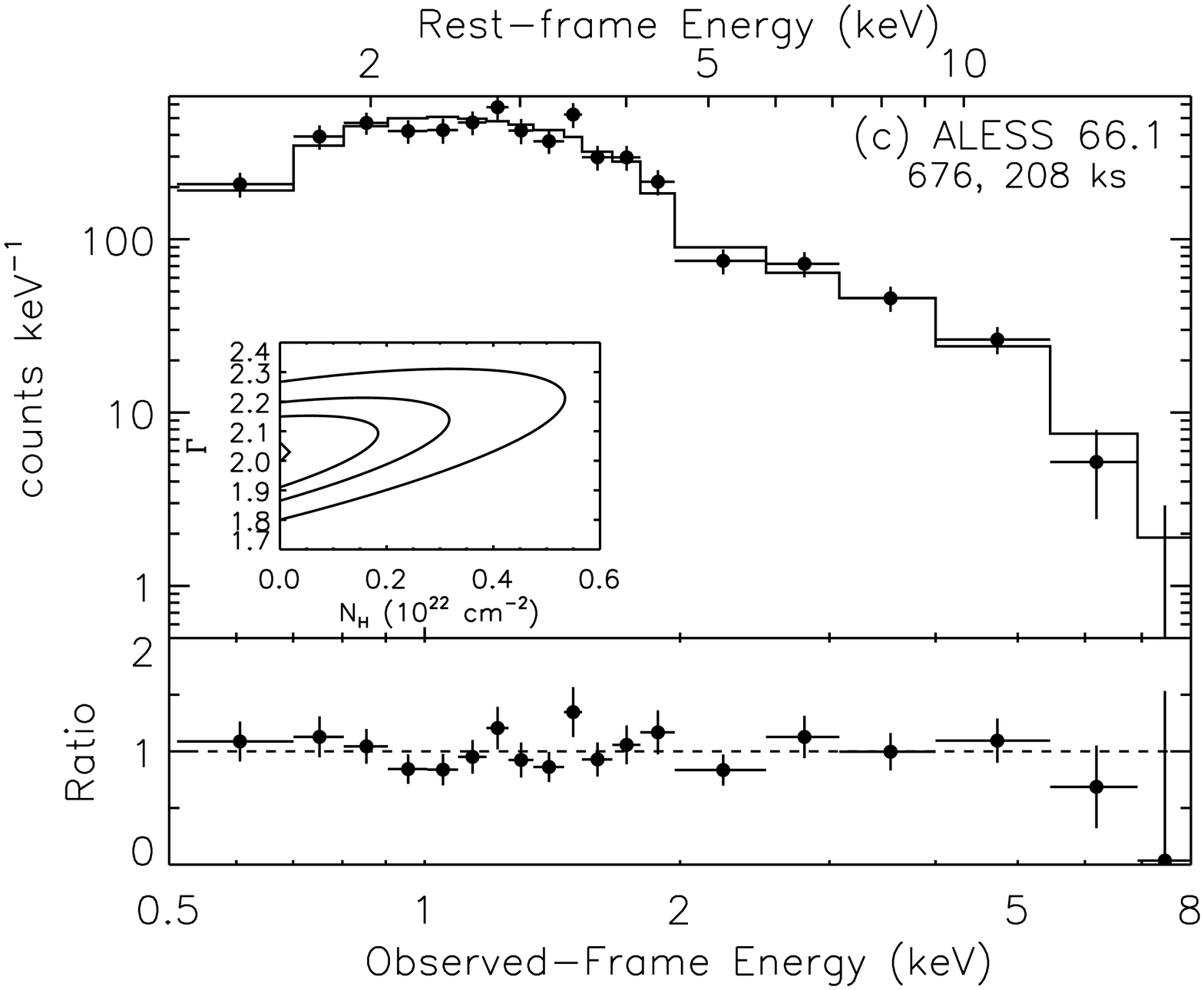}
\includegraphics[angle=0.,scale=0.35]{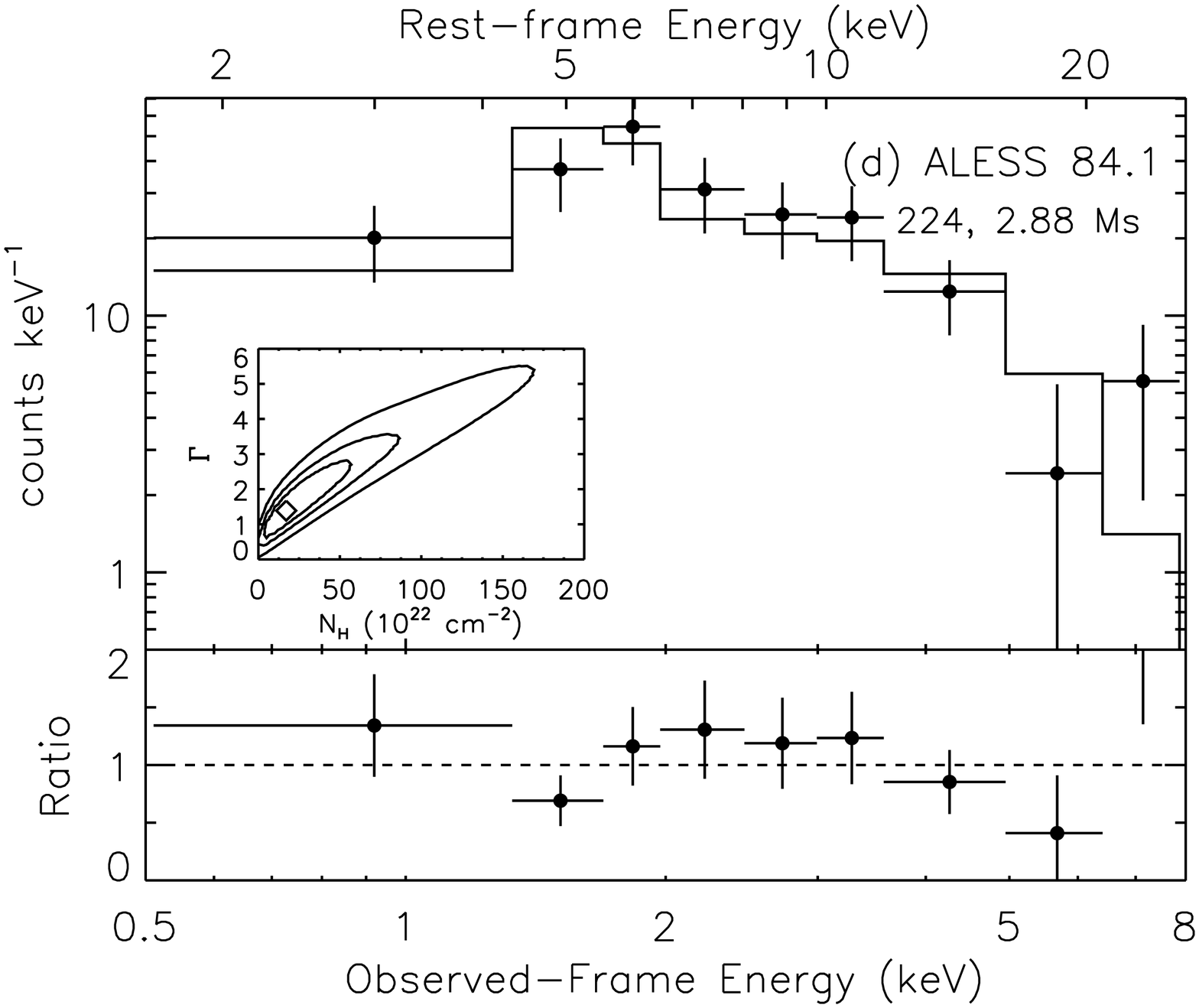}
\includegraphics[angle=0.,scale=0.35]{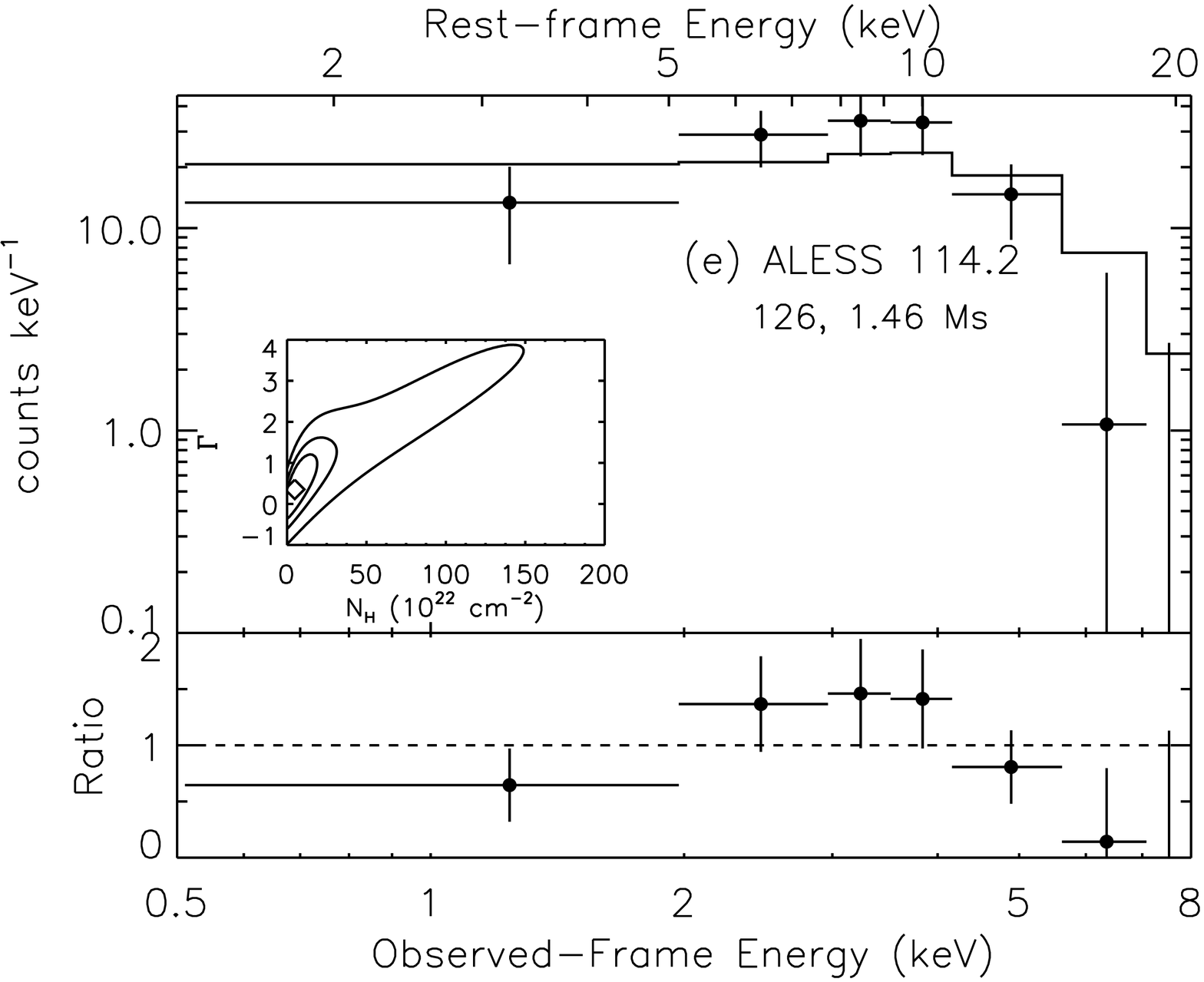}
\caption{\xray\ spectral fits for ALESS 11.1, 57.1, 66.1, 84.1, and
  114.2. Below the panel title the \fbrange\ (full-band) net photon
  counts and full-band effective exposure time are given, as also
  listed in Table~\ref{tab:spec}.
  The solid lines in the upper panels of each sub-plot are the
  best-fit models (a power-law with the effects of both Galactic and
  intrinsic absorption, {\tt XSPEC wabs*zwabs*zpow}), except for ALESS
  66.1 in sub-plot (c), whose best-fit model has no intrinsic
  absorption ({\tt wabs*zpow}). The upper $x$-axis gives the energy in
  the source rest frame. 
  The lower panel in each sub-plot shows the ratio between the data
  and the best-fit model. The inset figures are the contours for the
  intrinsic photon index $\Gamma_{\rm int}$ and column density $N_{\rm
    H}$ at the 68.3\%, 90\%, and 99\% confidence levels. The diamonds
  mark the best-fit values.
  The $\Gamma_{\rm int}$-$N_{\rm H}$ contours for ALESS 66.1 were
  computed using a {\tt wabs*zwabs*zpow} model (different from its
  best-fit model {\tt wabs*zpow}), which demonstrates that this source
  shows no detectable absorption.
  See Section~\ref{sec:xspec} for more details. \label{fig:spec}}
\end{figure*}

% spectral fitting
Among the 10 \xray\ detected SMGs, 5 have full-band net counts over
100 and therefore are qualified for spectral fitting. We fitted the
spectra of these 5 sources without binning, and we adopted the Cash
statistic (\citealt{Cash1979}; {\tt cstat} in {\tt XSPEC}) for finding
the best-fit parameters, which is well suited for fitting low-count
\xray\ sources and does not require any spectral binning
\citep{Nousek1989}. Figure~\ref{fig:spec} shows the spectra of the 5
sources with full-band net counts $>$100, ALESS 11.1, 57.1, 66.1,
84.1, and 114.2, with their best-fit {\tt wabs*zwabs*zpow} models, and
the inset figures show the 68.3\%, 90\% and 99\% confidence contours
for $\Gamma_{\rm int}$ vs.\ $N_{\rm H}$. For ALESS 66.1, the plotted
best-fit model is {\tt wabs*zpow}, since its spectral fitting
indicates no significant evidence for absorption, as illustrated by
its $\Gamma_{\rm int}$-$N_{\rm H}$ contours. ALESS 114.2 does not have
high photon counts (126 net counts in the full band) and exhibits high
background due to its large off-axis angle in the
\cdfs\ ($>9\arcmin$). Fixing its intrinsic photon index $\Gamma_{\rm
  int}$ at 1.8 (following X11; for typical AGNs) gives $N_{\rm H} \approx
2.4_{-1.2}^{+5.9} \times 10^{23}$~cm$^{-2}$.  The best-fit $\gammaint$
and $\nh$ values (and 90\% CI error bars) for these 5 sources are
listed in Table~\ref{tab:spec} (90\% CI upper limit for the $\nh$ of
ALESS 66.1).
  
% Fe K line fitting
We have also fitted the 4 obscured sources with $>$100 full-band net
counts (ALESS 11.1, 57.1, 84.1 and 114.2) with a model including an
Fe~K$\alpha$ line, {\tt (zpow*zwabs$+$zgau)*wabs}. We fixed the
rest-frame line
energy at 6.4 keV and width at 0.1 keV and only fitted for the
normalization (line strength). We then
calculated the equivalent width ({\tt XSPEC} command {\tt eqw}) and
its 90\% CI (using Markov chain Monte Carlo with the {\tt chain} command). We 
evaluated if the model including the Fe~K$\alpha$ line is
statistically a better model by computing the Bayesian Information
Criterion (BIC) and compared it with the BIC of the model without the
Fe~K$\alpha$ line ({\tt wabs*zwabs*zpow}).  Briefly, ${\rm BIC}=
  C+p\cdot \ln{n}$, where $C$ is the Cash statistic, $p$ is the number
of free parameters in the model, and $n$ is the number of data points in
the fit. The model with a smaller BIC value is the statistically
preferred model (see Section~3.7.3 of \citealt{Feigelson2012}). For
ALESS 11.1, 57.1, and 114.2, the model without the Fe~K$\alpha$ line is
favored, and they have rest-frame equivalent widths consistent with 0~keV within
90\% CI. The 90\% CI upper limits on the equivalent widths for ALESS
11.1, 57.1, and 114.2 are 0.15~keV, 0.67~keV, and 0.52~keV,
% xspec eqw reports observer frame values!
% they are: 0.04~keV, 0.17~keV, and 0.20~keV,
respectively.  For ALESS 84.1, however, the model with the Fe~K$\alpha$
line is slightly favored (BIC values being $494$ vs.~$496$ for the
model without the line), and the best-fit rest-frame equivalent width
%0.36 obs frame, 90\% CI of 0.07--0.66~keV
is $1.17$~keV, with a 90\% CI of 0.23--2.15~keV. Since the model
with the Fe~K$\alpha$ line is only slightly favored for one source, ALESS
84.1, for simplicity and comparison purposes, we report the spectral
analysis results using the model without the Fe~K$\alpha$ line component
for all sources.

% low count estimate
For the 5 sources with full-band net counts fewer than 100, we
estimated their $\nh$ values by running simulations in {\tt XSPEC}
using the {\tt wabs*zwabs*zpow} model with fixed $\gammaint=1.8$ and
varying $\nh$ until it reproduced the observed hardness ratio
(X11). For these 5 sources, spectral fittings does not provide more
constraints on the \xray\ properties than the simple method adopted
here. An illustration of this method is in Figure~\ref{fig:hard}
(similar to Figure~3 in \citealt{Alexander2005b}). The $\nh$ values
estimated this way are listed in Table~\ref{tab:spec} and are
distinguished from the ones derived from spectral fitting by having no
error bars. For ALESS 45.1 and 67.1, as their hardness ratios were
given as 90\% upper limits due to lack of photons in the hard band,
their $\nh$ values are therefore 90\% upper limits as well.

% Lxint
With the best-fit or estimated $\gammaint$ and $\nh$ values for each
source, we then estimated the rest-frame \fbrange\
absorption-corrected luminosity, $\lxint$, following Section~4.4 of X11, by first
deriving the intrinsic full-band flux $\fxint$ using the {\tt
  wabs*zwabs*zpow} model with $\gammaint$, $\nh$, and redshift, and
then calculating $\lxint$ using the equation $\lxint=4\pi d_{\rm L}^2
\fxint (1+z)^{\gammaint-2}$. This is corrected for both Galactic and
intrinsic absorption. Again, $\lxint$ estimates are 90\% upper limits
for ALESS 45.1 and 67.1 just as for their hardness ratios and $\nh$
values. As noted by X11, $\lxint$ values estimated for the lower count
sources typically agree within $\sim 30\%$ compared with those from
direct spectral fitting, but could potentially be subject to larger
uncertainties since spectral components such as reflection and
scattering can play an important role for heavily obscured
sources. This could also be true for the 5 sources with spectral fits,
but the precision should be sufficient for the purposes of our
study. For example, ALESS 73.1 is the known heavily obscured source
reported by \cite{Coppin2010} and \cite{Gilli2011}, who estimated
$L_{\rm 2-10keV}\approx 2.5\times 10^{44}$ \ergs\ --- a bit larger
than but in agreement with our estimate within a factor of two.

%------------------------------------------------------------------------------------------
\subsection{Multiwavelength Properties}\label{sec:multiv}

% overall
For the classification of AGNs among SMGs described in the next
section and also for the purpose of discussion, we need the rest-frame
1.4~GHz monochromatic luminosity ($L_{\rm 1.4GHz}$), the rest-frame
8--1000 \micron\ IR luminosity ($L_{\rm IR}$) and the 40--120
\micron\ FIR luminosity ($L_{\rm FIR}$), the stellar masses
($M_\star$), as well as the SFR. As we used different methods to
derive SFRs in different AGN classification schemes, the SFR estimates
are described in the relevant paragraphs in
Section~\ref{sec:class3}. The multiwavelength properties are listed in
Table~\ref{tab:multiv}.

% 1.4 GHz lum
The rest-frame 1.4 GHz monochromatic luminosity, $L_{\rm 1.4GHz}$, is calculated 
following \cite{Alexander2003}:
\beq
L_{\rm 1.4GHz} = 4\pi d_{\rm L}^2 f_{\rm 1.4GHz} 10^{-36}(1+z)^{\alpha-1},
\eeq
where $L_{\rm 1.4GHz}$ is in W Hz$^{-1}$, the observed 1.4 GHz radio
flux $f_{\rm 1.4GHz}$ is in $\mu$Jy, and the radio spectral index is
$\alpha=0.8$ (following A05). The radio counterparts and radio fluxes
of the \xray\ detected SMGs are from the catalog of \cite{Biggs2011}
(based on \citealt{Miller2008} VLA maps), identified
using a closest-counterpart matching method with $r_{\rm s}=1\arcsec$
(chosen to have a false match rate $<1\%$ and also verified by visual
examination; 39 SMGs are matched with radio sources).

% IR, FIR lum
The rest-frame IR ($8$--$1000\mu$m) luminosity $L_{\rm IR}$ and FIR
(40--120~\micron) luminosity were derived based on NIR-through-radio
SED fitting by \cite{Swinbank2013}. The SEDs were fitted using {\it
  Spitzer}, {\it Herschel}, ALMA, and VLA photometry at 3.6~$\mu$m,
4.5~$\mu$m, 5.8~$\mu$m, 8.0~$\mu$m, 24~$\mu$m, 250~$\mu$m, 350~$\mu$m,
500~$\mu$m, 870~$\mu$m and 1.4~GHz. The SED templates include the
star-forming galaxy templates from \cite{Chary2001} and that of
SMMJ2135$-$0102 (the Eyelash galaxy; \citealt{Swinbank2010}).

% mass
The stellar masses for the SMGs are from \cite{Simpson2013}. Briefly,
their stellar mass estimates are derived from the absolute $H$-band
photometry based on the optical-NIR SED fitting and a mass-to-light
ratio based on the best-fit star formation history (either burst or
constant) and stellar population synthesis models from
\cite{Bruzual2003}. Typical error bars for $M_\star$ are about a factor
of 2 (around a factor of $\sim$3--5 if taking into account model
uncertainties).

% reliability of mass
Although \cite{Simpson2013} only used galaxy templates in their SED
fitting, AGN contamination is probably not a concern here when
estimating stellar mass. Our \xray\ detected SMGs have a median
$\log{\lx}=43.0$ and all but ALESS 66.1 have $\log{\lx} \leq 43.7$
(Table~\ref{tab:spec}), which is the upper-limit cut chosen by
\cite{Xue2010} to minimize potential AGN contamination in the
optical-NIR bands. In Section 4.6.3 of \cite{Xue2010}, they studied
188 AGNs with $41.9 \leq \log{\lx} \leq 43.7$ and examined the AGN
contribution to their best-fit SED templates, the correlation of their
rest-frame absolute magnitudes/colors and \xray\ luminosities, and
their fractions of optical-NIR emission coming from the core regions
versus from the extended regions. They concluded that the AGN
contamination is minimal and does not affect the optical-NIR colors or
the mass estimates in a significant way. We note that, as shown in
Figure~\ref{fig:class4}, we do not see any correlation between $\fx$
and the IRAC 3.6 \micron\ magnitude/flux of the 9 SMGs with $\log{\lx}
\leq 43.7$, consistent with the findings of \cite{Xue2010}.

Also, \cite{Simpson2013} noted that only 3 (ALESS 57.1, 66.1, 75.1)
out of 77 SMGs have $\chi_{\rm red}^2>10$ due to 8~\micron\ excesses
indicative of AGN activity. For ALESS 57.1, the 8~\micron\ excess
feature is consistent with the fact that it is an obscured
AGN. Because of its low $\lx$ value and non-power-law spectral shape,
we do not consider that ALESS 57.1 is dominated by AGN in the
optical-NIR, and we take the stellar mass estimate as reliable but
caution the reader with this caveat. Since ALESS 66.1 is a known
optical quasar with high $\lx$ and it has the worst SED fit among all
sources in \cite{Simpson2013}, we take its estimated stellar mass as
less reliable and label it differently in the relevant plots involving
$M_\star$.

%%%%%%%%%%%%%%%%%%%%%%%%%%%%%%%%%%%%%%%%%%%%%%%%%%%%%%%%%%%%%%%%%%%%%%%%%%%%%%%%%%%%%%%%%%%%%%%%%%%%
\section{Classifications for the X-ray Detected SMGs}\label{sec:class}

% summary
In this section, we classify the 10 \xray\ detected SMGs to assess if
their \xray\ emission reveals the existence of AGNs or if they are
dominated by star formation in the \xray\ regime. To do so, we exploit
their \xray\ properties calculated in Section~\ref{sec:xspec} as well
as other characteristics derived from their multiwavelength data
(Section~\ref{sec:multiv}). We employed several independent
classification methods and cross-checked between them. These methods
and the derivation of the relevant multiwavelength properties used for
each method are described in each of the
subsections. Table~\ref{tab:classify} is a summary of the
classification methods we adopted, and Table~\ref{tab:multiv} lists
multiwavelength properties of the \xray\ detected SMGs and the
classification results. Some of these methods are closely related
(Method {\sc IIIa}, {\sc IIIb}, and {\sc IV}), but we have employed
all to enable cross-check between the results.

%------------------------------------------------------------------------
\subsection{Method I \& II.\ $\gammaeff$ and X-ray Luminosity}\label{sec:class12}

\begin{figure*}[!th]
  \center
\includegraphics[angle=0.,scale=0.7]{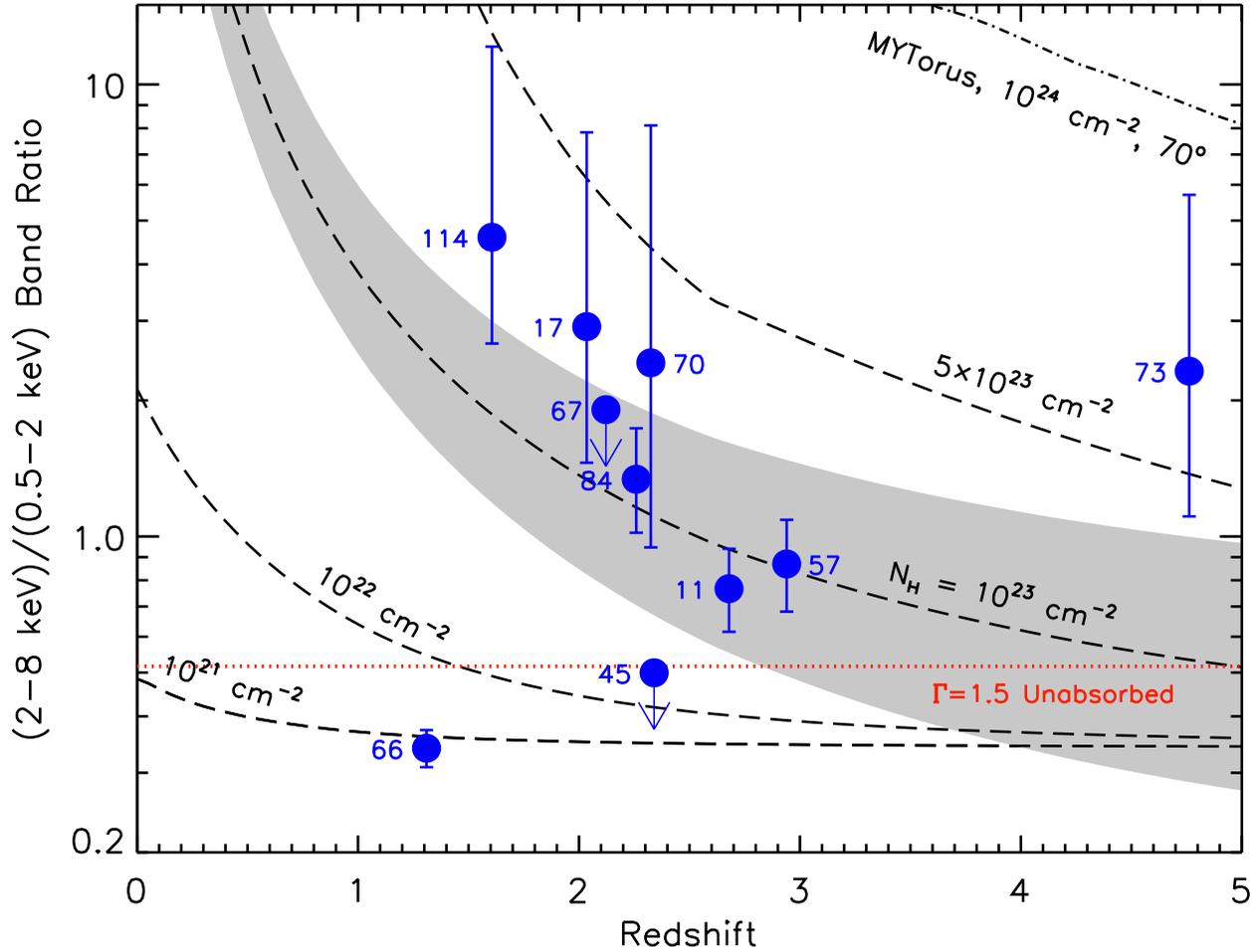}
\caption{The hardness ratio between the 2--8 keV (hard) and 0.5--2
  keV (soft) bands (i.e., \xray\ hardness) vs.\ redshift for our
  SMGs. Plotted are 1$\sigma$ error bars (68.3\% CIs) for the hardness ratios. See the
  notes in Table~\ref{tab:spec} and Section~\ref{sec:xspec} for details on the
  hardness ratios and their error bars (or 90\% CI upper limits for
  ALESS 45.1 and 67.1). 
  The dashed lines are tracks for AGN spectral models described by a
  power-law with the effects of both Galactic and intrinsic absorption
  with the intrinsic photon index $\Gamma_{\rm int}$ fixed to 1.8
  ({\tt XSPEC wabs*zwabs*zpow}). The shaded region is for models with
  a varying $\Gamma_{\rm int}=1.8\pm 0.5$ for $N_{\rm
    H}=10^{23}$~cm$^{-2}$. The dash dotted lines are tracks
  calculated using the MYTorus model (Murphy \& Yaqoob 2009), with
  $\Gamma=1.8$, $\nh=10^{24}$~cm$^{-2}$, and inclination angles of
  $50\degree$ and $70\degree$.
  The red dotted line marks the expected hardness ratio for
  $\Gamma=1.5$ and no intrinsic absorption, which could also describe
  a typical starbust/HMXB population (e.g., \citealt{Teng2005,Lehmer2008}).
  See Section~\ref{sec:xspec} for additional discussion.\label{fig:hard} }
\end{figure*}

\begin{figure*}[!th]
  \center
\includegraphics[angle=0.,scale=0.7]{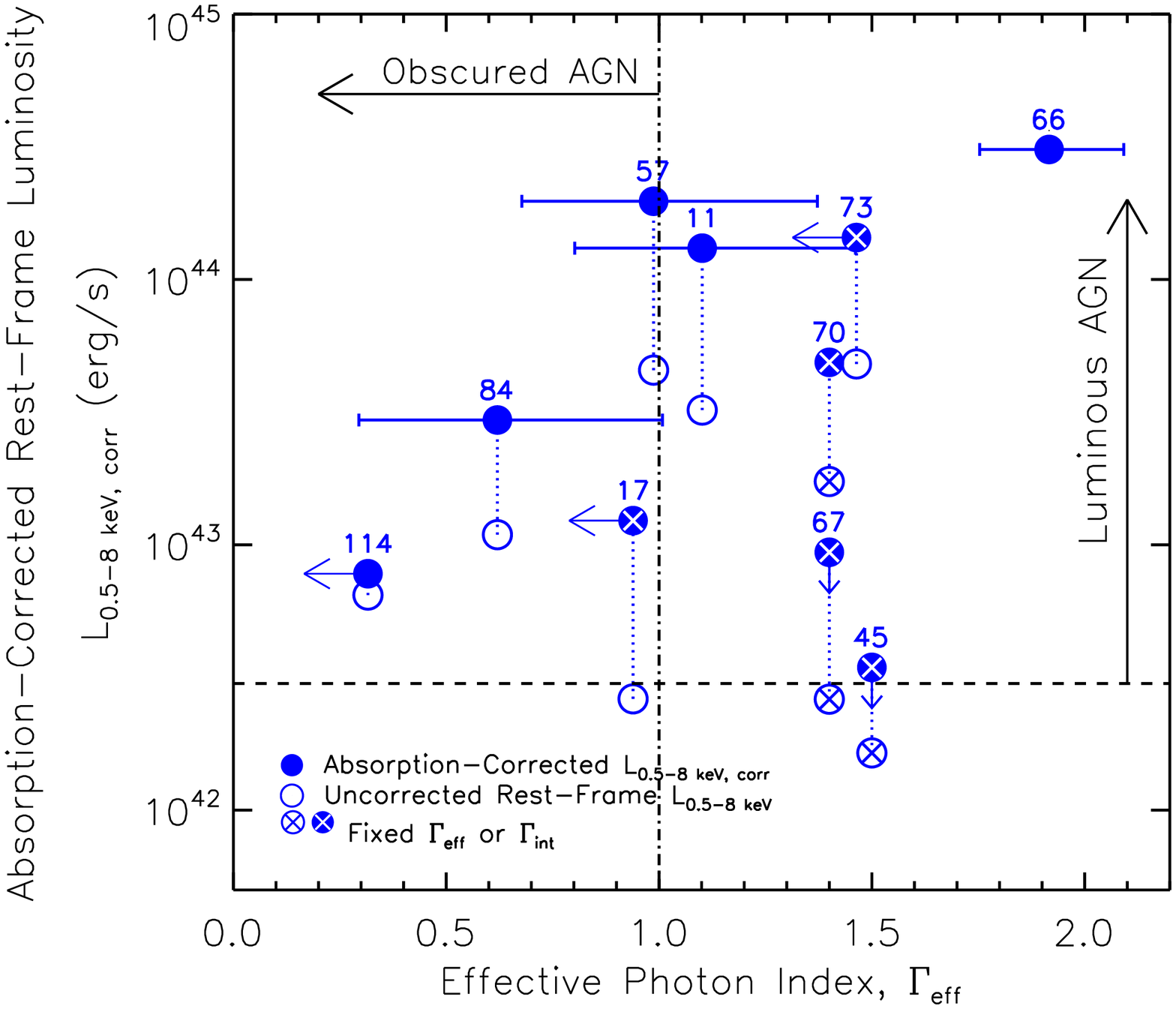}
\caption{{\bf Classification Method {\sc I \& II}}: Sources with
  observed effective photon index ($x$-axis) $\gammaeff< 1.0$ (to the
  left of the dash-dotted line) are classified as obscured AGNs
  (Method~{\sc I} in Table~\ref{tab:basic}), and sources with
  rest-frame \fbrange\ absorption-corrected luminosity ($y$-axis,
  filled circles) $\lxint>3\times 10^{42}$ erg s$^{-1}$ (above the
  dashed line) are classified as luminous AGNs (Method~{\sc II}; see
  Section~\ref{sec:class12}).
  The open circles connected to each source by a dotted line are the
  rest-frame \fbrange\ apparent luminosity $\lx$ (without intrinsic
  absorption correction). Error bars along the $x$-axis direction mark
  the 90\% confidence intervals for $\gammaeff$; arrows indicate 90\%
  upper limits (see the notes in Table~\ref{tab:spec} and
  Section~\ref{sec:xspec} for details). The $\gammaeff$ of ALESS 45.1
  is plotted at 1.5 only for display clarity.
  Crosses mark the $\lx$ (or $\lxint$) values with larger
  uncertainties due to fixed $\gammaeff$ (or $\gammaint$) as a result
  of having low counts.
  These two classification criteria do not rule out ALESS 45.1, 67.1
  or 70.1 as being AGNs, but to be conservative, we do not classify
  them as AGNs here (see Table \ref{tab:basic}).
  See Section~\ref{sec:class12} for more details.\label{fig:class12}}
\end{figure*}

% Method I, Gamma_eff
\textbf{Classification Method {\sc I}.\ $\gammaeff$:} Following
\cite{Alexander2005b} (A05 hereafter)
and X11, we classify sources with $\gammaeff<1.0$ as AGNs (see
Figure~\ref{fig:class12}). This hard signature of the \xray\ spectrum
is a feature of absorbed AGNs, as spectra having $\gammaeff<1.0$ are
empirically hard to explain with just the star forming component in a galaxy,
which typically has $\gammaeff \sim 1.5$ or even softer
(e.g., \citealt{Teng2005, Lehmer2008}). ALESS 17.1, 57.1, 84.1, and
114.2 are classified as (obscured) AGNs under this criterion. As we
adopted conservative $\gammaeff$ estimates for sources with relatively
low counts, some of the sources appear softer than indicated by
Figure~\ref{fig:hard} because their $\gammaeff$ values are upper
limits or are fixed to 1.4 (e.g., ALESS 73.1). For the calculation of
$\gammaeff$ and the error bars and upper limits, see
Section~\ref{sec:xspec}.

% Method II, \xray\ luminosity
\textbf{Classification Method {\sc II}.\ \xray\ Luminosity:} Following
the criterion adopted in, e.g., \cite{Bauer2004}, \cite{Lehmer2008},
and X11, we classify a source with rest-frame
\fbrange\ absorption-corrected luminosity $\lxint$ larger than
$3\times 10^{42}$ \ergs\ as an AGN host. This is based on studies of
local galaxies which found that all local star-forming galaxies have
lower \xray\ luminosities than $3\times10^{42}$
\ergs\ (e.g., \citealt{Zezas2001, Ranalli2003}; L10). The caveat is
that the SMGs are high-redshift star-forming galaxies, and it is
uncertain whether the criterion of $\lxint \geq 3\times 10^{42}$
established using local galaxies and AGNs would apply to these
high-redshift sources. This is why we have additional classification
methods (see the following sections) to ensure a reliable
identification of AGNs.

Figure~\ref{fig:class12} illustrates this method, with the $y$-axis
being $\lxint$. Filled circles mark the $\lxint$ values, while open
circles are $\lx$ values. Crosses mark the $\lx$ (or $\lxint$) values
with larger uncertainties due to fixed $\gammaeff$ (or $\gammaint$) as
a result of having low counts. For ALESS 45.1, 67.1 and 70.1 (with
crosses in open circles), $\gammaeff$ values are poorly constrained
due to low counts in both \xray\ bands, and thus $\gammaeff=1.4$ is
assumed (following X11), which means their $\lx$ values have larger
uncertainties. For sources with crosses in the filled circles, their
$\gammaint$ values were fixed at $1.8$ since they did not qualify for
spectral fitting due to low counts, and therefore their $\lxint$
values have larger uncertainties. Arrows on the $\lxint$ of ALESS 45.1
and 67.1 indicate that these are upper limits, because their hardness
ratios and $\nh$ values were given as 90\% CI upper limits (see
Fig.~\ref{fig:hard} and Table~\ref{tab:spec}).

All sources other than ALESS 17.1, 70.1, 67.1, and 45.1 are classified
as AGNs under method {\sc II}; the four sources were not classified as
AGNs because their $\lx$ and/or $\lxint$ were relatively poorly
constrained (crosses in Figure~\ref{fig:class12}) and the
well-constrained $\lx$ value was not above our threshold (for ALESS
17.1, but not the case for ALESS 73.1). We note here that the 6
sources classified as AGNs under this criterion all have $\lx$ values
larger than $3\times 10^{42}$ \ergs. Therefore, if we are conservative
and require $\lx>3\times 10^{42}$ \ergs\ (rather than $\lxint>3\times
10^{42}$ \ergs) as we are dealing with high-redshift star-forming
galaxies, the conclusion would still be the same.

%------------------------------------------------------------------------
\subsection{Method III.\ $\lx$ vs.\ SFR}\label{sec:class3}

% Method III, Lx vs SFR
The general idea of this method is to compare the rest-frame
\fbrange\ apparent luminosity, $\lx$, with the predicted amount as
expected from the level of star formation, $L_{\rm X,SF}$. If $\lx$ of an SMG
is $5\times$ or more than that expected from its star formation (i.e.,
$\lx \geq 5\times L_{\rm X,SF}$), it is classified as an AGN host
(similar to the criterion adopted by A05 and X11). As detailed below,
we estimated $L_{\rm X,SF}$ with two approaches, and they give
consistent classification results.

\begin{figure*}[!th]
  \center
\includegraphics[angle=0.,scale=0.39]{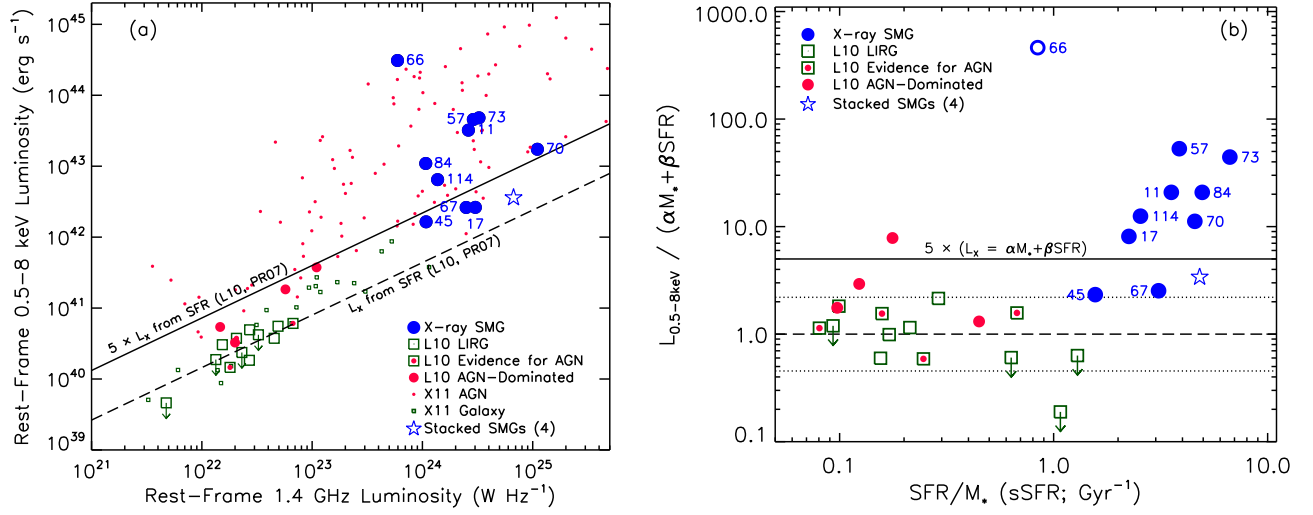}
\caption{{\bf Classification Method {\sc III}}:
  {\bf Panel (a)}: The rest-frame \fbrange\ apparent luminosity (no
  absorption correction) vs.\ the rest-frame 1.4~GHz monochromatic
  luminosity. The dashed line marks the $\lx$ to $L_{\rm 1.4\ GHz}$
  ratio as calculated by converting $L_{\rm 1.4\ GHz}$ into star
  formation rate (SFR) following PR07, and then converting SFR into
  $\lx$ following L10.
  {\bf Panel (b)}: The ratio between the rest-frame \fbrange\ apparent
  luminosity $\lx$ and the expected \fbrange\ luminosity from star
  formation (calculated following L10) vs.~the specific Star-Formation
  Rate (sSFR; $SFR/M_\star$, in per billion years, Gyr$^{-1}$). The
  dashed line marks the best-fit correlation of $\lx = \alpha M_\star
  + \beta \mbox{SFR}$ in L10, and the dotted lines show the 1$\sigma$
  scatter in this correlation.
  {\bf For both panels}: SMGs above the solid line (having
  $\lx>5\times L_{\rm X,SF}$) are classified as AGN hosts. As in
  previous plots, \xray\ detected SMGs are plotted as large blue dots
  and labeled with their short LESS IDs. ALESS 66.1 is plotted as an
  open circle here as its stellar-mass estimate has high uncertainty
  (see Section~\ref{sec:class3}). For comparison, the sources from L10
  are also plotted. The green squares are LIRGs from L10, red dots are
  L10 LIRGs that are dominated by AGNs in the \xray, and green
  squares with red dots in the center represent L10 LIRGs which show
  evidence of AGN activity. The small red dots are \xray\ sources
  classified as `AGNs' by X11, and the small green squares are sources
  classified as `Galaxies' by X11. The star marks the average radio
  luminosity and averaged stacked \xray\ luminosity for the 4
  radio-bright SMGs in our stacking sample (see
  Section~\ref{sec:stack}).
  All SMGs other than ALESS 17.1, 45.1, and 67.1 are classified as AGN
  hosts under Method~{\sc III}. See Section~\ref{sec:class3} and the
  notes in Table~\ref{tab:classify} for more details. \label{fig:class3}}
\end{figure*}

% IIIa
\textbf{Classification Method {\sc IIIa}} is to compare $\lx$ against
the rest-frame 1.4 GHz monochromatic luminosity, $L_{\rm 1.4GHz}$,
from which we derived a star formation rate (SFR) and $L_{\rm
  X,SF}$. Figure~\ref{fig:class3}a illustrates this classification
scheme. SMGs above the solid line ($\lx \geq 5\times L_{\rm X,SF}$)
are classified as hosting AGNs. The calculation of $L_{\rm 1.4GHz}$ is
described in Section~\ref{sec:multiv}. $L_{\rm X,SF}$ is derived using
a correlation between $L_{\rm 1.4GHz}$ and SFR for pure starbursts or
high-redshift star-forming galaxies (Equation~11 of
\citealt{Persic2007}, PR7 for short, and references therein), and then
converting SFR into $L_{\rm X,SF}$ following \cite{Lehmer2010} (L10
hereafter). The equations used are
\begin{eqnarray}
{\rm SFR} &=&  L_{\rm 1.4GHz}/ 8.93\times 10^{20};\label{eqn:sfrlr} \\
\log{(L_{\rm X,SF}/1.21)} &=& 39.49+0.74 \log{\rm (SFR/1.8)},\label{eqn:lxsfr}
\end{eqnarray}
where SFR is in $\msun$ yr$^{-1}$ and $L_{\rm 1.4GHz}$ is in W
Hz$^{-1}$. The factor $1.8$ division of the SFR is for converting the
Salpeter IMF adopted in this paper to the Kroupa IMF used by L10. The
factor $1.21$ is for converting the rest-frame 2--10~keV luminosity
$L_{\rm 2-10keV}$ adopted in L10 into the \fbrange\ luminosity $\lx$ in this
paper, and it was derived with {\tt XSPEC} using a simple power-law
model with $\Gamma=1.5$ (e.g., \citealt{Teng2005, Lehmer2008}). Though
there might be large scatter in conversions from $L_{\rm 1.4GHz}$ to
SFR and SFR to $\lx$, Figure~\ref{fig:class3}a clearly shows that our
adopted threshold of $\lx>5\times L_{\rm X,SF}$ appears to be
sufficient for identifying outliers in the correlation between $L_{\rm
  1.4GHz}$ and $\lx$. Method {\sc IIIa} classifies all but ALESS 17.1,
45.1 and 67.1 as AGN hosts. The results are discussed at the end of
this subsection.

% All radio detected; radio flux may have AGN contribution
It is notable that all of our \xray\ detected SMGs are also radio
detected, whereas among all of the 99 SMGs in the ALESS main catalog,
only 39 SMGs are matched with a radio source within $1\arcsec$ using
the \cite{Biggs2011} radio catalog. This is perhaps not surprising,
since \xray\ luminosity correlates with radio luminosity for starburst
galaxies like SMGs (e.g., \citealt{Schmitt2006}; PR07) and also the K
correction is similar in the \xray\ and radio bands compared to
submm. Furthermore, since AGNs can also contribute significantly in the
radio regime (e.g., \citealt{Roy1997,Donley2005,DelMoro2013}), it is
plausible that the \xray\ detected SMGs are generally brighter in
radio because of the extra radio flux contribution from AGNs. If this
is true, then the SFR derived based on radio fluxes are
over-estimated, which means that $L_{\rm X,SF}$ values are
over-estimated. Since classification method {\sc IIIa} would only
become more conservative due to this effect, we do not attempt to
correct for the AGN contribution in the radio band.

% IIIb
\textbf{Classification Method {\sc IIIb}} uses less direct but tighter
correlations to derive SFR and $L_{\rm X,SF}$ (see
Figure~\ref{fig:class3}b.
The SFRs were derived following Equation~3 in \cite{Kennicutt1998},
\beq
{\rm SFR} = 1.8\times 10^{-10} \times L_{\rm IR}/\lsun,\label{eqn:sfrlir}
\eeq
where SFR is in $\msun$~yr$^{-1}$ and the solar luminosity
$\lsun=3.9\times 10^{33}$ \ergs. We use the $L_{\rm IR}$ derived by
\cite{Swinbank2013} as described in Section~\ref{sec:multiv}.
We then derived $L_{\rm X,SF}$ from SFRs following L10:
\beq
L_{\rm X,SF} = 0.67\times (9.05\times10^{28}\cdot M_\star 
+ 1.62\times10^{39}\cdot {\rm SFR}),
\label{eqn:usemass}
\eeq
where SFR is in $\msun$ yr$^{-1}$ and the galaxy stellar mass,
$M_\star$, is in solar masses (see Section~\ref{sec:multiv}). For SMGs
which have very large SFRs, the contribution from the term
$9.05\times10^{28}\cdot M_\star$ is negligible ($<1\%$) except for a
couple of sources. The factor $0.67$ is for converting $L_{\rm
  2-10\ keV}$ into $\lx$ ($\times 1.21$), and for converting $M_\star$
as L10 adopted the Kroupa IMF ($/1.8$). This correlation has a smaller
scatter than the correlation adopted in Method {\sc IIIa} (see Table 4
in L10). Method {\sc IIIb} gives consistent classification results as
{\sc IIIa}, i.e., all but ALESS 17.1, 45.1 and 67.1 are classified as
AGN hosts.

% Discussion of classification results, 17.1
To summarize the classification results under Method {\sc III}, all
but ALESS 17.1, 45.1 and 67.1 are classified as AGN hosts consistently
under two different ways of calculating SFR and $L_{\rm X,SF}$. For
the case of ALESS 17.1, however, we argue that it probably hosts an
AGN based on its \xray\ spectral hardness (see
Section~\ref{sec:class12} and Fig.~\ref{fig:class12}) and the results
of classification Method {\sc IIIb}. The apparent \xray\ `deficit'
shown in Figure~\ref{fig:class3}a is likely due to radio contribution
from its AGN, and potentially also combined with the effect of gas and dust
obscuration, which is consistent with its low $\gammaeff$ and its high
hardness ratio and $N_{\rm H}$ value (see Fig.~\ref{fig:hard},
\ref{fig:class12}, and Table~\ref{tab:spec}).  The
absorption-corrected luminosity $\lxint$ for ALESS 17.1 would put it
above the classification threshold of Method {\sc IIIa} (the same
applies for ALESS 67.1 for both Method {\sc IIIa, b}, but its $\lxint$
is an upper limit). For similar reasons, this classification criterion
does not rule out the possibility of ALESS 45.1 and 67.1 hosting AGNs,
though their \xray\ spectral hardnesses and luminosities are
consistent with them being dominated by just star-formation activity
in the \xray\ band.

%------------------------------------------------------------------------
\subsection{Method IV.\ $f_{\rm 3.6\mu m}$ vs.\ $\fx$}\label{sec:class4}

% initiatives
We also use \xray--to--optical/NIR flux ratio as an AGN activity
indicator. In X11, sources with $\log (\fx/f_R)>-1$ are classified as
AGNs (e.g., \citealt{Maccacaro1988, Hornschemeier2001,
  Bauer2004}). However, as we are dealing with high-redshift ($z>1$)
SMGs whose observed $R$ band corresponds to the rest-frame UV band, it
is more appropriate to use a redder band to trace the stellar
component. We therefore chose the IRAC Channel 1 3.6 \micron\ band
(rest-frame $J$ band for a $z=2$ source) to replace the $R$ band, and
utilized the classifications of the X11 4~Ms \cdfs\ sources to
calibrate the classification threshold for $\log (\fx/f_{\rm 3.6\mu
  m})$. The $f_{\rm 3.6\mu m}$ data for the X11 \xray\ sources were
compiled by X11 based on the {\it Spitzer} SIMPLE catalog
\citep{Damen2011}, and the $f_{\rm 3.6\mu m}$ data for our \xray\
detected SMGs are from \cite{Simpson2013}.

\begin{figure*}[!th]
  \center
  \includegraphics[angle=0.,scale=0.7]{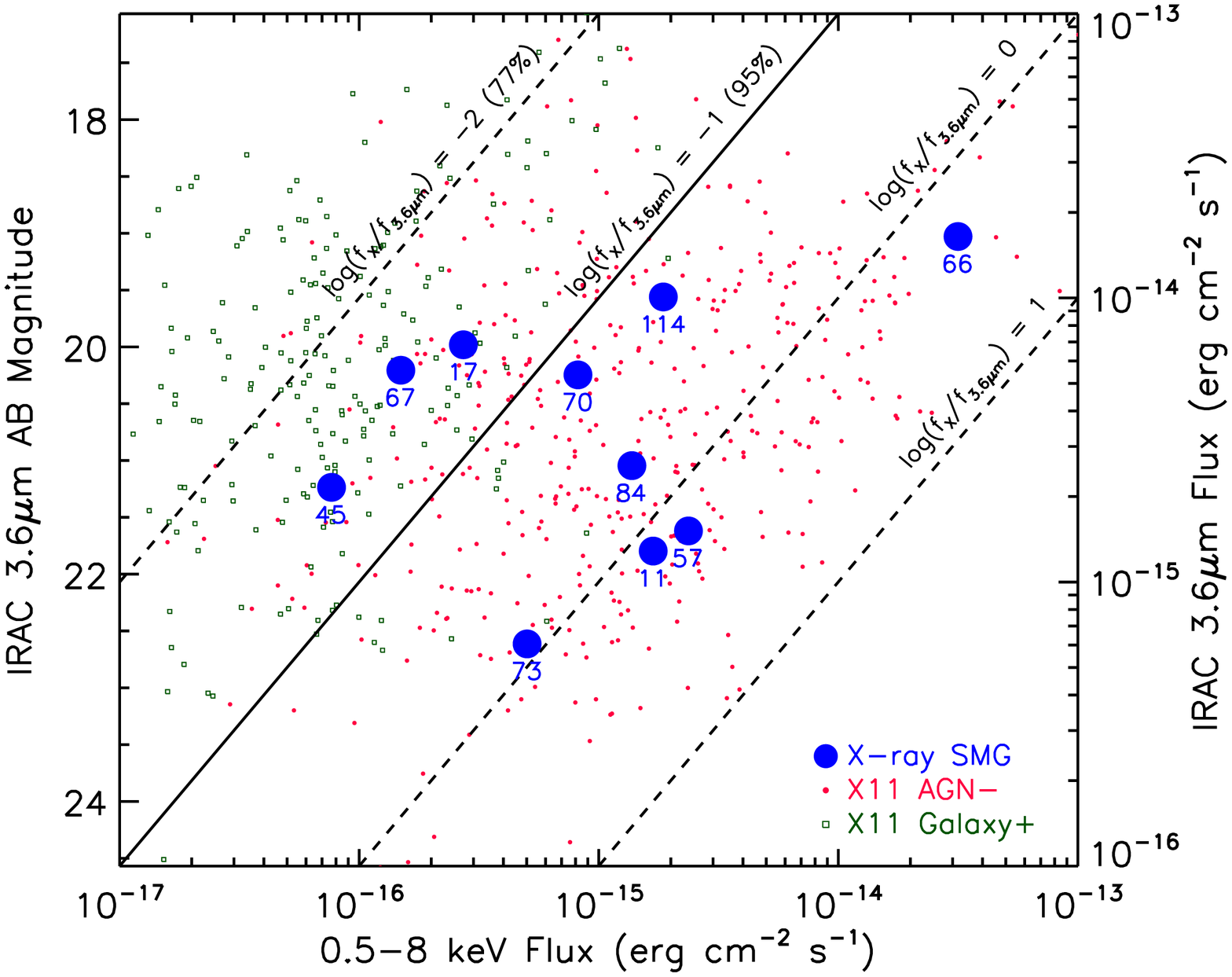}
\caption{{\bf Classification Method~{\sc IV}}: IRAC 3.6 $\mu$m
  magnitude/flux vs.~observed-frame \fbrange\ flux $\fx$ (with no
  absorption correction).
  The small red dots and green squares are for the \xray\ sources in
  X11. However, as detailed in Section~\ref{sec:class4}, to be
  conservative, the sources that have $z>1$ and were classified as
  `AGNs' {\it only} under the $f_R$--$\fx$ criterion in X11 were taken
  as `Galaxies' here (hence the label `AGN$-$' and `Galaxy$+$').
  To the lower right of the solid line ($\log{(\fx/f_{\rm 3.6\mu
      m})}>-1$), 95\% of the X11 sources on the plot are AGNs
  (similarly for the 77\% label). Therefore, we choose to classify the
  SMGs with $\log (\fx/f_{\rm 3.6\mu m})>-1$ as AGNs.
  For X11 sources that are only detected in the soft band or the hard
  band, the corresponding fluxes in the detection band are used
  instead of full-band flux upper limits.
  \label{fig:class4}}
\end{figure*}

% calibrate the classification threshold
Figure~\ref{fig:class4} illustrates this approach: all the X11 sources
having IRAC 3.6 \micron\ detections (643 sources; see details in
Section 4.4 of X11) are plotted here with symbols representing their
classifications --- AGNs as red dots and galaxies as green open
squares. The classifications are the same as in X11 with only one
exception: the 52 sources that have $z>1$ and were classified as `AGN'
{\it only} under the $\log (\fx/f_{\rm R})>-1$ criterion in X11 are
conservatively taken as galaxies here (hence the `AGN$-$' and
`Galaxy$+$' labels in the legend). This is for the purpose of calibrating
our \xray\--to--optical/NIR flux ratio threshold in a reliable and
conservative way. Also, if a source is only detected in the soft or
hard \xray\ band and thus only has an upper limit for the full-band
flux $\fx$ in X11, we used its soft- or hard-band flux for our
calibration (i.e. a lower limit of $\fx$). We then looked for the
$\log(\fx/f_{\rm 3.6\mu m})$ threshold above which a majority of the
X11 sources are AGNs. For this purpose and to avoid fine-tuning, we
chose $\log(\fx/f_{\rm 3.6\mu m})>-1$ as our classification threshold
(i.e., below the solid line in Figure~\ref{fig:class4}), as $\sim
95$\% of the X11 sources that satisfy this criterion are AGNs. 

% results and comments
Under this criterion, all but ALESS 17.1, 45.1 and 67.1 are classified
as AGN hosts, consistent with Method~{\sc III}.  We note here that
the somewhat conservative $\log(\fx/f_{\rm 3.6\mu m})$ threshold we
chose gives high reliability ($1-95\%=5\%$ mis-classification rate)
but relatively low completeness, as $26$\% of the X11 AGNs in
Figure~\ref{fig:class4} actually have $\log(\fx/f_{\rm 3.6\mu m})<-1$.

%------------------------------------------------------------------------
\subsection{Method V.\ X-ray Variability}\label{sec:class5}

% overall, cdfs vs ecdfs
\xray\ variability is one of the distinct characteristics of AGNs, and it is
an especially powerful tool for identifying highly obscured or
low-luminosity AGNs where the galaxy light may dominate even in the \xray\
regime. \cite{Young2012} studied 92 \xray\ galaxies in the 4~Ms \cdfs, and
found 20 \xray\ variable galaxies that are likely to host low-luminosity
AGNs. Here we follow the method in Section~3 of \cite{Young2012} and examine the
variability of the 8 sources in the \cdfs.\footnote{For ALESS 66.1 and
  70.1, which are only in the \ecdfs\ region, the total exposure time
  and time span for the \ecdfs\ observations are not sufficiently long for such
  studies.}

% following young 2012 method
First, we divided the \cdfs\ observations into four $\sim$1 Ms epochs,
which can give the amount of variability each source exhibits on
$\sim$month--year time scales in the observed frame.  Then through
Monte Carlo simulations, we calculated the probability $P$ that their
variability exceeds that expected from Poisson statistics. If $P$ is
less than 5\%, we conclude that the source is variable.

% results
We found that ALESS 84.1 is \xray\ variable ($P=0.025$), and it has a
maximum-to-minimum flux ratio of $3.1$ over the observed $10.8$ yr
time frame. We were unable to conclude similarly for any of the other
7 sources in the \cdfs. This does not rule out the possibility of them
being truly variable sources, since these sources have large off-axis
angles (high background counts) and/or low net source counts (see
Table~\ref{tab:spec}), which give us relatively low statistical power
when testing their variability.

%------------------------------------------------------------------------
\subsection{Summary of Classification Results}\label{sec:classall}

Combining the results of all the methods described above, 8 out of the
10 \xray\ detected SMGs show strong evidence of containing AGNs under
at least one classification method (see Table~\ref{tab:classify}). In
fact, 6 of these 8 sources are classified as AGN hosts consistently
under at least 3 methods. The only exception is ALESS 17.1, which is
identified as an AGN host only through its spectral hardness (Method
{\sc II}) because its \xray\ luminosity ($\lx$ and $\lxint$) appears to
be low due to heavy intrinsic obscuration ($\nh>10^{23}$ cm$^{-2}$;
see discussion in Section~\ref{sec:class3}).

We did not find strong AGN signatures in the \xray\ for ALESS 45.1 or
67.1 using any of the methods, though it is notable that their $\lx$
values exceed the expected $L_{\rm X,SF}$ by a factor of $\sim$3 (see
Section~\ref{sec:4567} for more details). They are treated as
starburst dominated systems instead of `\xray\ AGNs' in our AGN
fraction ($\fagn$) analysis in the next section. We note here that our
classification does not rule out the possibility of ALESS 45.1 or 67.1
hosting AGNs. Future panchromatic SED analysis and/or spectroscopic
study may reveal hidden or weak AGNs, or even AGNs in the
\xray\ undetected SMGs. For example, ALESS 45.1 actually lies within
the `Donley wedge' \citep{Donley2012}, which is an AGN selection
scheme using the four IRAC bands and is robust for screening out
high-redshift starburst galaxies (as compared to the `Lacy wedge' or
`Stern wedge'; \citealt{Lacy2004, Lacy2007, Stern2005}). However, such
analyses are beyond the scope of our paper as we focus on the
\xray\ properties of the SMGs. See more discussion in
Section~\ref{sec:4567}.

%%%%%%%%%%%%%%%%%%%%%%%%%%%%%%%%%%%%%%%%%%%%%%%%%%%%%%%%%%%%%%%%%%%%%%%%%%%%%%%%%%%%%%%%%%%%%%%%%%%%
\section{The AGN Fraction in Submm Galaxies}\label{sec:fagn}

% definition and clarification of terms
As concluded in the previous section, we classified 8 of the 10
\xray\ detected SMGs as AGN hosts (SMG-AGNs). That is, among the 91
ALESS SMGs in the \ecdfs\ region, 8 show \xray\ signatures of AGNs. As
mentioned in Section~\ref{sec:intro}, this does not mean the primary
energy sources of 8 SMGs are AGNs, though AGNs do dominate in the
\xray\ band (see discussion in Section~\ref{sec:vsQSO}). In this
section, we calculate the fraction of such `\xray\ AGNs' among this
population of SMGs in the \ecdfs. We refer to this fraction simply as
the AGN fraction, $\fagn$, hereafter. For a comparison between our
$\fagn$ results and the previous ones, see
Section~\ref{sec:comp:fagn}.

%------------------------------------------------------------------------
\subsection{Methodology}\label{sec:fagnmethod}

% why correcting for sensitivity inhomogeneity is important
We follow the method discussed in Section~3.1 of \cite{Lehmer2007} and
Section~5 of \cite{Silverman2008} to calculate $\fagn$, which takes
into account the spatial inhomogeneity of the \xray\ sensitivity limit across
the \ecdfs. To illustrate why such a consideration is necessary,
suppose that, by chance, all ALESS SMGs lay in the least sensitive region
in the \ecdfs\ (the lightest gray in Figure~\ref{fig:map}, with \xray\
sensitivity $>10^{-15}$ \ergcms). Then only 5 SMGs would have been
\xray\ detected and classified as AGNs (see the \xray\ flux listed in
Table~\ref{tab:basic}). Therefore, simply dividing the numbers of
SMG-AGNs and SMGs would bias $\fagn$ toward either larger or
smaller values depending on the spatial distribution of SMGs on an
inhomogeneous \xray\ sensitivity map.

% describe the method, for flux first
The AGN fraction, $\fagn$, we estimated here is the flux-- or
luminosity--dependent cumulative fraction. That is, the $\fagn$ value
for a certain \xray\ flux/luminosity represents the fraction of SMGs
that host AGNs with equal or greater \xray\ flux/luminosity. We first
describe how $\fagn$ is calculated as a function of full-band observed-frame
\xray\ flux ($\fx$; $f_{\rm X}$ for short).  Following
\cite{Silverman2008}, the cumulative AGN fraction for SMGs hosting
AGNs with \xray\ flux larger than or equal to $f_{\rm X,lim}$ is
\beq
  \fagn(f_{\rm X}\geq f_{\rm X,lim}) = \sum_{i=1}^{N}\frac{1}{N_{\rm SMG,i}},
\eeq
where $N$ is the number of SMG-AGNs with $f_{\rm X,i} \geq f_{\rm
  X,lim}$ ($f_{\rm X,i}$ being the \xray\ flux of the $i^{\rm th}$
SMG-AGN); and $N_{\rm SMG,i}$ represents the number of SMGs which lie
in a region with a sufficient \xray\ sensitivity limit such that they
would have been detected if hosting AGNs with $f_{\rm X} \geq f_{\rm
  X,i}$. Since we have 8 SMG-AGNs in the sample, naturally, we chose
the $f_{\rm X,lim}$ values to be each of the 8 $f_{\rm X,i}$ values in
turn. The error for $\fagn$ is calculated by adding in quadrature the
error for each $1/N_{\rm SMG,i}$ term (estimated following Gehrels
1986).  The results of $\fagn(f_{\rm X})$ are plotted in the
upper-left panel of Figure~\ref{fig:fagn}.

% fagn vs Lx or Lxint
The $\fagn$ values as a function of rest-frame \fbrange\ apparent
luminosity, $\lx$ ($L_{\rm X}$ for short), or rest-frame \fbrange\
absorption-corrected luminosity, $\lxint$ ($L_{\rm X,corr}$ for
short), can be calculated likewise. An additional step arises when
determining whether or not the ${\rm j^{th}}$ \xray\ undetected SMG at
redshift $z_{\rm j}$ would have been detected if hosting an AGN with
$L_{\rm X,i}$ or $L_{\rm X,corr,i}$. This additional step is to
translate $L_{\rm X,i}$ or $L_{\rm X,corr,i}$ into the (hypothetical)
observed flux (i.e., the same metric as the \xray\ sensitivity map),
assuming an AGN with $L_{\rm X,i}$ or $L_{\rm X,corr,i}$ in the ${\rm
  j^{th}}$ \xray\ undetected SMG.

% converting Lx
For translating $L_{\rm X,i}$ into flux, we used the equation for
calculating $\lx$ in Section~\ref{sec:xspec} with the redshift $z_{\rm
  j}$ and the $\gammaeff$ value of the ${\rm i^{th}}$ SMG-AGN,
$\Gamma_{\rm eff,i}$.
% converting Lxint
For translating $L_{\rm X,corr,i}$ into flux, we first calculated its
corresponding rest-frame apparent luminosity (absorbed), $L'_{\rm
  X,i,j}$, for the hypothetical AGN with $L_{\rm X,corr,i}$ hosted by
the ${\rm j^{th}}$ \xray\ undetected SMG. Such a conversion requires an
intrinsic column density ($N_{\rm H,j}$) assigned to each \xray\
undetected SMG. To do so, we drew each $N_{\rm H,j}$ value randomly
from the $\nh$ distribution described in Section~4 of
\cite{Rafferty2011}, which was formulated based on the results in
\cite{Tozzi2006}.\footnote{Briefly, this $\nh$ distribution includes a
  log-normal distribution for $20<\log{\nh/({\rm cm^{-2}})}<23$ which
  centers around $\log{\nh/({\rm cm^{-2}})} = 23.1$ with $\sigma
  \simeq 1.1$. It flattens out beyond $10^{23}\percm2$ and truncates
  at $10^{24}\percm2$, and it includes 10\% of objects with low column
  density ($\nh=10^{20}\percm2$).}
We then found $L'_{\rm X,i,j}$ by calculating the ratio $L_{\rm
  X,corr,i}/L'_{\rm X,i,j}$ through a simulation in {\tt XSPEC} using
the {\tt wabs*zwabs*zpow} model with $z_{\rm j}$, $N_{\rm H,j}$, and
the intrinsic photon index, $\Gamma_{\rm int,i}$, of the ${\rm
  i^{th}}$ SMG-AGN. Then we converted $L'_{\rm X,i,j}$ into the
observed flux in the same way as for converting $L_{\rm X,i}$ into
flux as described above. The results of $\fagn(L_{\rm X})$ and
$\fagn(L_{\rm X,corr})$ are shown in the middle- and lower-left panels
in Figure~\ref{fig:fagn}.

%------------------------------------------------------------------------
\subsection{The AGN Fractions}\label{sec:fagnresult}

% left panels of figure fagn, table fagn, the left most points
The black dots in the three left-hand panels of Figure~\ref{fig:fagn} show
the cumulative AGN fractions for the \ecdfs\ ALESS SMGs. The left-most
point in each panel is essentially the fraction of SMGs hosting AGNs
with flux/luminosity equal to or above the current faintest SMG-AGN in \xray\
ever detected in the \ecdfs. These $\fagn$ values are marked in the plots
and listed in Table~\ref{tab:fagn}. 

% submm flux cuts
The $\fagn$ analysis above is for all ALESS SMGs in the \ecdfs. This
sample of SMGs, however, may not constitute a flux-limited sample
of SMGs. The reason is that ALESS differs from a regular
flux-limited survey since it targeted the bright submm sources
discovered by LESS. As mentioned earlier, the LESS survey is in fact a
contiguous and flux-limited survey over the whole field of
\ecdfs. Thus, the LESS sources constitute a flux-limited sample for
$S_{\rm 870\mu m}\geq 3.5$--$4.5$~mJy (LESS de-boosted flux limit;
\citealt{Weiss2009}). Consequently, the ALESS SMGs with submm flux above
the LESS flux limit are actually a flux-limited sample. We hence
removed the SMGs (including SMG-AGNs) with $S_{\rm 870\mu m} <
3.5$~mJy in our sample and calculated the AGN fraction for the
flux-limited ($\geq 3.5$~mJy) SMG sample in the \ecdfs. The results
are shown in the right-hand panels of Figure~\ref{fig:fagn}, and the
$\fagn$ values for the left-most points are listed in
Table~\ref{tab:fagn}.

% some comments/observations
As expected, since brighter AGNs in general are rarer (e.g.,
\citealt{Xue2010, Aird2012}), $\fagn$ decreases as \xray\ flux or
luminosity increases for both cases with or without the submm flux
cut. Also, comparing the left panels with the right panels of
Figure~\ref{fig:fagn}, the AGN fractions for the $S_{\rm 870\mu m}\geq
3.5$~mJy SMG sample are larger than those for all ALESS SMGs overall. This
is not surprising given that 5 out of 8 (63\%) of the SMG-AGNs have
flux larger than 3.5 mJy, while a smaller fraction of SMGs without AGN
have $S_{\rm 870\mu m}\geq 3.5$ mJy (46 out of 91, 51\%). However, the
$\fagn$ values are actually consistent between the two SMG samples
within the error bars.

% fagn vs submm flux
A general trend of larger $\fagn$ for SMG groups with larger submm
flux $S_{\rm 870\mu m}$ was also observed when we computed $\fagn$ as
a function of $S_{\rm 870\mu m}$. It rises from about
$15^{+15}_{-5}$\% for $S_{\rm 870\mu m} \geq 1.3$ mJy (faintest ALESS
SMG) to about $34^{+37}_{-17}$\% for $S_{\rm 870\mu m} \geq 6$
mJy. However, the error bars on $\fagn$ are large due to the limited
sample size, especially at high submm flux, and the $\fagn$ values are
actually consistent with each other within 1$\sigma$ error bars,
including the two $\fagn$ values at the lowest and highest $S_{\rm
  870\mu m}$ ends. 

% discuss the dashed line
The dashed lines in each panel of Figure~\ref{fig:fagn} are $\fagn$
values if all 10 \xray\ detected SMGs, including ALESS 45.1 and 67.1,
are taken as AGN hosts. Note that this would only affect the $\fagn$
results at the low flux/luminosity end as ALESS 45.1 and 67.1 are very
faint in \xray. As we do not have strong evidence that ALESS 45.1 or
67.1 host AGN, this line can be viewed as the `fractions of SMGs that
have \xray\ flux/luminosity above certain values' instead of
cumulative AGN fractions. Future discoveries of SMG-AGNs with lower
\xray\ luminosities (probably involving disentangling the
contributions from the star formation and AGN) will be able to push
the $\fagn$ function to the lower $\fx$/$\lx$ ends.

%%%%%%%%%%%%%%%%%%%%%%%%%%%%%%%%%%%%%%%%%%%%%%%%%%%%%%%%%%%%%%%%%%%%%%%%%%%%%%%%%%%%%%%%%%%%%%%%%%%%
\section{Stacking of the X-ray Undetected SMGs}\label{sec:stack}

% Stacking source selection
Thanks to the high-precision positions of the SMGs provided by ALESS,
we are able to assess directly and reliably the average
\xray\ properties of the \xray\ undetected SMGs through stacking.
Previous works on \xray\ stacking of SMGs were typically based on the
positions of the IR/radio counterparts of SMGs (e.g.,
\citealt{Laird2010, Georgantopoulos2011, Lindner2012}), which, as
mentioned in Section~\ref{sec:intro} and \ref{sec:cat}, suffer from
the larger uncertainties in counterpart matching due to poor angular
resolution of single-dish submm surveys.

% selecting sources
To avoid poor \xray\ PSF regions and high background, we only stacked
sources within $7 \arcmin$ of the aim points of the 4~Ms \cdfs\ or 250
ks \ecdfs. SMGs within $2\times$ the 90\% encircled-energy aperture
radii of any \xray\ sources are also not included in our stacking
analysis. There are 50 sources that have small enough off-axis angles
and are far enough from any \xray\ sources, and only one of them lies
within the 4~Ms \cdfs\ region; it has an effective exposure time
$\sim$15 times larger than the other 49 sources in the \ecdfs. To
prevent this single source from biasing our stacking results, we did not
include it in the stacking. Therefore, there are 49 SMGs in total in
our stacking sample, all in the \ecdfs\ region.

% Stacking procedures and direct results
We follow the stacking procedures detailed in Section 3.1 of
\cite{Luo2011}. Briefly, we extracted the \xray\ counts within an
aperture of $1.5 \arcsec$ in radius centered around the submm position
of each of the 49 SMGs. The background counts for each SMG were
estimated by extracting counts within 1000 randomly placed $1.5
\arcsec$-radius apertures within $1 \arcmin$ of each SMG, avoiding
\xray\ sources and the central SMG, and then taking the average. The
total stacked counts ($S$) for these 49 SMGs are 33.5 in the soft band
and 54.0 in the hard band. The stacked background counts ($B$) are
13.6 (soft) and 40.2 (hard). Thus, the net counts are 20.0 (soft) and
13.8 (hard), and the signal to noise ratios S/N\footnote{We note here
  that the background counts are scaled to match the $1.5\arcsec$
  source extraction aperture. Total background counts are 1000$\times$
  larger since they are estimated using 1000 $1.5\arcsec$
  apertures. Thus the S/N can be calculated using Gaussian statistics
  as $(S-B)/(\sqrt{B\times1000}/\sqrt{1000})=(S-B)/\sqrt{B}$.}
($(S-B)/\sqrt{B}$) are 5.4$\sigma$ (soft) and 2.1$\sigma$ (hard),
which correspond to a probability of $p=3.7\times 10^{-6}$ for being
generated by Poisson noise for the soft band, and $p=0.021$ for the
hard band. The smoothed stacked images are shown in
Figure~\ref{fig:stack}. A summary of our stacking results can be found
in Table~\ref{tab:stack}.

% robustness test
We performed a robustness test by generating 1000 fake submm catalogs
at random RA and Dec with 49 sources each (avoiding \xray\ sources),
and stacking them the same way as described above. None of the 1000
cases has a S/N of $\sigma \geq 5.4$ ($<0.1$\%) in the soft band, and
21 cases ($2.1$\%) have S/N of $\sigma \geq 2.1$ in the hard band,
which are both consistent with our findings.

% radio excess, motivation
We then explore the possibility of identifying a sub-group of SMGs in
our stacking sample that have higher probabilities of hosting AGNs and
thus may contribute a significant fraction of the total stacked
signal. According to our $f_{\rm AGN}$ analyses in
Section~\ref{sec:fagn}, the expected number of SMG-AGNs in the entire
ALESS SMG sample is $99\times 16\% \approx 16$ for $\lx \geq
2.63\times 10^{42}$ \ergs\ (Table~\ref{tab:fagn}). That is, about 8
out of the 89 \xray\ undetected SMGs are expected to host AGNs with
$\lx \geq 2.63\times 10^{42}$ \ergs, which means for our stacking
sample of 49 \xray\ undetected SMGs, there would be a further $\sim$4
AGNs ($8\times49/89$) with $\lx \geq 2.63\times 10^{42}$ \ergs.

% radio excess, source selection
Recently, \cite{DelMoro2013} identified a group of distant
star-forming galaxies harboring AGNs with excess radio flux compared
to their rest-frame FIR flux and SED analyses (also see similar works
by \citealt{Roy1997} and \citealt{Donley2005}). Motivated by their
work, we stacked the 4 radio-detected SMGs which have the largest
rest-frame radio-to-IR luminosity ratios ($\rri=L_{\rm 1.4GHz}/L_{\rm
  IR}\times 10^{-12}$). These 4 SMGs (ALESS 2.1, 14.1,
75.1,\footnote{ALESS 75.1 exhibits an 8~\micron\ excess with respect to
  an SED fit using a star forming galaxy template
  \citep{Simpson2013}. This supports the idea that it is likely to
  host an AGN.} and 122.1) also have SFR estimated using $L_{\rm
  1.4GHz}$ (Equation~\ref{eqn:sfrlr}) $>3$ larger than SFR estimated
with $L_{\rm IR}$ (estimated using Equation~\ref{eqn:sfrlir} and
$L_{\rm IR}$ values from \citealt{Swinbank2013}). Figure~\ref{fig:rrs}
shows the histogram of the $\rri$ values of the 49 SMGs in our
stacking sample, with the inset showing the histogram for all ALESS
SMGs ($3\sigma$ flux upper limits were used when sources were not
detected in radio). The \xray\ stacking of these 4 SMGs yields a
$7.8\sigma$ detection ($S=8.9$, $B=1.0$, $p=1.5\times 10^{-6}$) in the
soft band and a $4.3\sigma$ detection ($S=10.9$, $B=3.2$, $p=6.1\times
10^{-4}$) in the hard band. These 4 SMGs (8\% of the 49 stacked SMGs)
contribute 46\% of the total full-band net counts of the entire
stacking sample (15.5 out of 33.4 photons), and their individual net
counts are among the top $\sim$20\% of the 49 stacked SMGs.

% radio-exc: properties
Using the BEHR package (\citealt{Park2006}; see
Section~\ref{sec:xspec1}), the hardness ratio is estimated to be
$1.0^{+0.8}_{-0.5}$ (68.3\% confidence interval, or ``1$\sigma$" error
bars), which corresponds to an effective photon index of
$\gammaeff=0.8^{+0.6}_{-0.5}$, following the conversion between
hardness ratio and $\gammaeff$ described in
Section~\ref{sec:xspec1}. After taking into account the
encircled-energy fraction for our extraction aperture for each SMG and
using $\gammaeff=0.8$, we estimated the average \xray\ flux per SMG to
be $3.6\times 10^{-16}$ \ergcms\ in the full band ($6.3\times
10^{-17}$ \ergcms\ in the soft band), with a $1\sigma$ error of $\sim
50$\% (flux and error calculation following L05 and X11). We note that
their average full-band flux is above the sensitivity limit of the
central $\sim$8$\arcmin$ region of the 4~Ms \cdfs.

% radio-exc Lx and comparison with Lx,sf, and they're likely AGNs
Using the full-band flux estimated above and the median redshift of
these 4 SMGs, $z=2.2$ (all are photometric redshifts from
\citealt{Simpson2013}), we estimate their average rest-frame
\fbrange\ apparent luminosity to be $\lx=3.6\times 10^{42}$
\ergs. This is more than $3\times$ the amount expected just from their
average star formation, $L_{\rm X,SF} \approx 1.1\times 10^{42}$
\ergs, estimated using their median stellar mass,
$\log{M_\star/\msun}=11.3$, and $\mbox{SFR}=960\ \msun$\peryr\ using
Equation~\ref{eqn:usemass} following L10. Considering their stacked
\xray\ luminosity and their hardness ($\gammaeff=0.8$, consistent with
being obscured AGNs; see Section~\ref{sec:class12}), these 4 SMGs are
likely to host AGNs (also see Figure~\ref{fig:class3}).

% the rest 45  
Stacking the remaining 45 SMGs (removing the 4 radio-bright SMGs)
yields a signal with $3.4\sigma$ in the soft band ($S=24.6$, $B=12.6$,
$p=0.0017$) and $1.0\sigma$ in the hard band ($S=43.1$, $B=37.0$,
$p=0.18$). Similar to the calculations above, we estimate the
effective photon index for the stacked signal of these 45 SMGs to be
$\gammaeff=1.6^{+2.3}_{-1.0}$, and the average full-band flux per SMG
to be $2.7\times 10^{-17}$ \ergcms. Using the median redshift of these
45 SMGs,\footnote{11 of these 45 SMGs have detections in no more than
  3 bands and thus their redshifts are drawn from the likely redshift
  distribution for these sources estimated by \cite{Simpson2013}.}
$z=2.9$, we estimate their average $\lx$ to be $1.2\times 10^{42}$
\ergs. The average $\lx$ is a factor of 2.4 larger than the amount of
\xray\ luminosity arising from just the star forming component of a
typical galaxy in the sample, $L_{\rm X,SF}=4.8\times 10^{41}$ \ergs,
estimated using their median stellar mass ($\log{M_\star/\msun}=10.9$)
and $\mbox{SFR}=430\ \msun$\peryr\ similarly as above. This hints that
additional AGNs might be in the sample, though we caution the reader
that the result is tentative due to the large uncertainty on the
estimated mean $\lx$. The error for $\lx$ is typically $\sim 70\%$ for
the lower 1$\sigma$ error and over 100\% for the upper 1$\sigma$ error
(estimated using both the error from the flux, $\sim$50\%, and the
error from $\gammaeff$ as listed above).

% that one sinlge CDFS source
As mentioned earlier, there is one SMG (ALESS 10.1; with a photometric
redshift of $z=2.0$ from \citealt{Simpson2013}) in the 4~Ms
\cdfs\ region that meets our stacking selection criteria but was not
included in our stacking sample. We note here that this SMG is also
likely to host an AGN. It has a soft band signal of $4.0\sigma$
($S=15.4$, $B=5.7$, $p=6.0\times 10^{-4}$), a hard band signal of
$2.2\sigma$ ($S=26.6$, $B=17.4$, $p=0.024$), and a full-band signal of
$3.9\sigma$ ($p=2.6\times 10^{-4}$). It was not included in our
\xray\ catalog because its false-positive probability is larger than
$10^{-5}$ (see Section~\ref{sec:xraycat}). We estimate its $\lx$ to be
$5.3\times 10^{43}$ \ergs, which is significantly beyond the expected
\xray\ luminosity arising just from its star formation ($L_{\rm
  X,SF}=6.0\times 10^{41}$ \ergs, estimated in the same way as above
using its redshift, $z=2.0$, its stellar mass,
$\log{M_\star/\msun}=10.6$, and $\mbox{SFR}=550\ \msun$\peryr).

% Null test of stacking irac sources?

%%%%%%%%%%%%%%%%%%%%%%%%%%%%%%%%%%%%%%%%%%%%%%%%%%%%%%%%%%%%%%%%%%%%%%%%%%%%%%%%%%%%%%%%%%%%%%%%%%%%
% Discussion Section
\section{Discussion}\label{sec:discuss}
% Section Discussion Overview
We have presented the \xray\ properties and AGN fractions of the
\xray\ detected ALESS SMGs in the \ecdfs\ region, as well as our
\xray\ stacking results, in the previous sections. We compare our
results with those of the previous studies in Section~\ref{sec:comp}
below, and compare the \xray\ and relevant multiwavelength properties
and the AGN fraction of our SMG sample with other populations in
Section~\ref{sec:otherpop}. We revisit the two \xray\ detected SMGs
that are not classified as AGNs, ALESS 45.1 and 67.1, and discuss the
origin of their \xray\ emission in Section~\ref{sec:4567}. In
Section~\ref{sec:future}, we discuss possible future works.

%---------------------------------------------------------------------------------------------------
\subsection{Comparison with Previous Studies}\label{sec:comp}

% Overall, mostly
In this section, we compare our study and results with four previous
studies of \xray\ AGNs in SMGs by A05,
\citeauthor{Laird2010}~(\citeyear{Laird2010}; La10),
\citeauthor{Georgantopoulos2011}~(\citeyear{Georgantopoulos2011};
G11), and \citeauthor{Johnson2013}~(\citeyear{Johnson2013}; J13). We
first compare the AGN fraction, \xray\ detection rate, and stacking
results, and then we explore the reasons behind the similarities and
differences between the results by comparing the submm/\xray\ catalogs
and methodologies used in the previous studies and ours.

%-------------------------------------------------------------
\subsubsection{The AGN Fraction among SMGs}\label{sec:comp:fagn}

Our AGN fraction, $\fagn$, results are presented in
Figure~\ref{fig:fagn} and Table~\ref{tab:fagn}. A direct comparison
between our results and those of A05, La10, G11 and J13 is also shown
in Figure~\ref{fig:fagn}, where the AGN fraction estimates from these
four studies are plotted with red letters. The $x$-axis
values of the red letters are the \xray\ flux or luminosity of the
faintest SMG-AGNs in these four previous studies (all
converted into the 0.5--8.0~keV energy range assuming $\gammaeff=1.4$
or $\gammaint=1.8$ when necessary). The quoted error bars on their AGN
fraction estimates are typically a few to ten percent, which is
roughly reflected by the size of the red letters. The error bars are
not plotted for clarity of presentation and also because they are likely to be
underestimated due to the large uncertainty in counterpart matching
and their simple methods of computing $\fagn$ values (see
Section~\ref{sec:comp:method} for details). Taking these factors into
account, our $\fagn$ estimates agree with the estimates in previous
studies, especially for the $\fagn$ results for SMGs with $S_{\rm
  870\mu m}>3.5$~mJy shown in the right panels of
Figure~\ref{fig:fagn}, considering the differences in the submm
catalogs (see Section~\ref{sec:comp:method}).

To be specific, the estimated $\fagn$ values are: $38^{+12}_{-10}\%$
in A05 (75$\pm19\%$ for the radio-selected SMGs);
$(20$--$29)\pm7$\% in La10 (29\% if 3 ambiguous sources are counted as
AGNs; the plotted value is 24.5\%); 18$\pm7$\% in G11 (for the sources
in the \cdfs\ only and no \ecdfs\ sources, which is more appropriate
for comparison here); and $\sim$28\% in J13.

The AGN fraction reported by A05 appears higher than the rest (though
consistent within 1$\sigma$ error bars), potentially due to the fact
that many of their SMGs were discovered via specific targeting of
radio sources. Radio-detected SMGs seem to have a higher AGN fraction
than the general SMG population, suggesting that the A05 result is
potentially biased towards high AGN fraction, as also noted by A05,
L10, and G11. The findings of A05 (a high AGN fraction among radio SMGs
with heavy obscuration) are also consistent with the stacking
analysis by \cite{Lindner2012}, who studied radio-selected SMGs and
found Fe~K$\alpha$ line emission with high equivalent width.

Among the 20 radio SMGs in \cite{Alexander2005a} and A05 (selected for
being radio sources), 15 (75$\pm19$\%) are identified as AGNs; and
among the 21 radio-detected SMGs in the ALESS catalog and in the
\cdfs\ region, 6 (29$\pm12$\%) are identified to be AGNs. The reason
why our number (29\%) is lower than that of A05 (75\%) could be that
the radio data associated with our work are deeper (ours:
\citealt{Biggs2011} down to $\sim$15~$\mu$Jy at $3\sigma$; A05:
\citealt{Richards2000,Chapman2004} down to $\sim$24~$\mu$Jy at
$3\sigma$), which would imply that brighter radio sources are more
likely to host AGN.

To test this, we performed an $\fagn$ analysis on the radio-detected
SMGs. For all radio-detected SMGs in our sample (36 SMGs), the AGN
fraction is typically a factor of 2--3 larger than that for the entire
ALESS sample at any given flux/luminosity limit. If we apply a radio
flux cut of 40~$\mu$Jy (to match with the radio depth of A05), then
the AGN fraction is a factor of 2.5--4 larger than that of all ALESS
SMGs, and also larger than the AGN fraction for that of all
radio-detected SMGs. Thus, it seems that AGN fraction increases with
increasing radio flux. However, the results are tentative due to
limited source statistics.

We emphasize that our AGN-fraction results are more reliable, and
our reported error bars are more realistic values than previous
studies because of our superb submm/\xray\ data and the analysis
method adopted. We explain in details in
Section~\ref{sec:comp:method}.

%-------------------------------------------------------------
\subsubsection{The X-ray Detection Rate of SMGs}

% Our X-ray detection rate (dashed line in fagn plot), flux
As mentioned in Section~\ref{sec:fagnresult}, the dashed lines
in Figure~\ref{fig:fagn} can be taken as the `fraction of SMGs that
have \xray\ flux/luminosity above certain values' instead of AGN
fractions. As indicated by the upper-left panel of
Figure~\ref{fig:fagn}, the fraction of ALESS SMGs that have
$\fx>7.7\times 10^{-17}$ \ergcms\ (the $\fx$ of ALESS 45.1) could be
as much as 50\% or even higher (1$\sigma$ lower limit; the actual
fraction is estimated to be 100\% due to the small number of
\xray\ detected SMGs at the faint end).

Our \xray\ detection rate estimate for the SMG population (dashed line
in Figure~\ref{fig:fagn}) is consistent within 1--2$\sigma$ with the
previous studies, especially considering the different \xray\ depths
and methodologies. For the previous studies with the \xray\ data from
the 2~Ms \cdfn: A05 reported an \xray\ detection rate of
$85^{+15}_{-20}\%$ among their radio-detected SMG sample, and La10 reported
45$\pm8$\%. For the studies with the \xray\ data from the 4~Ms \cdfs:
G11 reported 26$\pm9\%$, and J13 reported 34\% (both for the sources
within \cdfs\ only). We note that even for studies in the same
\xray\ field, the numbers are likely for different \xray\ depths,
because their SMG samples differ and their faintest \xray\ detected
SMGs are of different \xray\ fluxes.

Overall, we confirm the high \xray\ detection rate of SMGs as reported
in previous studies.

%---------------------------------------------------------------------------------------------------
\subsubsection{Comparison of Stacking Results}\label{sec:comp:stack}

This work is the first to perform \xray\ stacking directly at the
precisely known position of SMGs. All previous studies were aided by
the positions of the multiwavelength counterparts of SMGs and thus
suffer larger uncertainties. Thus, we caution the reader that the
comparison presented in this section is not a completely direct one.

The works of A05 and \cite{Lindner2012} perhaps suffer the least from
positional uncertainties and mismatches, as they were both
based on radio-selected SMGs. A05 stacked 3 SMGs that are
\xray\ undetected in their sample, and found marginal detections in the
soft and full bands, with an estimated rest-frame 0.5--8~keV
luminosity of $\lx=2\times10^{42}$ \ergs\ at median redshift
$z=2.1$. This is consistent with our result stacking all 49 SMGs in
our stacking sample, with $\lx=1.2\times10^{42}$ \ergs\ at median
$z=2.6$, especially considering the error bars for $\lx$
($\gtrsim 70\%$, see Section~\ref{sec:stack}). The work by
\cite{Lindner2012} stacked 38 SMGs in the Lockman Hole, and also
obtains a detection in the soft band. They estimated $\lx$ to be
$4.4\times10^{42}$ \ergs\ based on their \xray\ stacking in the
rest-frame of the SMGs (converted from rest-frame 2--8~keV luminosity
assuming $\gammaeff=1.5$). Considering the median SFR of SMGs is
around 500 $\msun$ \peryr (for the 49 SMGs in our stacking sample)
with expected $\lx=5.5\times10^{41}$ \ergs, the stacked mean $\lx$
values from this work, A05, and \cite{Lindner2012} all have
\xray\ excesses relative to the amount expected form just star
formation. Especially for A05 and \cite{Lindner2012}, which are both
on radio-selected SMGs, the excesses appear to be larger.

If such \xray\ excesses are real, this suggests that there are
AGNs in the stacking samples. In Section~\ref{sec:stack}, we mentioned
that there are potentially 4 AGN candidates among our stacking sample
of 49 SMGs selected via rest-frame radio-luminosity
excesses. \cite{Lindner2012} also conclude that there is strong
evidence supporting an AGN presence in their sample based on the
detection of strong Fe~K$\alpha$ line emission (with high equivalent
width $>1$~keV), which corroborates the results of the rest-frame
spectral analyses by A05 on the most obscured group of SMG-AGNs (see
Figure~7 of A05). These stacking results are broadly consistent with
the fact the radio-selected SMGs seem to have a large AGN fraction, as
mentioned in Section~\ref{sec:comp:fagn}.

The stacking analyses by La10 and G11 report $\lx=6\times10^{41}$
\ergs\ and $\lx=4\times10^{41}$ \ergs, respectively (both at $z=2.1$;
both converted into 0.5--8~keV rest-frame energies assuming
$\gammaeff=1.5$). These are lower than the results of A05 and
\cite{Lindner2012} and also our stacking results, though still
consistent within 1$\sigma$ error bars considering the large
uncertainty associated with $\lx$. However, their stacking analyses
were based on IRAC (La10) and MIPS (G11) positions, which could suffer
source mismatches and larger positional uncertainties (see the next
subsection). It is perhaps not surprising that these two stacking
analyses yield such a level of stacked \xray\ luminosity, regardless
of correct matching onto SMGs, since the IRAC or MIPS detected
galaxies are usually moderately or even highly star-forming
galaxies. It is also plausible that the $\lx$ values in La10 and G11 are
lower than ours because IRAC/MIPS selected sources have a lower SFR on
average and perhaps also a lower AGN fraction.

%-------------------------------------------------------------
\subsubsection{Comparison of Catalogs and Methodologies}\label{sec:comp:method}

% overall
We have mentioned above that compared to previous studies, our work
provides the most reliable estimates of AGN fraction (as well as
\xray\ fraction) and realistic error bars thanks to our superior
submm/\xray\ catalogs and the methodology adopted. We describe in
detail these differences between our work and previous studies below.

% advantage of submm catalogs
First, the superb angular resolution ($\sim$1$\arcsec$) and great
sensitivity ($\lesssim1.5$~mJy at 3.5$\sigma$) of ALMA has enabled us
to identify unambiguously the \xray\ counterparts to the SMGs with
high confidence for the first time among similar works. Due to a
larger false-positive rate, source blending, and the poor positional
precision in the previous generation submm surveys, earlier studies
suffer from large uncertainty in their analyses. As an illustration of
this issue, we compare with the submm sources and their
\xray\ counterparts identified in G11: among the 14
\xray\ counterparts in G11, 7 are not recovered in our
study,\footnote{Among the 7 sources in G11 that we do not recover, 2
  (W-4 and 40) are due to bad quality ALMA maps, 1 (W-108) has no ALMA
  detection in the field, and 4 (W-9, 59, 92, and 101) are false
  counterpart matches.} and we find 3 more true \xray\ counterparts
that are not in G11 (ALESS 17.1, 70.1, and 73.1). We note these 3
additional matches are all submm bright ($>$4~mJy), and therefore this
addition is not because our submm catalog (ALESS) is deeper than
previous ones, which are typically at a limit of 3--4~mJy. ALESS 17.1
and 73.1 are discovered with \xray\ counterparts due to the improvement
of \xray\ catalog depth (the 4~Ms \chandra\ catalog by X11 in this
work, compared with the 2~Ms in \citealt{Luo2008} used in G11). ALESS
70.1 is matched with an \xray\ source due to its improved submm
position. Such a large discrepancy in source matching could indicate
that our agreement on the AGN fraction with G11 (and possibly also
other similar studies) could be, at least partially, coincidental.

% X-ray catalog depth
Secondly, the \xray\ catalog we used includes data derived from the deepest
\xray\ survey --- the 4~Ms \cdfs, while previous studies mostly used
the 2~Ms \cdfn\ (A05; La10) or the 2~Ms \cdfs\ (G11). We also went
beyond the main catalogs of L05 and X11 and included the sources from
the supplementary catalogs and even the additional sources from the
candidate {\tt WAVDETECT} catalogs used by L05 and X11 (see
Section~\ref{sec:xraycat} for details). This has enabled us to provide as
many reliable \xray\ counterparts to the SMGs as possible. 

% careful analyses of fAGN
Finally, we adopted the `1/$N$' method for our AGN fraction or
\xray\ detection rate analyses, which corrects for the bias introduced
by the inhomogeneous \xray\ sensitivity coverage in the \ecdfs\ region
(see Section~\ref{sec:fagnmethod}). In comparison, the AGN fractions
reported in the previous studies are derived from simply dividing the
number of AGNs by the number of SMGs. Our method has enabled us to
report $\fagn$ values in terms of fractions of AGNs above certain
\xray\ flux/luminosity limits, which is a more well-defined and
meaningful way to report such quantities when there are large
variations in \xray\ sensitivity.

%---------------------------------------------------------------------------------------------------
\subsection{Comparison between the SMG-AGNs \\
  and Other Populations}\label{sec:otherpop}

In this section, we compare the SMG-AGNs with other SMGs, galaxies and
AGNs/quasars. Section~\ref{sec:cmd} provides a context for the SMGs
and SMG-AGNs among other galaxies and AGNs at similar redshifts by
presenting them in the commonly used color-mass
diagram. Section~\ref{sec:vsQSO} discusses the similarities and
differences between the SMG-AGNs and the general \xray\
AGNs/QSOs. Section~\ref{sec:vsULIRG} compares the AGN fractions in
SMGs and local ULIRGs, with cautions on the differences between these
two populations.

%----------------------------------------------------
\subsubsection{SMGs in the Color-Mass Diagram}\label{sec:cmd}

% cmd plot
% color is a bad indicator of sfr, consistent with xue 2010 and
% rozario 2013
We first place the SMGs in the context of other galaxies and the
\xray\ detected AGNs in the \cdfs\ region by plotting the effective
color vs.~mass diagram (CMD; Figure~\ref{fig:cmd}). The $x$-axis is
the stellar mass estimated as detailed in Section~\ref{sec:class3}.
Following \cite{Xue2012}, the $y$-axis is the effective color defined
as $C_{\rm eff}=(U-V)_{\rm rest} + 0.31z + 0.08 M_V + 0.51$
\citep{Bell2004}. The absolute magnitudes are from the SED fitting by
\cite{Simpson2013}. This effective color is defined based on the
dividing line that separates galaxies into the blue cloud and the red
sequence. We chose to use $C_{\rm eff}$ instead of, for example, the
rest-frame $U-V$ color so that the comparison galaxies are shown in
the plot in their corresponding evolutionary sequences. Galaxies with
$C_{\rm eff}<-0.05$ are within the blue cloud, and galaxies with
$C_{\rm eff}>0.05$ belong to the red sequence, while the ones with
$-0.05\leq C_{\rm eff}\leq 0.05$ are in the green valley (see
\citealt{Bell2004} and Section~3 of \citealt{Xue2012} for more
details). As usual, SMGs are plotted as blue dots, and the
\xray\ detected SMGs are labeled with their short LESS IDs. Also
plotted are the \xray\ detected AGNs (red dots; red crosses for
obscured AGNs with $\gammaeff<1.0$) and the \xray\ detected galaxies
(green squares) in the 4~Ms \cdfs\ region \citep{Xue2011}.

% massive and mass comparison
As mentioned in previous sections, SMGs are strongly star-forming
galaxies with SFR $>100$--$1000$ $\msun$ \peryr, and they are at the
massive end of the galaxy stellar-mass distribution
(Figure~\ref{fig:cmd}). Just like the \xray\ detected AGNs (and
obscured AGNs) from X11, the SMGs occupy the region from the blue
cloud all the way up to the top of the red sequence in the CMD plot,
while the \xray\ detected SMGs or SMG-AGNs are mostly in the green
valley and the red sequence. Notably, the SMG-AGNs are more massive
than the rest of the SMG population, having a mean stellar mass of
$1.8\pm0.5 \times 10^{11}\ \msun$, while the mean is $8.0\pm1.3 \times
10^{10}\ \msun$ for the rest. A K-S test for the mass distributions of
these two populations has revealed that they are likely from different
parent distributions ($p=0.01$). An AGN fraction analysis with a
stellar-mass cut of $M_\star\geq 10^{11}\ \msun$ shows that $\fagn$
for the massive SMGs is $33^{+30}_{-12}\%$ for AGNs with $\lxint \geq
7.8 \times 10^{42}$ \ergs, higher than the $\fagn$ value for all ALESS
SMGs ($17^{+16}_{-6}\%$, as listed in
Table~\ref{tab:fagn}).\footnote{We also calculated the AGN fraction
  with a rest-frame absolute $H$-band magnitude cut of $H \leq -24.5$
  for the SMGs (or $H$-band luminosity $\geq 5\times10^{11}\ L_\odot$)
  and obtained $\fagn=31^{+32}_{-14}\%$.} This is perhaps expected, as
multiple studies have found that AGNs preferentially reside in the
more massive galaxies
\citep[e.g.,][]{Xue2010,Rafferty2011,Mullaney2012}.
% Note: Results are very similar for H band cut < -24.5 absolute mag
% (L > 5d11 Lsolar), 31+32-14%.

% color is a bad SFR indicator
The large FIR and submm fluxes of SMGs indicate that they
are dust rich, so extinction is expected. Therefore, instead of being
blue, many of the SMGs appear red --- probably caused by the rich dust
content in these extreme star-forming galaxies. Some of them still
appear to be blue, which could be due to unobscured star formation
content dominating the emitted light because of dust inhomogeneity or
viewing angle differences. Rest-frame colors are unreliable and
insufficient indicators of SFR or galaxy evolutionary stage (quiesent
or star forming; also see \citealt{Xue2010}; \citealt{Rosario2013};
and references therein). Our results are consistent with the findings
in Figure~11 and Section~5.2.2 of \cite{Xue2010}.

%----------------------------------------------------
\subsubsection{SMG-AGNs and the General X-ray AGNs and QSOs}\label{sec:vsQSO}

% nH
The \xray\ properties of our SMG-AGN sample are similar to those in
the previous studies of A05, La10, and G11. They are similar to the
general moderately luminous \xray\ AGN population in X11, having $\gammaint \sim
1.8$, $\lx \sim 10^{42}$--$10^{45}$ \ergs, and $\nh \sim
10^{20}$--$10^{24}\ \percm2$ (see Table~\ref{tab:spec}). The sample
lacks lower luminosity AGNs with $\lx<10^{42}$ \ergs, which is not
surprising given the redshifts of our SMG sample and the
\xray\ sensitivity coverage. Our SMG-AGN sample exhibits a large
obscured fraction, with $\sim 60\pm20\%$ of the sources having
$\nh>10^{23}\ \percm2$ ($73\pm 15\%$ in A05), consistent with the findings in
previous studies (A05; La10; G11;
\citealt{Lutz2010,Hill2011,Bielby2012}).

% quasars
Besides the unobscured quasar ALESS 66.1, three other SMG-AGNs, ALESS
11.1, 57.1, and 73.1 could be considered as obscured quasars as they
have $\lxint>10^{44}$ \ergs. They are not \xray\ bright because they
are fairly obscured --- all having $\nh>10^{23}\ \percm2$ (with ALESS
73.1 being a Compton-thick AGN candidate; \citealt{Gilli2011}).
  
% Figure fx vs. submm flux 
All of our SMG-AGNs (and SMGs) except ALESS 66.1 have \xray\ fluxes
that are orders of magnitude lower than those of typical quasars. This is
illustrated by Figure~\ref{fig:fxsubmm} (similar to Figure~1 in
A05). The gray dots with arrows are \xray\ undetected ALESS SMGs
plotted with their 3$\sigma$ upper-limit \xray\ fluxes. Most of the SMGs are
probably powered by host galaxy starbursts in the submm band.

% Figure LFIR vs. Lxint
Figure~\ref{fig:lfir} provides context for the SMGs via some
well-studied starburst (squares labeled with green `S') and active
(squares labeled with red `A') galaxies, as well as quasars (dashed
line and shaded region; figure following Figure~8 of A05). As noted by
A05, according to the literature starburst and active galaxies on the
plot, the dividing line between starburst- and AGN-dominated systems
appears to be $\lx\simeq 0.004\times L_{\rm FIR}$ (the solid line), and the
A05 SMG classification is consistent with such a notion.

It is not necessarily true, however, that the FIR luminosity of an
SMG-AGN is dominated by the AGN component just because it lies in the
gray region in Figure~\ref{fig:lfir} (defined by the typical quasars
from \citealt{Elvis1994}).  As increasing rest-frame FIR data
become available for quasars both locally and at high redshift, it has
become clear that for the FIR luminous or moderately luminous quasars,
the star formation component dominates or at least contributes
significantly in the FIR band \citep[e.g.,][]{Lutz2008, Wang2010,
  Dai2012, Carrera2013}. Therefore, the FIR luminosity of unobscured
quasars like ALESS 66.1 could still have a substantial contribution from
the host-galaxy star formation. The star formation probably
contributes more significantly in the FIR for the `obscured quasars',
ALESS 11.1, 57.1, and 73.1, which have higher FIR luminosities and
lower $\lxint/L_{\rm FIR}$ ratios. As for ALESS 84.1, 70.1, 17.1, and
114.2, which have $\lxint/L_{\rm FIR}$ ratios of $\sim 0.001$ and
smaller, they are almost certainly starburst-dominated systems and
have very little or nearly no AGN contribution in the FIR band. Overall,
our SMG-AGN sample spans a large range in terms of $\lxint/L_{\rm
  FIR}$ ratios, varying by more than a factor of 10. This reveals the
heterogeneity in the SED compositions of SMGs (see more in
\citealt{Swinbank2013}).

%----------------------------------------------------
\subsubsection{The AGN Fraction in SMGs and ULIRGs}\label{sec:vsULIRG}

% Why are we doing the comparison
Most of the ALESS SMGs (84/99) can be considered as high-redshift
ULIRGs as they meet the definition of having $L_{\rm IR}>10^{12}\ L_\odot$
 \citep{Sanders1996}. From this perspective, it is interesting
to compare the AGN fraction in local ULIRGs and that in these
`distant cousins' of ULIRGs.

% Comparison of fAGN
\mbox{X-ray} and multiwavelength studies of the AGNs in local ULIRGs
have shown that the AGN fraction among these local starbursts is
generally very large \citep[$>50\%$;
e.g.,][]{Teng2010,Iwasawa2011,U2012,Koss2013}. At a first glance, the
AGN fraction of SMGs seems to be lower. However, the studies of the
local ULIRGs can usually reveal AGNs with low luminosities thanks to
the proximity of the local ULIRGs and the available spectroscopic
indicators of AGNs. For example, \cite{Iwasawa2011} reached an
\xray\ luminosity threshold of the order $\lx\sim10^{41}$ \ergs, while
the faintest \xray\ source in our sample has $\lx>10^{42}$ \ergs. We
use the data from \cite{Iwasawa2011} to perform a more direct comparison
of the AGN fractions of the ULIRGs and SMGs. \cite{Iwasawa2011} was based on
\chandra\ observations of local ULIRGs, and their sample of local ULIRG is
complete down to $\log{L_{\rm IR}/L_\cdot}=11.73$, . Among the 23 ULIRGs in
the \cite{Iwasawa2011} sample, 3 are AGNs with $\lx>2.5\times
10^{42}$ \ergs\ (to match the SMG-AGN with the smallest $\lx$ in our
sample), so the corresponding AGN fraction is $13^{+11}_{-7}\%$. For
our SMG sample, there are 81 ULIRGs, including the 8 SMG-AGNs, so the
AGN fraction among the ULIRG-SMG sample is $\sim 10^{+5}_{-3}\%$ ---
consistent with the AGN fraction of local ULIRGs according to a
two-tailed Fisher's exact test ($p=0.7$;
\citealt{Preacher2001}).\footnote{We note that the comparison here is
  not completely fair, as the \xray\ detections for the local ULIRGs
  could have a higher completeness rate at $\lx>2.5\times 10^{42}$
  \ergs\ than those for the $z\sim 2$ SMGs. On the other hand, the
  \xray\ observations of local ULIRGs are less sensitive in terms of
  identifying heavily obscured AGNs than for the high-redshift
  galaxies (which are being covered in a higher rest-frame
  \xray\ energy band; see Table~\ref{tab:spec}). These two effects are
  probably not pronounced here, but as one is biasing toward
  detecting more AGNs and the other is biasing against it, it could be
  coincidental that the AGN fraction of local ULIRGs agrees with that
  of ULIRG-SMGs at $z\sim2$.}

% Difference!
We want to emphasize here that the SMGs and the local ULIRGs are two
different populations, as mentioned in Section~\ref{sec:intro}. The
local ULIRGs are all starburst galaxies with SFR/sSFR significantly
above the local `galaxy main sequence' of star formation. Whereas the
SMGs, or the ULIRGs at $z\sim2$, are a heterogeneous sample of star
forming galaxies \citep[e.g.,][]{Daddi2007,Sargent2012}. Though a
majority of the ALESS SMGs with redshift, SFR and stellar mass
estimates (80/96, 83\%) have SFR/sSFR above the median of the main
sequence galaxies \citep[e.g.,][]{Elbaz2011,Rodighiero2011,Whitaker2012}, only
a fraction of SMGs (35/96, 36\%) are significantly above the main
sequence and therefore can be classified as starburst (with SFR/sSFR
values a factor of 4 above the median main
sequence; \citealt{Elbaz2011, Rodighiero2011}).

% compare AGN fraction in starbursts
We now compare the AGN fractions in the two starburst populations at
different redshifts: the local ULIRGs (again using data
from \citealt{Iwasawa2011}) and the starburst SMGs. Four SMG-AGNs in
our sample are starbursts (having $>4\times$ the SFR/sSFR of the
main-sequence galaxies; \citealt{Rodighiero2011}), which yields an AGN
fraction of $4/35=11^{+8}_{-5}\%$. This is consistent with the AGN
fraction for the local starbursts/ULIRGs ($13^{+11}_{-7}\%$). Larger
samples of both local ULIRGs and starburst SMGs would be able to
provide better statistical constraints on the AGN fractions and their
differences (if any) between the local and distant starbursts.

% Note:
% Iwasawa starburst median SFR = 100 Msun/yr
% SMG starburst median SFR = 1000 Msun/yr

%---------------------------------------------------------------------------------------------------
\subsection{The Origin of the X-ray Emission\\
  from ALESS 45.1 and 67.1}\label{sec:4567}

% aless 45.1 and 67.1
Among the 10 \xray\ detected SMGs, only two were not classified as
AGNs, ALESS 45.1 and 67.1. However, their rest-frame
\xray\ luminosities $\lx$ exceed $\sim3\times$ the amount expected
from their star formation components (see Figure~\ref{fig:class3}). As
mentioned in Section~\ref{sec:classall}, the classification methods we
applied for identifying AGNs cannot rule out the presence of AGN
in either of these two sources. Therefore, it is possible
that there are AGNs in ALESS 45.1 and 67.1 and that they contribute
non-negligibly in the \xray\ band, which could be the reason
why they have some \xray\ excess. This is also perhaps supported by the fact
that ALESS 45.1 lies within the NIR `Donley Wedge' for selecting AGNs
\citep{Donley2012}.

% Basu-Zych et al. result
Alternatively, the \xray\ excesses of ALESS 45.1 and 67.1 could be
explained by evolution of the \xray/SFR relation. Recently,
\cite{Basu-Zych2013} performed deep \xray\ stacking of Lyman break
galaxies in the \cdfs, and found a weak dependence of $L_{\rm
  X,SF}$ on redshift [$\log{L_{\rm
      HX,SF}}=0.93\log{(1+z)}+0.65\log{\rm SFR}+39.80$ for a Kroupa
  IMF and 2--10~keV X-rays]. This implies that the estimated $L_{\rm
  X,SF}$ would increase by a factor of $\simeq$3 for sources at $z=2$
compared with local star-forming galaxies. While this is a plausible
explanation for the \xray\ excesses of ALESS 45.1 and 67.1, it does
not affect the AGN classification of other \xray\ detected SMGs in our
sample. For example, after taking into account this possible
evolutionary effect in the X-ray/SFR relation, the $\lx$ values of the
8 SMG-AGNs are still a factor of $\gtrsim$3 larger than their expected
$L_{\rm X,SF}$ (see Figure~\ref{fig:class3}). Even if we still adopt
the threshold of $L_{\rm X,SF}/\lx > 5$, this will only affect ALESS
70.1, 84.1, and 114.1, which are independently classified as AGN hosts
through other methods. We did not adopt this redshift-dependent
X-ray/SFR relation in our study as the result is tentative due to the
nature of \xray\ stacking.

% SED
AGNs with moderate-to-low \xray\ luminosity in highly star-forming
systems can be hard to identify through \xray. For example, as
illustrated by the ULIRGs from L10 plotted in Figure~\ref{fig:class3},
some of these ULIRGs are identified to have AGN activities but they
have \xray\ luminosities at a similar level with what is expected from
their star formation (plotted as squares with centered red
dots). Future works of panchromatic SED analyses on the \xray\
detected SMGs with models of separate AGN and host-galaxy
contributions will certainly shine more light on this issue and help
to determine how much AGN is responsible in the \xray\ in ALESS 45.1
and 67.1 and SMGs alike.

%%%%%%%%%%%%%%%%%%%%%%%%%%%%%%%%%%%%%%%%%%%%%%%%%%%%%%%%%%%%%%%%%%%%%%%%%%%%%%%%%%%%%%%%%%%%%%%%%%%%
\section{Conclusions and Future Work}\label{sec:future}

%---------------------------------------------------------------------------------------------------
\subsection{Summary of Main Conclusions}

In this paper, we have studied the \xray\ properties and the AGN
content of the \xray\ detected SMGs. The exquisite angular resolution
and sensitivity of ALMA combined with the deep \xray\ coverage in the
\cdfs/\ecdfs\ region have provided new insights into \xray\ SMGs and
SMG-AGNs. The major results are summarized as follows:

\begin{enumerate}
  \item Among the 91 SMGs within the \ecdfs\ region, 10 are found to
    have \xray\ counterparts. The submm catalog used in this work, the
    ALESS main catalog, is based on ALMA observations in the
    \ecdfs\ region \citep{Hodge2013, Karim2013}, which have a typical
    angular resolution of $<1.5\arcsec$ or even $<0.5\arcsec$ and an
    RMS of $0.6$~mJy. The \xray\ catalog used is derived from 4~Ms
    \chandra\ observations in the \cdfs\ region (X11) and 250~ks observations in
    the \ecdfs\ (L05). We employed the likelihood-ratio matching
    method for finding \xray\ counterparts of SMGs, and we estimated a
    false-match probability of less than 3\%. This work is the first
    among similar works to identify unambiguously the
    \xray\ counterparts of SMGs. See Section~\ref{sec:cat}.
  \item Spectral analyses of the 10 \xray\ detected SMGs reveal that 6
    have moderate to heavy obscuration, with
    $\nh>10^{23}\ \percm2$. Their rest-frame 0.5--8~keV luminosities
    range from $10^{42.2}$ \ergs\ to $10^{44.5}$ \ergs. Through
    \xray\ spectral fitting (for sources with sufficient \xray\ photon
    counts) and other spectral analysis methods (for low-count
    sources), we estimated their \xray\ spectral hardness ratios,
    effective photon indices $\gammaeff$, intrinsic photon indices
    $\gammaint$, and rest-frame absorption-corrected \mbox{0.5--8~keV}
    luminosities $\lxint$. See Section~\ref{sec:prop}.
  \item We classified 8 out of 10 \xray\ detected SMGs as AGN
    hosts. We employed multiple classification methods using their
    \xray\ properties, multiwavelength properties, and
    variability. The similarities or differences in the physical
    motivations behind these methods enabled the cross-checking of the
    classification results. See Section~\ref{sec:class}.
  \item We estimated an AGN fraction of $17^{+16}_{-6}\%$ among the
    SMGs for AGNs with rest-frame absorption-corrected luminosity
    $\lxint \geq 7.8\times10^{42}$ \ergs. We also estimated AGN
    fractions and \xray\ detection rates for a series of different
    \xray\ flux or luminosity limits. Our method takes into account
    the spacial inhomogeneity in the \xray\ sensitivity. See
    Section~\ref{sec:fagn}.
  \item We stacked 49 \xray\ undetected sources within $7 \arcmin$ of
    the aim points of the 4~Ms \cdfs\ or 250 ks \ecdfs, and detected
    significant \xray\ signals in the full and soft \xray\ bands, with
    a marginal detection in the hard band. The mean \xray\ luminosity
    is roughly consistent with the amount expected from the star
    formation components of SMGs. We also identified 4 potential AGN
    candidates among these 49 SMGs through their rest-frame
    radio-luminosity excesses, and they yield significant
    \xray\ signals in both the soft and hard bands with a hard
    \xray\ spectral index ($\gammaeff<1$) and a higher mean
    \xray\ luminosity than the expected level from their star
    formation. See Section~\ref{sec:stack}.
  \item We compared our results with the previous works by A05, La10,
    G11, and J13 (and also \citealt{Lindner2012} for stacking), and
    found that our AGN fractions agree with the previous estimates
    within 1$\sigma$ error bars (though in some cases, this could be
    coincidental). Compared to all previous works, our results are the
    most robust and do not suffer from significant uncertainties as a result of
    poor submm positional accuracy and ambiguous or even erroneous
    counterpart matching. The subgroup of radio-selected SMGs (in A05
    and \citealt{Lindner2012}) appears to have a larger AGN fraction
    than the general SMG population. See Section~\ref{sec:comp}.
  \item We also discussed the similarities and differences between
    SMGs and other galaxy or AGN populations. We found that SMG-AGNs
    have larger stellar masses than the general SMG population.
    The massive SMGs with $M_\star \geq 10^{11}\ \msun$ have a higher
    AGN fraction ($33^{+30}_{-12}\%$ for AGNs with $\lxint \geq 7.8
    \times 10^{42}$ \ergs) than all ALESS SMGs ($17^{+16}_{-6}\%$ for
    the same $\lxint$ limit). See Section~\ref{sec:otherpop}.
  \item We also commented briefly upon the possible origins of the
    \xray\ emission from the two \xray\ SMGs that are not classified
    as AGNs in this work, ALESS 45.1 and 67.1, in
    Section~\ref{sec:4567}.
\end{enumerate}

%---------------------------------------------------------------------------------------------------
\subsection{Future Work}

\begin{figure*}[!th]
%\center
%\includegraphics[angle=0.,scale=0.65]{../table_plot/fagn_plot.eps}
  \includegraphics[angle=0.,scale=0.65]{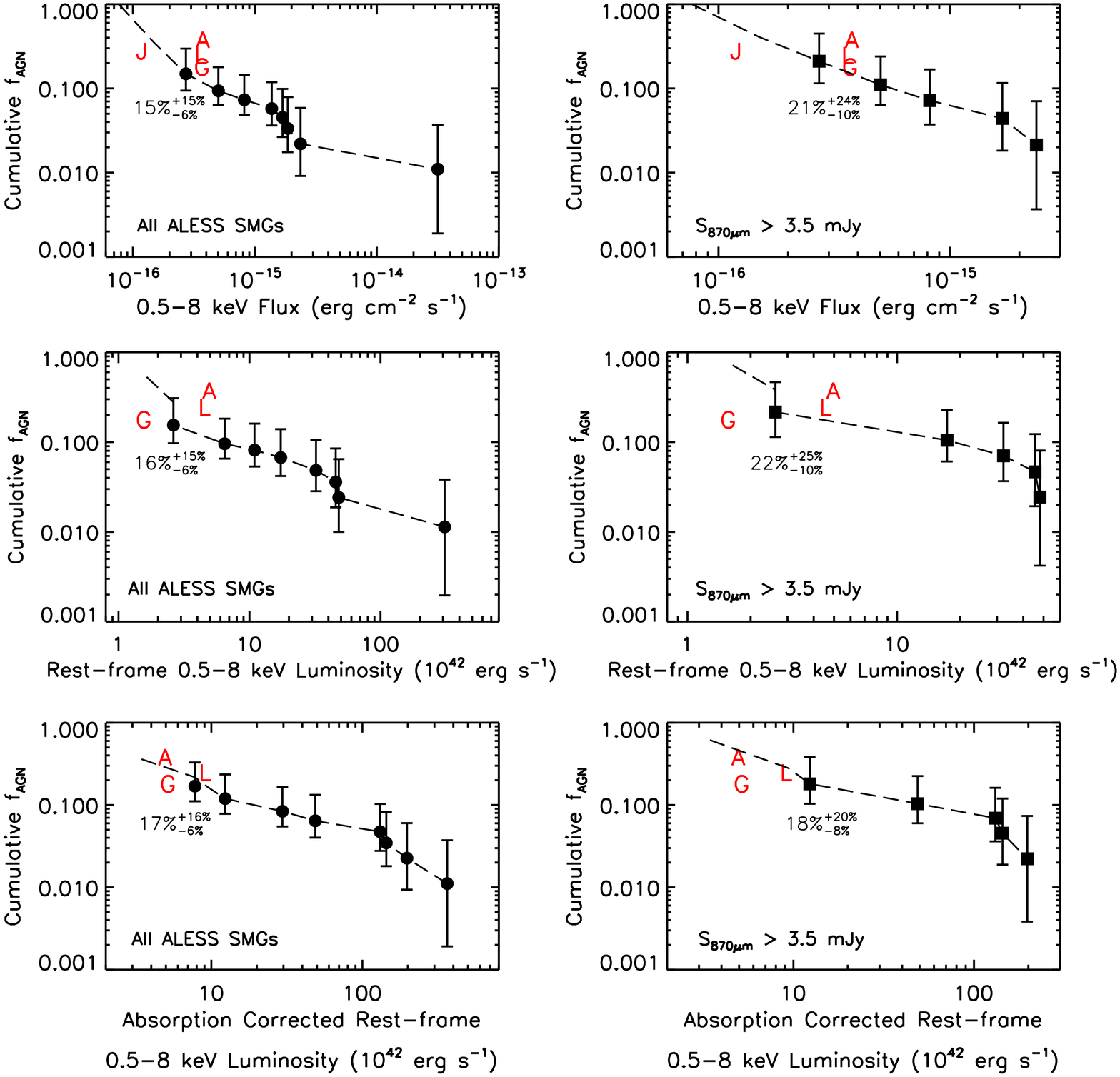}
\caption{Cumulative AGN fraction $\fagn$ for SMGs as a function of the
  observed-frame \fbrange\ flux ($\fx$), rest-frame \fbrange\
  apparent luminosity ($\lx$), or rest-frame \fbrange\
  absorption-corrected luminosity ($\lxint$). Each point represents
  the AGN fraction in the SMG sample for AGNs with a certain \xray\
  flux/luminosity value or larger. The percentage number and error
  bars given are for the left-most point on each plot (also listed in
  Table~\ref{tab:fagn}).
  The plotted flux or luminosity values are those of the 8 SMG-AGNs,
  as listed in Tables~\ref{tab:basic} and \ref{tab:spec}. The dashed
  lines are the $\fagn$ results if we classify all 10 \xray\ SMGs
  (i.e., including ALESS 45.1 and 67.1) as AGNs.
  The three panels on the left (with filled circles) show $f_{\rm
    AGN}$ calculated using all ALESS main-catalog sources, while the
  three on the right (with filled squares) show $f_{\rm AGN}$
  calculated using only ALESS sources that have $S_{\rm 870 \mu
    \mbox{m}}>3.5$~mJy (a flux-limited sample; see
  Section~\ref{sec:fagnmethod}). Error bars are 1$\sigma$ errors
  calculated using the methods in Gehrels (1986).
  The red letters mark the $\fagn$ estimates by
  \citeauthor{Alexander2005b}~(\citeyear{Alexander2005b}; `A'),
  \citeauthor{Laird2010}~(\citeyear{Laird2010}; `L'),
  \citeauthor{Georgantopoulos2011}~(\citeyear{Georgantopoulos2011};
  `G'), and \citeauthor{Johnson2013}~(\citeyear{Johnson2013}; `J',
  lowest \xray\ luminosity values are not available thus not plotted
  in the four lower panels).
  See Section~\ref{sec:fagn} and Section~\ref{sec:comp:fagn} for more
  details.\label{fig:fagn}}
\end{figure*}

\begin{figure*}[!th]
  \center
\includegraphics[angle=0.,scale=0.6]{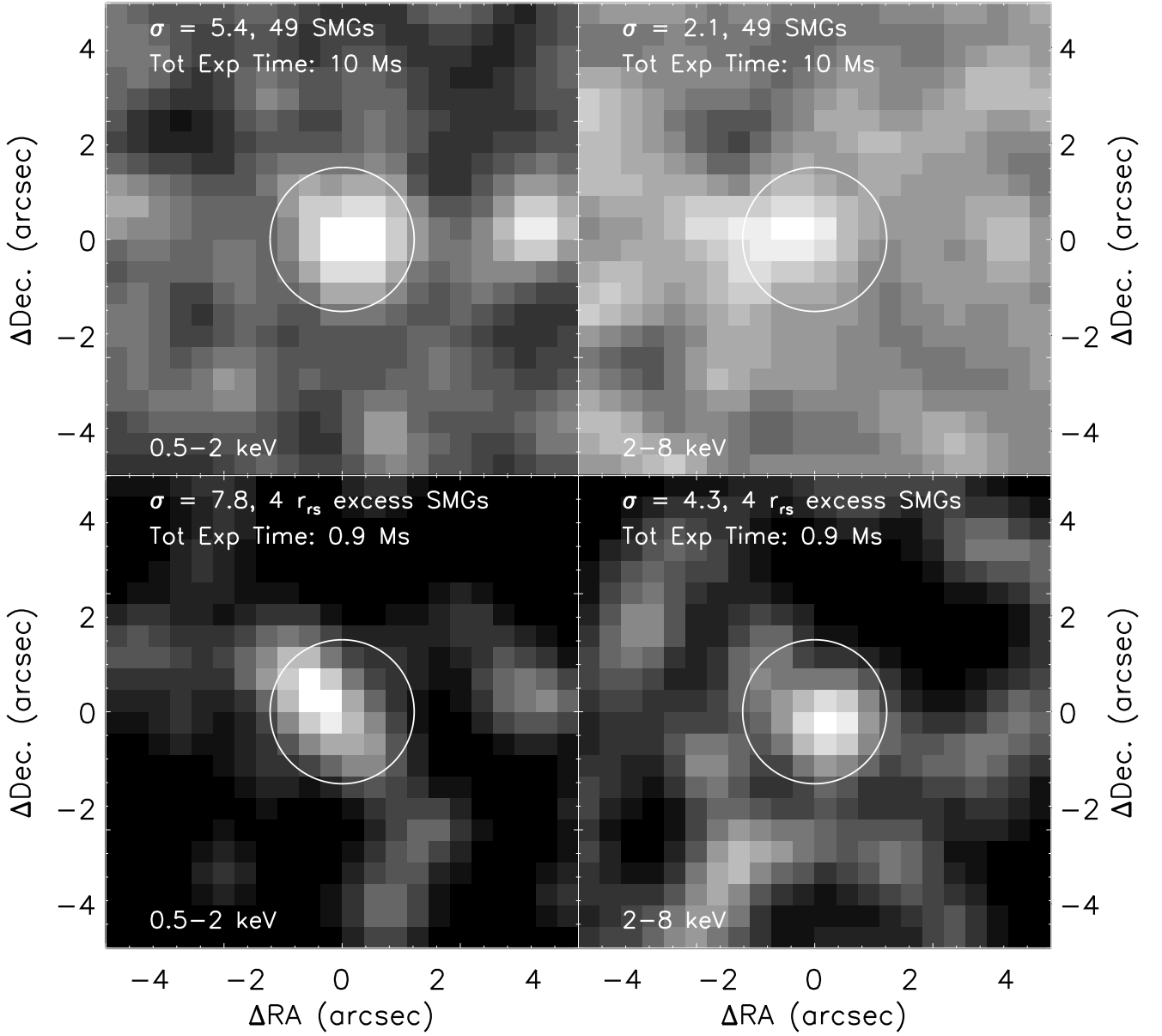}  
\caption{Stacked \xray\ images (Gaussian smoothed with a FWHM of 3
  pixels) for the 49 SMGs within $7 \arcmin$ of any of the \ecdfs\ aim
  points (top) and for the 4 SMGs with $r_{\rm rs}$ excesses (bottom).
  On the left are the stacked images in the soft band (0.5--2
  keV), and on the right are the hard band (2--8 keV). The circles show
  the size of the extraction aperture ($1.5 \arcsec$ radius) for the
  source, which is the same for the individual background region (1000
  such background regions were used to estimate the mean background
  for each band).
  See Section~\ref{sec:stack} for details. \label{fig:stack}}
\end{figure*}

\begin{figure*}[!th]
  \center
\includegraphics[angle=0.,scale=0.6]{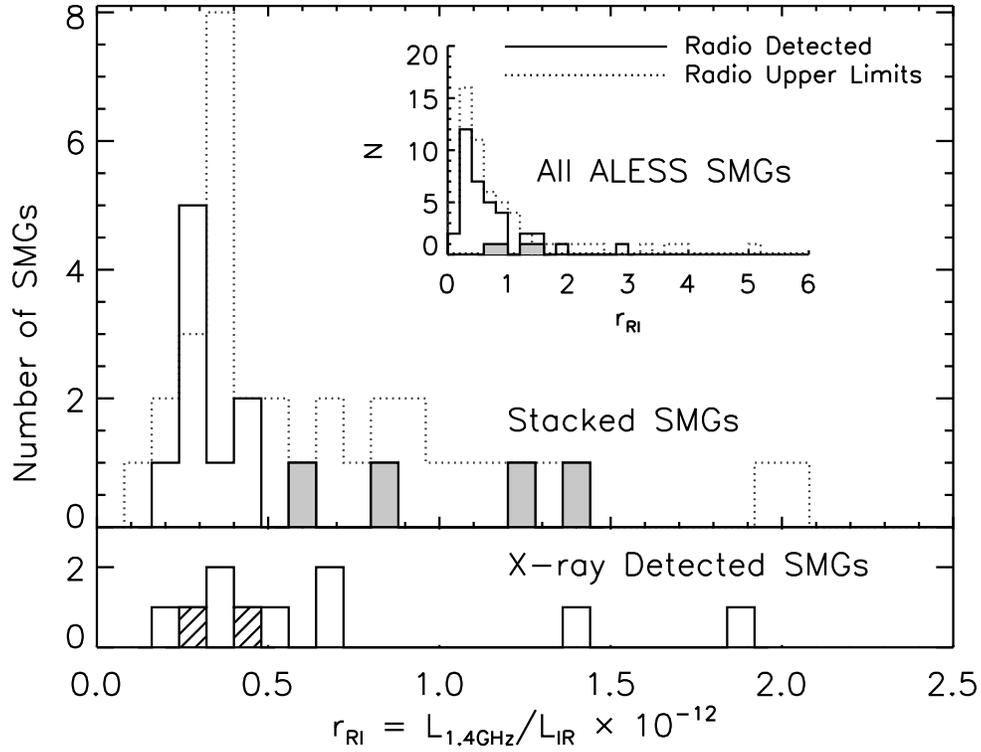}
\caption{Histogram of the ratios between the rest-frame 1.4 GHz
  monochromatic luminosity, $L_{\rm 1.4GHz}$, and the rest-frame IR
  luminosity, $L_{\rm IR}$ (8--1000 \micron). The inset is for all
  ALESS SMGs and the lower panel for the 10 \xray\ detected ones
  (non-AGN in hatched marks). Overall, the SMG-AGNs have high $\rri$
  values, especially considering that the majority of SMGs only have
  radio upper limits but not detections (dotted histograms). Just like
  the X-ray detected SMGs, the four sources that are selected as
  potential AGN candidates in the stacking analysis (shaded in gray)
  are high in $\rri$ compared to either the rest of SMGs in the
  stacking sample or all ALESS SMGs.
  See Section~\ref{sec:stack} for details. \label{fig:rrs}}
\end{figure*}

\begin{figure*}[!th]
  \center
\includegraphics[angle=0.,scale=0.7]{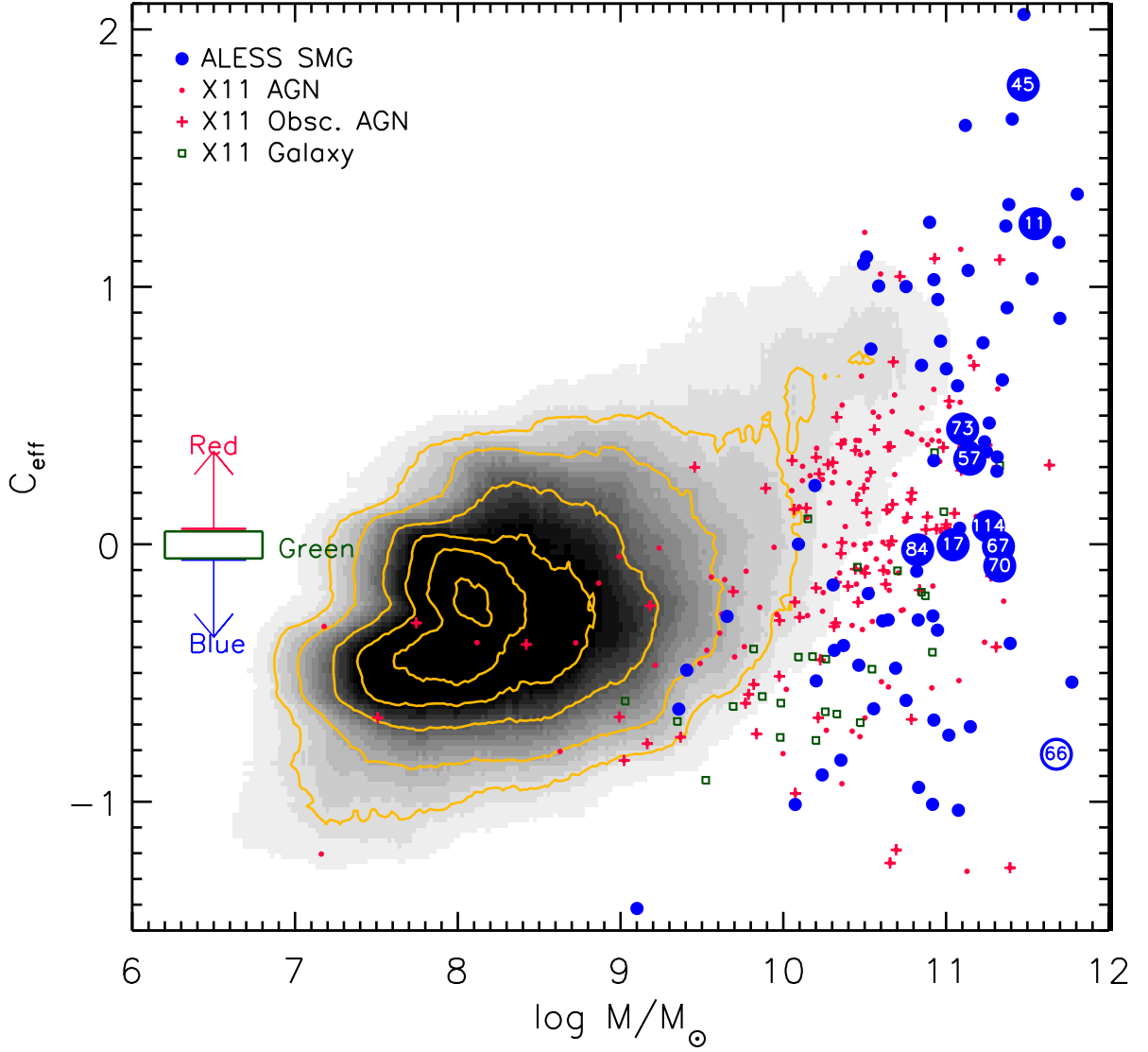}  
\caption{Effective color vs.~stellar mass. The effective color is
  defined as $C_{\rm eff}=(U-V)_{\rm rest} + 0.31z + 0.08 M_V + 0.51$, as 
  in \cite{Bell2004}. The divisions defined by $C_{\rm eff}$ values
  for the red sequence, blue cloud, and green valley are illustrated
  on the left side by the red/blue arrows and green square.
  The grayscale density map and contours (levels for 0.5, 1--5 sources
  per pixel) are for $\sim 20,000$ galaxies with $1 \leq z \leq 5$
  (a similar redshift range as for the ALESS SMGs) within the GOODS-S
  region. The galaxies are from the $z$-band selected catalog of
  \cite{Dahlen2010}, with their $C_{\rm eff}$ values and stellar masses
  calculated by \cite{Xue2012}.
  The red dots and crosses are the X11 AGNs, and green squares mark the
  X11 galaxies (both with $1 \leq z \leq 5$). The red crosses are AGNs
  that are classified as obscured AGN by X11 as they have
  $\gammaeff<1$.  Other symbols are the same as in previous figures:
  blue dots for SMGs, with larger dots for the \xray\ detected SMGs,
  labeled with their short LESS IDs. Again ALESS 66.1 is plotted as
  an open circle as its stellar mass has large uncertainty
  (see Section~\ref{sec:multiv}). 
  See Section~\ref{sec:cmd} for more discussion. \label{fig:cmd}}
\end{figure*}

\begin{figure*}[!th]
  \center
\includegraphics[angle=0.,scale=0.7]{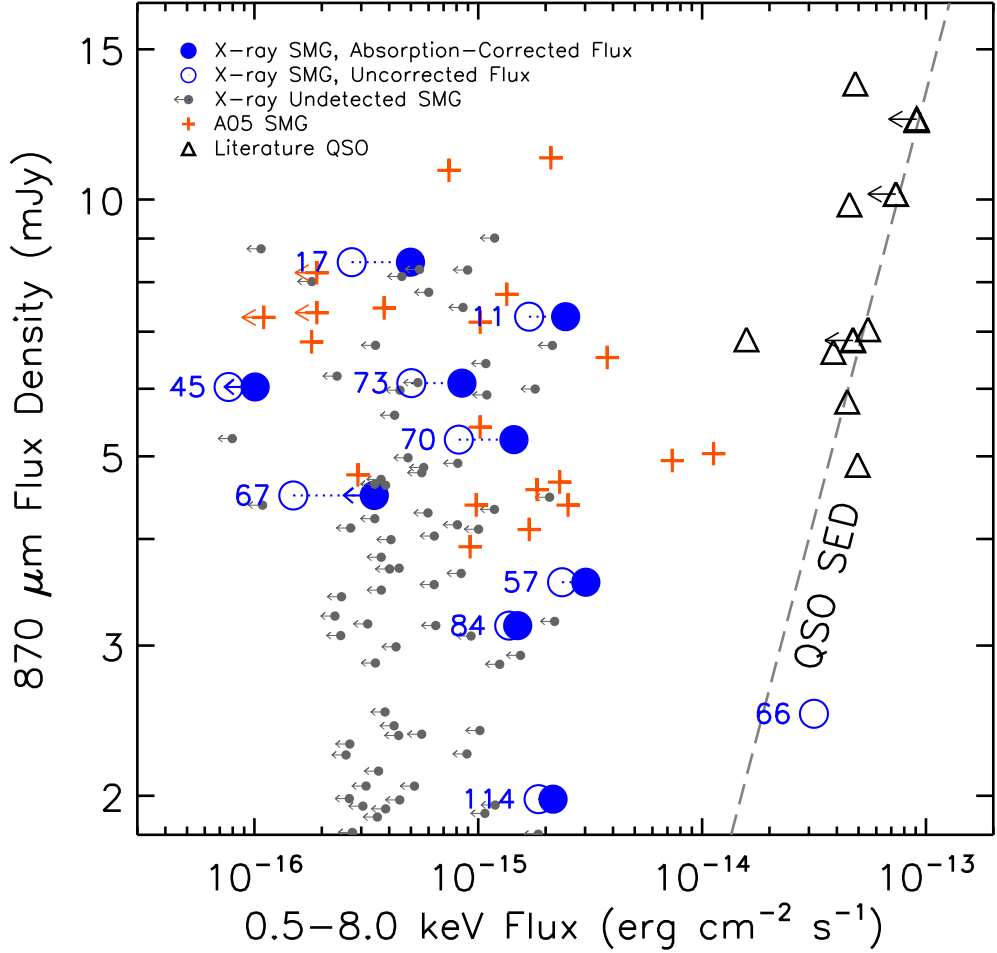}  
\caption{Observed 870 $\mu$m flux density vs.\ \fbrange\ flux for our
  \xray\ detected SMGs (large open blue circles), the \xray\ detected SMGs in
  A05 (orange crosses), and quasars from the literature (black triangles).
  The large blue dots connected with dotted lines are for the
  absorption-corrected \fbrange\ flux (observed frame).  The dots
  labeled with leftward arrows indicate 90\% confidence upper limits as the hardness
  ratios for ALESS 45.1 and 67.1 are not well constrained and are
  given as 90\% confidence upper limits (see Section~\ref{sec:xspec}
  and Table~\ref{tab:spec}). 
  The small gray dots with leftward arrows are the \xray-undetected
  SMGs with their \xray\ flux upper limits (being the \fbrange\ \xray\
  sensitivity for their positions in the \ecdfs\ region).
  The dashed line marks the 870 $\mu$m flux density to \xray\ flux
  ratio for a typical 3C273-like unobscured quasar (Fabian et
  al.~2000). Arrows on the triangles indicate sources with \xray\ flux
  upper limits. The data for the quasars are from \cite{Page2001},
  \cite{Vignali2001}, and \cite{Isaak2002} as also plotted in
  A05. The $S_{870 \mu\mbox{m}}$ of the A05 sources are converted from
  their $S_{850 \mu\mbox{m}}$ values with an assumed submm spectral
  index of $\alpha=3$ ($S \propto \nu^\alpha$; see, e.g.,
  \citealt{Carilli2000}), and similarly for the quasars with
  $\alpha=1$. 
  Except for ALESS 66.1, all of the \xray\ detected SMGs show significant
  870 $\mu$m flux excesses compared with typical unobscured quasars,
  indicating that their submm emission is likely host dominated. We
  remind the reader here that the flux distributions of the
  \xray\ detected SMGs and the rest are not statistically different,
  as mentioned in the caption of Figure~\ref{fig:hist}.
  See Section~\ref{sec:vsQSO} for more discussion.\label{fig:fxsubmm}}
\end{figure*}

\begin{figure*}[!th]
%\center
%\includegraphics[angle=0.,scale=0.9]{../table_plot/lfir_lx_ratio.eps}
  \includegraphics[angle=0.,scale=0.9]{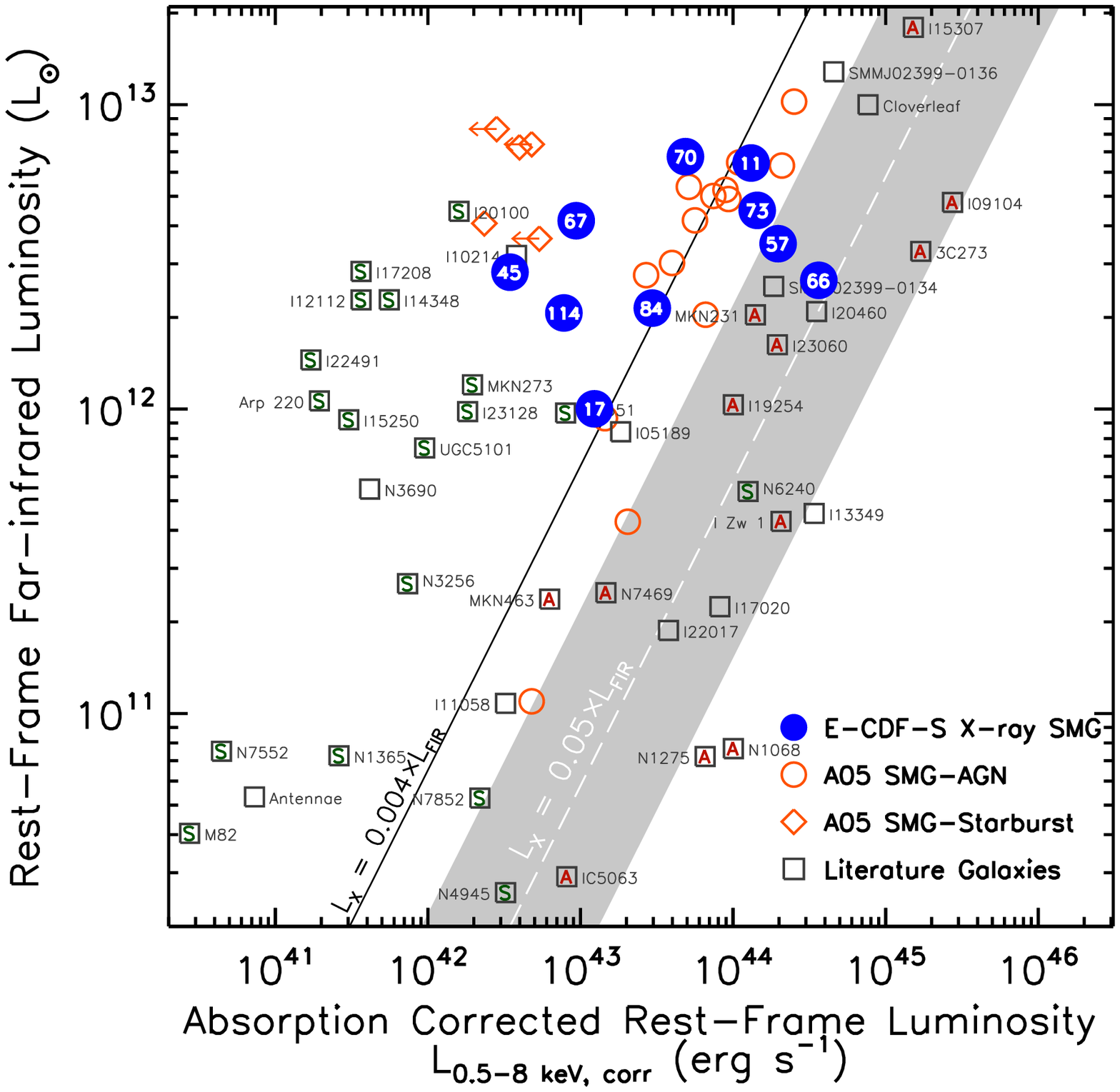}
\caption{Rest-frame far-infrared luminosity ($L_{\rm FIR}$) vs.
  rest-frame \fbrange\ absorption-corrected luminosity
  ($\lxint$). The large blue dots mark the \xray\ detected SMGs in this
  work with their short LESS IDs in the middle. The small orange open
  circles (diamonds) are for SMG-AGN (SMG-Starbursts) in A05.
  The open squares are galaxies whose \xray\ and far-IR luminosities
  were calculated by A05 based on data compiled from the literature
  (see caption of Fig. 8 in A05 and references therein). They are
  classified either as AGN-dominated (labeled with red letter `A') or
  star formation-dominated (labeled with green letter `S') by
  \cite{Rigopoulou1999} and \cite{Tran2001}.
  The dashed line and gray region is the median luminosity ratio for
  the quasars in \cite{Elvis1994} and its standard deviation.
  See Section~\ref{sec:vsQSO} for more discussion.\label{fig:lfir}}
\end{figure*}

% Niel's rewrite
There are several ways that the work described in this paper
could be productively advanced. First, while our study benefits 
greatly from reliable SMG vs. \xray\ source matching, the sample 
size of matched objects is small and limits detailed statistical 
analyses. Further ALMA coverage and/or deeper \xray\ surveys in 
the \ecdfs\ and other well-studied survey fields can effectively
enlarge the sample size, improving constraints upon the AGN
fraction over a wide range of AGN luminosity. It is clear from our
analyses (e.g., Figure~\ref{fig:fagn}) that highly sensitive \xray\ 
observations are preferable when constraining the AGN fraction, even 
at relatively high AGN luminosities, since many of the SMGs contain 
obscured AGNs which can be intrinsically luminous but faint in flux. 
For example, if the four sub-fields of the \ecdfs\ each had 4~Ms 
\chandra\ coverage like the \cdfs\ proper, we estimate that an 
additional $\gtrsim 36$ SMGs would be \xray\ detected. 

Second, targeted ALMA observations should also be performed to measure
the star-formation properties of \xray\ sources, both AGNs and
starburst galaxies, in the very well-studied central regions of the
\cdfs; these will generally have \hbox{$S_{\rm 870\mu m} \lesssim
  3.5$--4.5~mJy}. This complementary targeted approach with ALMA
observing of known \xray\ sources will allow considerably lower SFRs
to be probed in systems that already have unmatched \xray\ spectral
characterization from the existing 4~Ms \chandra\ exposure (soon to be
raised to 7~Ms) and 3~Ms \xmm\ exposure.

Finally, additional approaches should continue to be employed to search 
for any AGNs in SMGs that have been missed even in highly sensitive 
\xray\ data. These include near-infrared and submm spectroscopy, 
as well as radio spectral/morphology measurements. Such approaches
can reveal AGNs that are highly Compton-thick or intrinsically 
\xray\ weak, both of which must be included when developing a complete 
census of the growing SMBHs in SMGs.

%%%%%%%%%%%%%%%%%%%%%%%%%%%%%%%%%%%%%%%%%%%%%%%%%%%%%%%%%%%%%%%%%%%%%%%%%%%%%%%%%%%%%%%%%%%%%%%%%%%%
\acknowledgements
\hspace{0.5in}

The ALMA observations were carried out under program
ADS/JAO.ALMA\#2011.0.00294.S. ALMA is a partnership of ESO
(representing its member states), NSF (USA), and NINS (Japan),
together with NRC (Canada) and NSC and ASIAA (Taiwan), in cooperation
with the Republic of Chile. The Joint ALMA Observatory is operated by
ESO, AUI/NRAO and NAOJ. This publication also makes use of data from
the ESO VLT under program ID 183.A-0666.  The data used in this paper
are available from the ALMA and \chandra\ data archives.

We gratefully acknowledge financial support from SAO grant AR3-14015X
(SXW, WNB, BL), HST grant GO-12866.01-A (SXW, WNB, BL), NASA ADP grant
NNX11AJ59G (SXW, WNB, BL), and ACIS Instrument Team contract SV4-74018
(SXW, WNB, BL). IRS acknowledges support from STFC (ST/I001573/1), a
Leverhulme Fellowship, the ERC Advanced Investigator program DUSTYGAL
321334, and a Royal Society/Wolfson Merit Award.  We also acknowledge
financial support from the Leverhulme Trust (DMA), the STFC (DMA),
STFC studentship (ALRD). AK acknowledges support from STFC as well as
the Collaborative Research Council 956 funded by the Deutsche
Forschungsgemeinschaft (DFG). YQX acknowledges the financial support
of the Thousand Young Talents (QingNianQianRen) program
(KJ2030220004), USTC startup funding (ZC9850290195), and the National
Natural Science Foundation of China through NSFC-11243008.

We thank the referee for a constructive report. We thank Monica Young
for her help with the variability analysis.

%%%%%%%%%%%%%%%%%%%%%%%%%%%%%%%%%%%%%%%%%%%%%%%%%%%%%%%%%%%%%%%%%%%%%%%%%%%%%%%%%%%%%%%%%%%%%%%%%%%%
% References:
%\newpage

%%%%%%%%%%%%%%%%%%%%%%%%%%%%%%%%%%%%%%%%%%%%%%%%%%%%%%%%%%%%%%%%%%%%%%%%%%%%%%%%%%%%%%%%%%%%%%%%
% BEGIN FIGURES AND STUFF

\newpage

%----------------------------------------------------------------
% Figure: Map - ECDFS & CDFS sensitivity map with ALESS
% SMGs over plotted.
%----------------------------------------------------------------

%----------------------------------------------------------------
% Figure: positional offset between xray and submm (histogram and
% inset), number of false matches vs. search radius 
%----------------------------------------------------------------

%----------------------------------------------------------------
% Figure: thumbnails for x-ray, submm
%----------------------------------------------------------------

%----------------------------------------------------------------
% Figure: flux histogram for ALESS sources and SMG-AGNs
%----------------------------------------------------------------

%%%%%%%%%%%%%%%%%%%%%%%%%%%%%%%%%%%%%%%%%%%%%%%%%%%%%%%%%%%%%%%%%%%%%%%%%%%%%%%%%%%%%%%%%%%%%%%%%%%%

%---------------------------------------------------------------- 
% Figure: spec for 5 sources
%----------------------------------------------------------------

%----------------------------------------------------------------
% Figure: hardness vs redshift
%----------------------------------------------------------------

%----------------------------------------------------------------
% Figure: classification 1 & 2: Gamma vs Lxint
%----------------------------------------------------------------

%----------------------------------------------------------------
% Figure: classification 3a: Lx vs Lradio1.4
%----------------------------------------------------------------

%----------------------------------------------------------------
% Figure: classification 4: fx vs f3.6
%----------------------------------------------------------------

%----------------------------------------------------------------
% Figure: fAGN vs fx lx lxint
%----------------------------------------------------------------

%----------------------------------------------------------------
% Figure: Stacked Images
%----------------------------------------------------------------

%----------------------------------------------------------------
% Figure: rrs histogram for stacking subsample selection
%----------------------------------------------------------------

%----------------------------------------------------------------
% Figure: color-mass diagram for smgs
%----------------------------------------------------------------

%----------------------------------------------------------------
% Figure: submm flux vs xray flux (both obs and int), SED
%----------------------------------------------------------------

%----------------------------------------------------------------
% Figure: Lxint vs LFIR 

%----------------------------------------------------------------

%%%%%%%%%%%%%%%%%%%%%%%%%%%%%%%%%%%%%%%%%%%%%%%%%%%%%%%%%%%%%%%%%%%%%%%%%%%%%%%%%%%%%%%%%%%%%%%%%%%%
% Tables

%----------------------------------------------------------------
% Table: Properties of Matched Submm and \xray\ Sources
\renewcommand{\arraystretch}{1.3} % more row spacing for the table
\begin{turnpage}
\begin{center}
\begin{deluxetable*}{lllllccc}
%\rotate
\tabletypesize{\scriptsize}
\tablewidth{0pt}
\tablecaption{Properties of Matched Submm and \xray\ Sources\label{tab:basic}}
\tablehead{
%-------------  
% FIRST ROW  
% submm stuff  
\colhead{} & \colhead{} &
\colhead{$S_{870 \mu{\rm m}}$} &
% xray stuff
\multicolumn{2}{c}{\xray\ Position\tablenotemark{b}}  & \colhead{\cdfs} & \colhead{\ecdfs} &
\colhead{$f_{X}/10^{-15}$} \\ \cline{4-5}
%-------------
% SECOND ROW
% submm stuff  
\colhead{ALESS SMG Full Name\tablenotemark{a}} & \colhead{ID\tablenotemark{a}} &
\colhead{(mJy)} &
% xray stuff
\colhead{$\alpha_{\rm J2000.0}$}  &  \colhead{$\delta_{\rm J2000.0}$}
& \colhead{Cat \& ID\tablenotemark{b}} & \colhead{Cat \& ID\tablenotemark{b}} &
\colhead{erg cm$^{-2}$ s$^{-1}$\tablenotemark{c}} 
}
\startdata
ALESS J$033213.85-275600.3  $ & 011.1 & $  7.3 \pm 0.4 $ & 03 32 13.85 & $ -27 $ 56 00.44 & M 197 & M 332 &  1.69 \\
ALESS J$033207.30-275120.8  $ & 017.1 & $  8.4 \pm 0.5 $ & 03 32 07.34 & $ -27 $ 51 20.57 & M 131 & \nodata &  0.27 \\
ALESS J$033225.26-275230.5  $ & 045.1 & $  6.0 \pm 0.5 $ & 03 32 25.26 & $ -27 $ 52 30.83 & M 348 & \nodata &  0.08 \\
ALESS J$033151.92-275327.1  $ & 057.1 & $  3.6 \pm 0.6 $ & 03 31 51.95 & $ -27 $ 53 27.25 & M  34 & M 203 &  2.37 \\
ALESS J$033331.93-275409.5  $ & 066.1 & $  2.5 \pm 0.5 $ & 03 33 31.93 & $ -27 $ 54 10.58 & \nodata & M 725 & 31.60 \\
ALESS J$033243.20-275514.3  $ & 067.1 & $  4.5 \pm 0.4 $ & 03 32 43.21 & $ -27 $ 55 15.20 & A  & \nodata &  0.15 \\
ALESS J$033144.02-273835.5  $ & 070.1 & $  5.2 \pm 0.5 $ & 03 31 44.05 & $ -27 $ 38 35.98 & \nodata & M 146 &  0.82 \\
ALESS J$033229.29-275619.7  $ & 073.1 & $  6.1 \pm 0.5 $ & 03 32 29.27 & $ -27 $ 56 19.83 & M 403 & \nodata &  0.50 \\
ALESS J$033154.50-275105.6  $ & 084.1 & $  3.2 \pm 0.6 $ & 03 31 54.52 & $ -27 $ 51 05.70 & M  48 & \nodata &  1.38 \\
ALESS J$033151.11-274437.3  $ & 114.2 & $  2.0 \pm 0.5 $ & 03 31 51.11 & $ -27 $ 44 37.48 & M  31 & M 194 &  1.86 
\enddata
\tablenotetext{a}{The official IAU full names (numbers being J$2000.0$
  R.A.\ and Dec.) and the short ALESS ID numbers for the
  \xray\ detected SMGs. The first 3 digits of the short ID give the ID of
  the targeted LESS source at the center of each sub-field, and the
  last digit is the sub-ID for the ALESS sources detected in each
  sub-field. All of the SMGs listed are from the main catalog of
  ALESS, and none was classified as extended source. The sources are
  labeled with their short LESS IDs in the plots throughout the paper;
  for example, ALESS 011.1 is labeled as `11'.}
\tablenotetext{b}{\xray\ catalogs and matched \xray\ ID numbers for the
  \xray\ counterparts of SMGs. `M' stands for the \cdfs\ or
  \ecdfs\ main catalog, while `A' stands for the additional catalog
  which consists of sources not in the \cdfs\ or \ecdfs\ main or
  supplementary catalogs but detected by {\tt WAVDETECT} with a
  false-positive probability threshold of $10^{-5}$ (see
  Section~\ref{sec:xraycat}). Positions and ID numbers are the same as in X11
  for \cdfs\ or L05 for sources only in the \ecdfs. The \xray\ counterpart
  of ALESS 67.1 is in the 2~Ms \cdfs\ main catalog (XID 362) of Luo et
  al. (2008). }
\tablenotetext{c}{Full-band (0.5--8.0~keV) \xray\ flux, as reported in
  X11 or L05 for sources only in the \ecdfs.}
\end{deluxetable*}
\end{center}
\end{turnpage}
%----------------------------------------------------------------

%\newpage
%----------------------------------------------------------------
% Table: Xray properties of SMGs
\renewcommand{\arraystretch}{1.6} % more row spacing for the table
\begin{turnpage}
\begin{center}
\begin{deluxetable*}{@{\hspace{-5mm}}c@{\hspace{-5mm}}lrcrccccccccc}
%\rotate
\tabletypesize{\scriptsize}
\tablewidth{0pt}
\tablecaption{\xray\ Properties of \xray\ Detected SMGs\label{tab:spec}}
\tablehead{
%-------------  
% FIRST ROW
%  
\colhead{ALESS} & \colhead{} & \colhead{\xray} &
\colhead{Off-Axis} & \colhead{Exp.} & \colhead{FB} &
\colhead{Bkg} & \colhead{Rest-frame} &
\colhead{Hardness} & \colhead{} & \colhead{$\log{L_{\rm 0.5-8keV}}$\tablenotemark{e}} &
\colhead{} & \colhead{$N_{\rm H}/10^{22}$} & \colhead{$\log{L_{\rm 0.5-8keV,corr}}$} \\
%-------------  
% SECOND ROW
\colhead{ID} & \colhead{$z$\tablenotemark{a}~~~} & \colhead{Cat \& ID\tablenotemark{b}} &
\colhead{Angle\tablenotemark{c}} & \colhead{Time\tablenotemark{d}} &
\colhead{Counts\tablenotemark{e}} & \colhead{Counts\tablenotemark{e}} &
\colhead{Energy (keV)} &
\colhead{Ratio\tablenotemark{f}} & \colhead{$\Gamma_{\rm eff}$\tablenotemark{f}} &
\colhead{$\mbox{erg\ s}^{-1}$} &
\colhead{$\Gamma_{\rm int}$\tablenotemark{h}} &
\colhead{cm$^{-2}$\tablenotemark{h}} &
\colhead{$\mbox{erg\ s}^{-1}$\tablenotemark{i}}
}
\startdata
011.1 & 2.679$^1$ & CDF-S M 197 & $ 8.2\arcmin $ &   2.16 Ms & 265 & 297 & $ 1.84$--$29.43 $ & $ 0.77^{+0.17}_{-0.15} $ & $ 1.10^{+0.36}_{-0.30} $ & 43.5 & $ 1.89^{+0.64}_{-0.56} $ & $ 22.4^{+11.4}_{-12.4} $ & 44.1 \\ 
017.1 & 2.035$^1$ & CDF-S M 131 & $ 5.4\arcmin $ &   3.04 Ms &  46 &  83 & $ 1.52$--$24.28 $ & $ 2.91^{+4.92}_{-1.46} $ & $ <0.94 $ & 42.4 & (1.80) & 25.7 & 43.1 \\ 
045.1 & $2.34^{+0.26}_{-0.67}$$^2$ & CDF-S M 348 & $ 4.1\arcmin $ &   3.36 Ms &  21 &  37 & $ 1.67$--$26.72 $ & $ <0.86 $ & (1.40) & 42.2 & (1.80) & $ < 6.6 $ & $<42.5$ \\ 
057.1 & 2.940$^3$ & CDF-S M  34 & $ 9.4\arcmin $ &   1.44 Ms & 237 & 306 & $ 1.97$--$31.52 $ & $ 0.87^{+0.22}_{-0.19} $ & $ 0.99^{+0.38}_{-0.31} $ & 43.7 & $ 1.88^{+0.78}_{-0.61} $ & $ 24.7^{+19.7}_{-13.7} $ & 44.3 \\ 
066.1 & 1.310$^1$ & E-CDF-S M 725 & $ 7.1\arcmin $ & 208 ks & 676 &  53 & $ 1.15$--$18.48 $ & $ 0.34^{+0.03}_{-0.03} $ & $ 1.92^{+0.18}_{-0.16} $ & 44.5 & $ 2.03^{+0.11}_{-0.14} $ & $ < 0.2 $ & $<44.6$ \\ 
067.1 & 2.122$^1$ & CDF-S A  & $ 7.6\arcmin $ &   3.05 Ms &  38 & 310 & $ 1.56$--$24.98 $ & $ <6.51 $ & (1.40) & 42.4 & (1.80) & $ <56.7 $ & $<43.0$ \\ 
070.1 & 2.325$^1$ & E-CDF-S M 146 & $ 3.5\arcmin $ & 222 ks &  15 &   7 & $ 1.66$--$26.60 $ & $ 2.42^{+5.69}_{-1.48} $ & (1.40) & 43.2 & (1.80) & 27.8 & 43.7 \\ 
073.1 & 4.762$^4$ & CDF-S M 403 & $ 7.9\arcmin $ &   2.85 Ms &  82 & 321 & $ 2.88$--$46.10 $ & $ 2.32^{+3.38}_{-1.21} $ & $ <1.46 $ & 43.7 & (1.80) & 85.4\tablenotemark{j} & 44.2 \\ 
084.1 & 2.259$^1$ & CDF-S M  48 & $ 7.9\arcmin $ &   2.88 Ms & 224 &  89 & $ 1.63$--$26.07 $ & $ 1.34^{+0.40}_{-0.32} $ & $ 0.62^{+0.39}_{-0.33} $ & 43.0 & $ 1.39^{+1.56}_{-0.84} $ & $ 17.1^{+44.4}_{-15.1} $ & 43.5 \\ 
114.2 & 1.606$^1$ & CDF-S M  31 & $ 9.0\arcmin $ &   1.46 Ms & 126 & 310 & $ 1.30$--$20.85 $ & $ 4.59^{+7.53}_{-1.92} $ & $ <0.32 $ & 42.8 & $ 0.35^{+0.93}_{-0.78} $ & $  4.8^{+14.7}_{- 4.8} $ & 42.9
\enddata
% redshifts
\tablenotetext{a}{Redshifts for the multiwavelength counterparts of
  these \xray\ detected SMGs. Except for ALESS 45.1, all the redshifts
  listed are spectroscopic redshifts. The superscript on each redshift
  indicates its reference: [1] zLESS spec-z \citep{Danielson2013};
  [2] \cite{Simpson2013}; [3] \cite{Zheng2004} spec-z; [4] \cite{Vanzella2008}
  spec-z.}
% catalog ID
\tablenotetext{b}{For sources in both the 4~Ms \cdfs\ and 250~ks
  \ecdfs\ catalogs (see Table~\ref{tab:basic}), we use the \xray\ data
  from the \cdfs\ as they have longer exposure time and more
  counts. Hence, their \cdfs\ IDs are listed here.}
% off axis
\tablenotetext{c}{Angular distance in arcminutes from the source to the
  average aim point of the \cdfs\ (X11), or to the aim point of the
  \ecdfs\ sub-field where the source was detected in (L05). }
% exp time
\tablenotetext{d}{Full-band (0.5--8.0~keV) effective exposure time in
  mega-seconds (Ms) or kilo-seconds (ks) as in X11 (for \cdfs\ sources)
  or L05 (for \ecdfs\ sources).}
% counts
\tablenotetext{e}{Net counts and background counts within source
  aperture in the full band as calculated by X11 and L05.}
% band ratio, gamma_eff
\tablenotetext{f}{Observed hardness ratio (photon counts ratio between
  hard 2--8 keV band and soft 0.5--2 keV band) and effective photon
  index. The hardness ratio is estimated using the {\tt BEHR} package by
  \cite{Park2006}, and $\gammaeff$ is derived from hardness ratio
  following X11. For the hardness ratios, the error bars are 1$\sigma$
  (68.3\% posterior CI), and 90\% posterior CI upper limits are provided if the mode of the
  posterior distribution is nearly 0, meaning the hardness ratio is badly
  constrained. For $\gammaeff$, the error bars are 90\%, and following
  criteria in L05 and X11, for sources with low counts in soft band or
  hard band or both, $\gammaeff$ values are given as 90\% confidence upper limits
  or 90\% confidence lower limits or set to be 1.4, respectively.}
% lx
\tablenotetext{g}{Rest-frame \fbrange\ apparent luminosity ($\lx$),
  calculated using observed \fbrange\ flux, redshift, and
  $\Gamma_{\rm eff}$ following X11. These have not been
  corrected for any absorption effects. See Section~\ref{sec:xspec1}}
% nh
\tablenotetext{h}{Intrinsic photon index $\Gamma_{\rm int}$ and
  intrinsic column density $N_{\rm H}$ derived from \xray\ spectral
  analyses (see Section~\ref{sec:xspec2}).  The sources whose
  $\Gamma_{\rm int}$ values are `(1.80)' are the ones with full-band
  counts less than 100 and thus not qualified for a proper spectral
  fitting in {\tt XSPEC}. Their $N_{\rm H}$ values were derived using
  {\tt XSPEC} simulations using a {\tt wabs*zwabs*zpow} model with
  $\Gamma_{\rm int}$ fixed at 1.8 and varying $N_{\rm H}$ values until
  the model produces the observed hardness ratio. See
  Fig.~\ref{fig:hard} for a simple illustration of this. The
  $\Gamma_{\rm int}$ and $N_{\rm H}$ values of the other five sources
  are from \xray\ spectral fits (see Fig.~\ref{fig:spec}). Error bars
  reported are for the 90\% confidence intervals. For sources with
  upper limits on hardness ratios (ALESS 45.1 and 67.1) and ALESS 66.1
  which appears to be unabsorbed, 90\% confidence upper limits are
  given.}
% lxint
\tablenotetext{i}{Rest-frame \fbrange\ absorption-corrected
  luminosity ($\lxint$), corrected for both intrinsic absorption and Galactic
  absorption, with Galactic column density $8.8\times
  10^{19}$~cm$^{-2}$ for the \ecdfs\ line of sight. Calculated
  following the method in Section~4.4 of X11. Again 90\% confidence upper limits are
  given for ALESS 45.1, 66.1 and 67.1.}
% 73 nH
\tablenotetext{j}{\cite{Gilli2011} estimated the column density for
ALESS 73.1 to be $>10^{24}$ cm$^{-2}$. They used 3 different models
({\tt XSPEC plcabs, pexrav}, and the MYTorus model by Murphy \& Yaqoob
2009) and found consistent results. } 
\end{deluxetable*}
\end{center}
\end{turnpage}
%----------------------------------------------------------------

%----------------------------------------------------------------
% Table: Multiwavelength properties and classification
\renewcommand{\arraystretch}{1.6} % more row spacing for the table
\begin{turnpage}
\begin{center}
\begin{deluxetable*}{ccccccrcccccc}
%\rotate 
\tabletypesize{\scriptsize}
\tablewidth{0pt}
\tablecaption{Multiwavelength Properties and Classification of \xray\
  Detected SMGs\label{tab:multiv}}
\tablehead{
%-------------  
% FIRST ROW
% basic stuff
\colhead{ALESS} &   % 1
% ir-radio
\colhead{$3.6\mu$m\tablenotemark{a}} & \colhead{$\log{L_{\rm IR}}$\tablenotemark{b}} &
\colhead{$\log{L_{\rm FIR}}$\tablenotemark{b}} & \colhead{$\log{L_{\rm 1.4GHz}}$\tablenotemark{c}} & % 4
% mass sfr
\colhead{${M_\star}$\tablenotemark{d}} & \colhead{SFR\tablenotemark{e}} & % 2
% classification
\multicolumn{5}{c}{Classification\tablenotemark{f}} & \colhead{Has} \\ \cline{8-12} % 6
%-------------
% SECOND ROW
% basic stuff
\colhead{ID} &  % 1
% ir-radio
\colhead{AB mag} & \colhead{${L_\odot}$} &
\colhead{${L_\odot}$} & \colhead{${\rm W\ Hz^{-1}}$} & % 4
% mass sfr
\colhead{$10^{11}\ {M_\odot}$} & \colhead{$M_\odot$\peryr} & % 2
% classification
\colhead{\sc I} & \colhead{\sc II} & \colhead{\sc III} &
\colhead{\sc IV} & \colhead{\sc V} & \colhead{AGN?\tablenotemark{f}} % 6
}
\startdata
011.1 &  21.8 & $12.90^{+ 0.12}_{- 0.04}$ & $12.81^{+ 0.12}_{- 0.04}$ & 24.42 & $ 3.5\pm 1.3$ & $ 1420^{+ 340}_{- 130}$ & N & Y & Y & Y & N & Y  \\
017.1 &  20.0 & $12.21^{+ 0.03}_{- 0.11}$ & $12.00^{+ 0.03}_{- 0.13}$ & 24.48 & $ 1.1\pm 0.5$ & $  290^{+  20}_{-  80}$ & Y & N & N & N & N & Y  \\
045.1 &  21.2 & $12.55^{+ 0.07}_{- 0.02}$ & $12.45^{+ 0.07}_{- 0.02}$ & 24.03 & $ 3.0\pm 1.1$ & $  630^{+  90}_{-  40}$ & N & N & N & N & N & ?  \\
057.1 &  21.6 & $12.64^{+ 0.08}_{- 0.07}$ & $12.54^{+ 0.08}_{- 0.08}$ & 24.46 & $ 1.4\pm 1.0$ & $  790^{+ 130}_{- 140}$ & Y & Y & Y & Y & N & Y  \\
066.1 &  19.0 & $12.51^{+ 0.10}_{- 0.06}$ & $12.42^{+ 0.11}_{- 0.06}$ & 23.77 & $ 4.8\pm 3.4$ & $  580^{+ 120}_{-  80}$ & N & Y & Y & Y & N & Y  \\
067.1 &  20.2 & $12.72^{+ 0.12}_{- 0.06}$ & $12.62^{+ 0.12}_{- 0.06}$ & 24.40 & $ 2.1\pm 1.5$ & $  950^{+ 230}_{- 130}$ & N & N & N & N & N & ?  \\
070.1 &  20.2 & $12.90^{+ 0.07}_{- 0.04}$ & $12.83^{+ 0.07}_{- 0.05}$ & 25.04 & $ 2.1\pm 1.5$ & $ 1420^{+ 220}_{- 150}$ & N & N & Y & Y & N & Y  \\
073.1 &  22.6 & $12.75^{+ 0.09}_{- 0.12}$ & $12.65^{+ 0.09}_{- 0.14}$ & 24.51 & $ 1.3\pm 0.3$ & $ 1000^{+ 190}_{- 320}$ & N & Y & Y & Y & N & Y  \\
084.1 &  21.0 & $12.43^{+ 0.13}_{- 0.05}$ & $12.33^{+ 0.14}_{- 0.05}$ & 24.03 & $ 0.7\pm 0.5$ & $  480^{+ 130}_{-  60}$ & Y & Y & Y & Y & Y & Y  \\
114.2 &  19.6 & $12.42^{+ 0.05}_{- 0.14}$ & $12.32^{+ 0.05}_{- 0.15}$ & 24.14 & $ 1.8\pm 0.6$ & $  470^{+  50}_{- 190}$ & Y & Y & Y & Y & N & Y
\enddata
\tablenotetext{a}{AB magnitudes at $3.6\mu$m (IRAC Channel 1) from
  \cite{Simpson2013}.}
\tablenotetext{b}{8--1000 $\mu$m luminosity ($L_{\rm IR}$) and 40--120
  $\mu$m luminosity ($L_{\rm FIR}$) from \cite{Swinbank2013} based
  on IR-through-radio SED fitting.}
\tablenotetext{c}{Rest-frame 1.4 GHz monochromatic luminosity,
  following Equation~(2) in \cite{Alexander2003} for $\alpha=0.8$
  using the radio flux density from \cite{Biggs2011}. Radio
  counterparts were matched using a search radius of $1 \arcsec$. See
  Section~\ref{sec:multiv}.} 
\tablenotetext{d}{Stellar mass from \cite{Simpson2013} estimated by
  optical-NIR SED fitting. See Section~\ref{sec:multiv}.}
\tablenotetext{e}{Star Formation Rate (SFR) estimated following
  \cite{Kennicutt1998}, using the correlation between SFR and $L_{\rm IR}$
   (from \cite{Swinbank2013}; see Section~\ref{sec:multiv} \&
  \ref{sec:class3}).} 
\tablenotetext{f}{Classification for the source to determine whether
  it hosts an AGN. `Y' means it is classified as an AGN under a
  specific classification scheme, or for the last column, means the
  source is treated as an AGN in our AGN fraction analyses. `?' in the
  last column means that we do not have sufficient evidence to
  classify the source as AGN (ones that dominate the \xray\ band), but
  cannot completely rule out the possibility either (see discussion in
  Section~\ref{sec:4567}). See Section~\ref{sec:class},
  Table~\ref{tab:classify} and Figures \ref{fig:class12},
  \ref{fig:class3}, and \ref{fig:class4} for details of the
  classification methods.}
\end{deluxetable*}
\end{center}
\end{turnpage}
%----------------------------------------------------------------

%----------------------------------------------------------------
% Table: Classification Methods
\renewcommand{\arraystretch}{1.7} % more row spacing for the table
\begin{turnpage}
\begin{center}
\begin{deluxetable*}{cccr}
\tabletypesize{\scriptsize}
\tablewidth{0pt}
\tablecaption{Classification Methods\label{tab:classify}}
\tablehead{
  \colhead{Method} & \colhead{Relevant Figure}  & \colhead{Classified as AGN if
  \ldots} & \colhead{Reference}
}
\startdata
{\sc I} & \ref{fig:class12} & $\Gamma_{\rm eff} < 1.0$ &
\cite{Alexander2005b, Xue2011} \\
{\sc II} & \ref{fig:class12} & $L_{\rm 0.5-8\ keV,corr}>3\times
10^{42}$ erg s$^{-1}$ & \cite{Bauer2004, Lehmer2008, Xue2011} \\
{\sc IIIa} & \ref{fig:class3}a & $\lx>5\times \lx$ as &
\cite{Persic2007, Lehmer2010}; \\
 & & expected from $L_{\rm 1.4\ GHz}$ (SFR) & \cite{Alexander2005b, Xue2011} \\
{\sc IIIb} & \ref{fig:class3}b & $\lx>5\times (\lx = \alpha
M_{\star}+\beta$SFR$)$ & \cite{Lehmer2010} \\
{\sc IV} & \ref{fig:class4} & $\log{(f_{\rm X}/f_{\rm 3.6\mu m})} >
-1$ & This paper --- see Section~\ref{sec:class4}\\
{\sc V} & \nodata & \xray\ Variable & \cite{Young2012}
\enddata
\end{deluxetable*}
\end{center}
\end{turnpage}
%----------------------------------------------------------------

%----------------------------------------------------------------
% Table: fAGN values
\renewcommand{\arraystretch}{1.7} % more row spacing for the table
\begin{turnpage}
\begin{center}
\begin{deluxetable*}{cccc}
%\rotate
\tabletypesize{\scriptsize}
\tablewidth{0pt}
\tablecaption{AGN Fractions\label{tab:fagn}}
\tablehead{
  \colhead{} & \colhead{$f_{\rm X}>f_{\rm X,min}$\tablenotemark{a}} &
  \colhead{$L_{\rm X}>L_{\rm X,min}$\tablenotemark{a}} &
  \colhead{$L_{\rm X,corr}>L_{\rm X,corr,min}$\tablenotemark{a}}
}
\startdata
All ALESS $\fagn$ (\%) &
$ 14.9^{+14.8}_{- 5.5} $ & 
$ 15.5^{+15.3}_{- 5.7} $ & 
$ 17.0^{+15.9}_{- 5.9} $ \\
$S_{\rm 870\mu m}>3.5$ mJy $\fagn$ &
$ 21.0^{+23.7}_{- 9.5} $ & 
$ 21.6^{+24.9}_{-10.2} $ & 
$ 18.1^{+20.1}_{- 7.7} $ 
\enddata
\tablenotetext{a}{The values for these flux or luminosity limits are
  $f_{\rm X,min}=2.7\times 10^{-16}$ erg cm$^{-2}$ s$^{-1}$, $L_{\rm
    X,min} = 2.6\times 10^{42}$ erg s$^{-1}$, and $L_{\rm X,corr,min}
  = 7.8\times 10^{42}$ erg s$^{-1}$ (or $L_{\rm X,corr,min} =
  1.2\times 10^{43}$ erg s$^{-1}$ for $S_{\rm 870\mu m}>3.5$
  mJy). These are the \xray\ flux/luminosity values of the faintest
  SMG-AGNs in our sample. See Section~\ref{sec:fagn} for details.}
\end{deluxetable*}
\end{center}
\end{turnpage}
%----------------------------------------------------------------

%----------------------------------------------------------------
% Table: Stacking results
\renewcommand{\arraystretch}{1.7} % more row spacing for the table
\begin{turnpage}
\begin{center}
\begin{deluxetable*}{ccccccccc}
%\rotate
\tabletypesize{\scriptsize}
\tablewidth{0pt}
\tablecaption{Stacking Results\label{tab:stack}}
\tablehead{
  \colhead{Stacking} & \colhead{Number} &
  \colhead{Full-Band} & \colhead{} &
  \colhead{Hardness} & \colhead{} &
  \colhead{Mean Full-Band Flux} & \colhead{Median} &
  \colhead{Mean Rest-Frame Luminosity} \\
  \colhead{Sample} & \colhead{of SMGs} &
  \colhead{Net Counts} & \colhead{S/N} &
  \colhead{Ratio} & \colhead{$\gammaeff$} &
  \colhead{$\fx$ (\ergcms)} & \colhead{Redshift} &
  \colhead{$\lx$ (\ergs)}
}
\startdata
All               & 49 & $33.2^{+10.5}_{-9.5}$ & $4.5\sigma$ &
$0.6^{+0.6}_{-0.4}$ & $1.3^{+1.2}_{-0.6}$ & $5.5\times10^{-17}$ &
2.6 & $1.2\times10^{42}$ \\

$\rri \geq 0.5$\tablenotemark{a} &  4 & $15.5^{+5.6}_{-4.4}$  & $7.4\sigma$ &
$1.0^{+0.8}_{-0.5}$ & $0.8^{+0.6}_{-0.5}$ & $3.6\times10^{-16}$ &
2.2 & $3.6\times10^{42}$ \\

$\rri < 0.5$      & 45 & $17.7^{+9.4}_{-8.4}$  & $2.5\sigma$ &
$0.4^{+0.8}_{-0.4}$ & $1.6^{+2.3}_{-1.0}$ & $2.7\times10^{-17}$ &
2.9 & $1.2\times10^{42}$ 
\enddata
\tablenotetext{a}{$\rri$ is the rest-frame radio-to-IR luminosity
  ratio, $L_{\rm 1.4GHz}/L_{\rm IR}$. See Section~\ref{sec:stack} for
  more details.}
\end{deluxetable*}
\end{center}
\end{turnpage}
%----------------------------------------------------------------

\end{CJK*}

\end{document}